\documentclass[preprint,times,12pt]{elsarticle}

\usepackage{amssymb}
\usepackage{amsthm}
\usepackage{amsmath}
\usepackage{hyperref}
\usepackage{longtable}
\usepackage{float}
\usepackage{subcaption}
\usepackage[margin=1in]{geometry}

\journal{Environmental Research: Energy}

\begin{document}

\begin{frontmatter}



\title{Reducing transmission expansion by co-optimizing sizing of wind, solar, storage
and grid connection capacity}

\author[4,1]{Aneesha Manocha\corref{cor1}}
%%\ead{aneesha.manocha@gmail.com}

\cortext[cor1]{Corresponding Author}

\author[1,3]{Gabriel Mantegna}

\author[2]{Neha Patankar}

\author[1,3]{Jesse D. Jenkins}

\affiliation[4]{organization={Energy Resources Group},
addressline={University of California, Berkeley},
city={Berkeley},
state={CA},
country={USA}}

\affiliation[1]{organization={Andlinger Center for Energy and the Environment},
addressline={Princeton University},
city={Princeton},
state={NJ},
country={USA}}

\affiliation[2]{organization={Department of Systems Science and Industrial Engineering},
addressline={State University of New York at Binghamton University},
city={Binghamton},
state={NY},
country={USA}}

\affiliation[3]{organization={Department of Mechanical and Aerospace Engineering},
addressline={Princeton University},
city={Princeton},
state={NJ},
country={USA}}


\begin{abstract}

Expanding transmission capacity is likely a bottleneck that will restrict variable renewable energy (VRE) deployment required to achieve ambitious emission reduction goals. Interconnection and inter-zonal transmission buildout may be displaced by the optimal sizing of VRE to grid connection capacity and by the co-location of VRE and battery resources behind interconnection. However, neither of these capabilities is commonly captured in macro-energy system models. We develop two new functionalities to explore the substitutability of storage for transmission and the optimal capacity and siting decisions of renewable energy and battery resources through 2030 in the Western Interconnection of the United States. Our findings indicate that modeling optimized interconnection and storage co-location better captures the full value of energy storage and its ability to substitute for transmission. Optimizing interconnection capacity and co-location can reduce total grid connection and shorter-distance transmission capacity expansion on the order of 10\% at storage penetration equivalent to 2.5-10\% of peak system demand. The decline in interconnection capacity corresponds with greater ratios of VRE to grid connection capacity (an average of 1.5-1.6 megawatt (MW) PV:1 MW inverter capacity, 1.2-1.3 MW wind:1 MW interconnection). Co-locating storage with VREs also results in a 10-15\% increase in wind capacity, as wind sites tend to require longer and more costly interconnection. Finally, co-located storage exhibits higher value than standalone storage in our model setup ($\sim$22-25\%). Given the coarse representation of transmission networks in our modeling, this outcome likely overstates the real-world importance of storage co-location with VREs. However, it highlights how siting storage in grid-constrained locations can maximize the value of storage and reduce transmission expansion.

\end{abstract}

%%\includegraphics[scale=0.4]{images/colocated-methods.png}


%%\item Strategic battery siting and increased buildout reduce required transmission upgrades. 

%%\item Grid connection declines when co-optimizing and co-locating VREs and batteries.

%%\item Co-optimizing and co-locating VREs and storage yield larger ratios of VRE to grid connection capacity.

%%\item Co-located storage exhibits greater value than standalone by substituting for transmission interconnection.



\end{frontmatter}


\section{Introduction}
\label{intro}

    The most cost-effective scenarios for the deep decarbonization of the electricity sector involve significant expansion of wind and solar photovoltaic (PV) capacity and associated buildout of high-voltage electricity transmission to connect renewable energy projects to demand centers concentrated in populated areas \cite{williams, BROWN2021115, RODRIGUEZSARASTY2021112210, FREW201665}. For example, Larson et al. 2021 estimate the need for approximately 400-750 gigawatts (GW) of wind and solar PV capacity deployment and a roughly 60\% increase in transmission capacity by 2030 to put the United States on track to reach net-zero greenhouse gas emissions by 2050 \cite{andlinger_center_for_energy_and_the_environment_acee_net-zero_2020}. In practice, however, long- and short-distance electricity transmission and interconnection projects take years to site, permit, and build. Many projects face significant delays due to public opposition from various communities or fail to acquire necessary permits from all relevant jurisdictions along their routes \cite{klass, BUIJS20111794, mit_grid, trans_europe, jenkins_net_zero}. As such, expanding both long- and short-distance transmission capacity is likely to constitute a bottleneck that could constrain the rate of decarbonization.
    
    Given the practical challenge and economic cost of transmission expansion, it is prudent to design variable renewable energy (VRE) projects to effectively utilize transmission connections. Co-optimizing VRE and grid connection sizing (optimally oversizing the solar PV or wind to inverter or transmission interconnection capacity needed) and co-locating VRE and battery resources (siting VRE and battery technologies at the same physical interconnection location) can lower capital costs and increase the value of projects \cite{nrel_pv_battery_2018, nrel_pv_battery_2021, nrel_pv_battery_capacity_expansion_2020, GORMAN2020106739, osti_1808491, osti_eval}. Solar projects entering the interconnection queue are already seeing larger ratios of solar to inverter and interconnection capacity, while wind projects are witnessing the decline of specific power \cite{osti_1818277, osti_1888246}. Meanwhile, as of the end of 2023, over half of all battery storage active in the U.S. interconnection queues (or 525 GW) were `hybrid' projects co-located with solar PV or wind at the same site \cite{lbnl}.

    Prior studies have investigated the potential and value of co-located resources compared to independently-sited resources \cite{GORMAN2020106739, GORMAN2022105832, VIRGUEZ2021116120}. Other studies have demonstrated the potential substitutability of energy storage and transmission capacity \cite{BROWN2018720, VICTORIA2019111977, LIU20191308, CONLON20191085,  MALLAPRAGADA2020115390}. However, most power system capacity expansion models do not optimize VRE to interconnection sizing or allow for the co-location of VREs and batteries, likely biasing the estimation of necessary transmission reinforcements produced by such models \cite{GORMAN2020106739, jenkins_enhanced_2017}. VREs have been traditionally modeled by fixing the grid connection and inverter capacity built for the capacity of solar PV or wind technologies \cite{nrel_pv_battery_2018, nrel_pv_battery_capacity_expansion_2020, osti_1808491}. For example, \cite{BROWN2018720, MALLAPRAGADA2020115390, CHILD201980} all model solar PV with a fixed inverter loading ratio (ILR) (the ratio of direct current (DC) solar capacity to alternating current (AC) inverter and grid connection capacity) of 1.3:1 and assume all wind projects have a grid connection equal to their generating capacity. None of these studies consider the co-location of storage with wind and solar capacity. On a site-by-site basis, the optimal ratio of VRE to grid connection capacity and co-located storage capacity may vary significantly from these assumed values based on resource quality, distance of site to demand centers, availability of nearby substations, and the evolving generation resource portfolio \cite{ MARTINSDESCHAMPS2019106, SCHLEIFER2021100015}. Without being able to optimally site and size these resources, conventional modeling approaches likely overestimate the requisite transmission expansion, undervalue energy storage, and bias estimates of optimal generation and storage portfolios. More detailed modeling can permit us to capture granular trade-offs and trends that optimizing VRE sizing and co-locating resources may have in selecting the locations and type of technologies to build in decarbonization pathways.

    While \cite{nrel_pv_battery_capacity_expansion_2020} enables the modeling of `hybrid,' DC-coupled solar PV and storage resources in a capacity expansion planning context, no study to our knowledge analyzes 1) the independent sizing of both the inverter and grid connection capacities relative to each solar PV and wind generator, and 2) the ability to site any combination of solar PV, wind, and/or storage resources (both short- and long-duration) behind a single interconnection point. Thus, the goal of this study is to investigate whether optimized sizing of renewable energy and transmission interconnection capacity and co-location of VRE resources and storage capacity can decrease transmission reinforcements needed for a low-carbon future and alter the optimal generation and storage capacity decisions. 
    
    In this work, we improve an existing, Julia-based, open-source electricity capacity expansion model, GenX, to optimize VRE to interconnection sizing and co-locate VRE and battery resources \cite{genx_github}. The comprehensive and granular modeling of VREs enables the exploration of trade-offs between transmission and battery capacity and the optimal locations of VRE siting. We test the model framework with a fourteen-zone case study of the Western Interconnection (WECC) for 2030, analyzing the impacts of optimizing resource sizing and co-location on VRE, transmission, and battery buildout in near-term decarbonization goals. We run three main scenarios (see Table \ref{tab:scenarios}): 1) fixed interconnection (fixed ratios of VRE to grid connection capacity with no co-location of storage), as is common practice in most capacity expansion models, 2) optimized interconnection (model selects optimal ratios of VRE to grid connection capacity for each wind and solar resource cluster in the VRE supply curve with no co-location of storage), and 3) co-located storage (model can co-optimize and co-locate VREs, battery resources, and grid interconnection capacity for each cluster in the VRE supply curve) under four system-wide battery buildout requirements (3.75, 5, 7.5, and 15 GW) and two VRE and battery cost projections (mid-range and low-cost scenarios for solar, wind, and batteries). We additionally downscale the aggregated results, processing the GenX outputs to 1) 4 kilometers (km) by 4 km candidate project areas (CPAs) based on the levelized cost of electricity inclusive of interconnection costs, and 2) intra-zonal interconnection capacity (short-distance site-to-substation and intra-zonal transmission-to-metro lines within a modeled zone) and inter-zonal long-distance transmission capacity (spanning two or more modeled zones) \cite{repeat, downscaling}. 

    Section \ref{methods} dives into the methodology of the study, highlighting the model formulation of co-located resources, case study overview, assumptions, cost analysis, and scenarios. Section \ref{results} is structured to analyze two different scenarios. First, we investigate how enabling the model to select how much inverter and grid connection capacity to build for each VRE site will alter the optimal sizing of these resources, technology buildout, and transmission expansion. Second, we examine how adding the option to co-locate solar PV or wind with batteries changes resource siting decisions, capacity deployment, and transmission buildout. We also analyze the value of co-location and how batteries are sited in near-term decarbonization strategies. Section \ref{conclusion} synthesizes our main takeaways on the substitutability of storage and transmission assets and resource mix trade-offs with the ability to co-locate resources and optimally size interconnection. Section \ref{conclusion} also discusses the implications of this detailed modeling for electricity system planning and macro-energy system modeling practices.

\section{Methods}
\label{methods}

    \subsection{GenX Model Formulation}
        \begin{figure}[t]
            \makebox[\textwidth]{\includegraphics[scale=0.4]{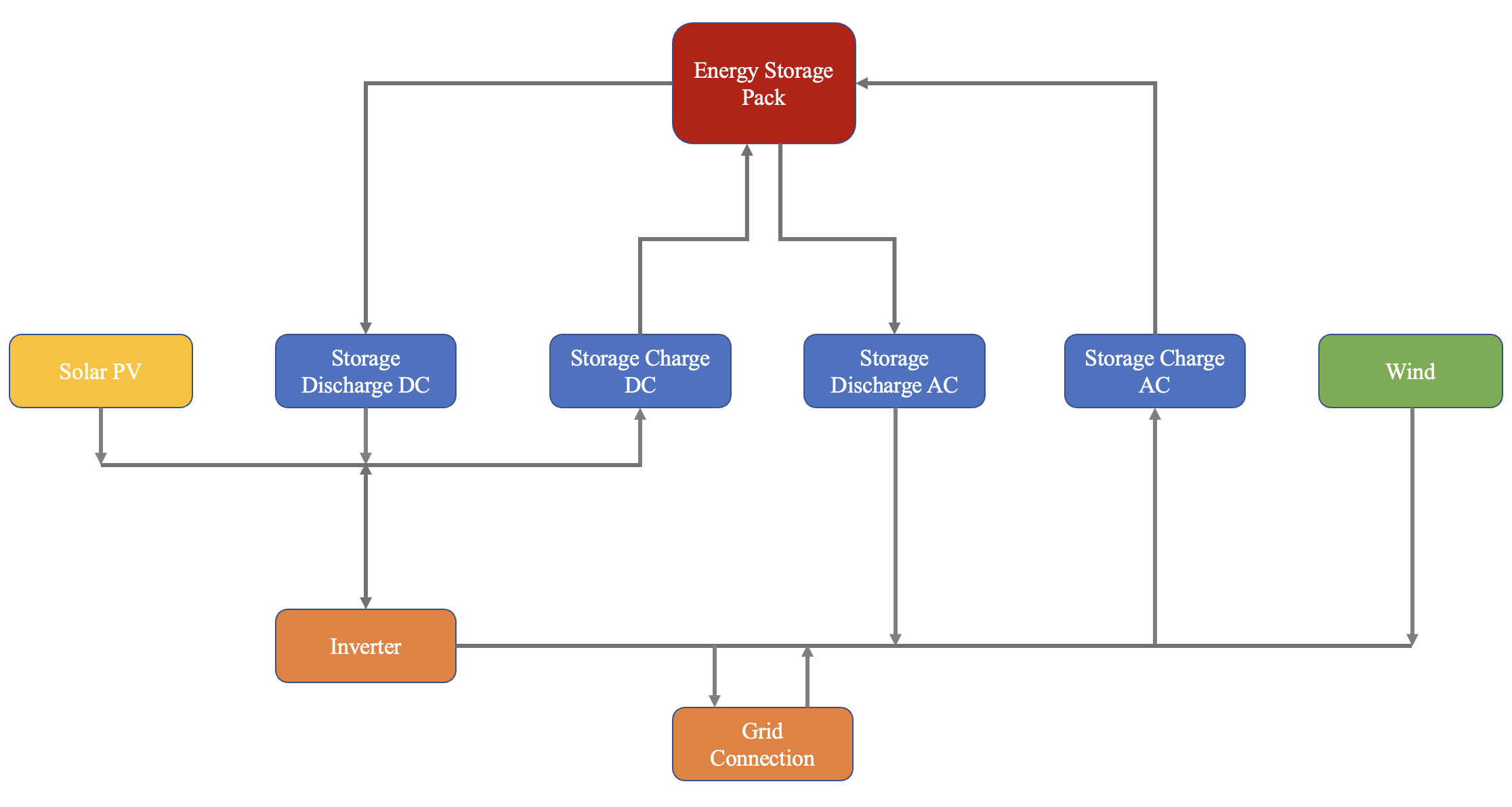}}
            \caption{\textbf{GenX Co-Location and Optimized Interconnection Module Overview for Each Configurable Resource.}\\ This figure was adapted from \cite{nrel_pv_battery_2021}.}
            \label{fig::methods}
        \end{figure}

        GenX is an open-source, power system capacity expansion model created to analyze low-carbon electricity systems. GenX is configurable to model different resolutions along the dimensions of time, unit commitment constraints, and geospatial and network representation, as suited for various studies. More information about GenX and the implementation can be found on GitHub \cite{genx_github} and in an existing working paper \cite{jenkins_enhanced_2017}.

        While GenX and other similar capacity expansion models are capable of modeling thermal generators, VRE resources, hydroelectric generators, energy storage technologies, and many other conventional and advanced resources, most have been unable to model 1) optimized interconnection sizing for VREs, and 2) co-located VRE and energy storage technologies \cite{GORMAN2020106739, jenkins_enhanced_2017}. We implemented a new technology module that provides more detailed modeling of VREs and energy storage in GenX using the Julia programming language and JuMP package for mathematical programming \cite{genx_github, Julia, DunningHuchetteLubin2017}. Fig. \ref{fig::methods} showcases how each component interacts in the configurable module. The mathematical formulation of constraints added to this module is listed and described below. All definitions of sets, constants, and decision variables can be found in \ref{model_notation}, respectively in Tables \ref{indices}, \ref{constants}, and \ref{variables}.

        There are nine configurable capacity decision variables for a co-located resource: solar PV, wind, inverter, grid interconnection, storage DC discharge, storage DC charge, storage AC discharge, storage AC charge, and storage energy capacity components. This modular representation of storage capacity components is flexible and permits the modeling of any type of storage resource, including electrochemical, chemical, thermal, and mechanical systems and resources with symmetric charge and discharge capacities (e.g. electrochemical batteries) and asymmetric charge and discharge capacities (e.g. hydrogen electrolysis, compressed hydrogen storage, and fuel cell or combustion turbine). Grid connection and inverter decision variables are added to this new formulation to create a more detailed representation and separation of the DC and AC components. In the context of this module, the terms `grid connection' and `interconnection capacity' are used interchangeably and reference both the spur lines (short-distance site-to-substation lines) and the intra-zonal transmission-to-metro lines required to connect VRE sites to demand centers. The inverter is sized independently from the `grid connection' or `interconnection capacity' when a resource incorporates a solar PV or storage DC component. Furthermore, there are six operational decision variables for a co-located resource: solar PV generation, wind generation, storage DC discharge, storage DC charge, storage AC discharge, and storage AC charge components.

        Renewable energy and storage resources can be co-located in two ways: 1) DC-coupled, and 2) AC-coupled. DC-coupled co-located solar PV and battery resources sit behind a single bidirectional inverter, where the battery can charge directly from the solar PV technology or from the grid. Meanwhile, AC-coupled co-located solar and battery resources both sit behind their respective inverters and share a single interconnection point. More information on the benefits and drawbacks of each configuration is described by \cite{nrel_pv_battery_2018, nrel_pv_battery_2021}. The co-located module in GenX enables both configurations of DC-coupled and AC-coupled resources to be built. In this paper, we focus on electrochemical battery storage and assume DC-coupled solar and storage systems.

        \textbf{Constraints}: The following constraints are created for the set of resources within the co-location module. For each constraint, let the set $C$ denote each component at a co-located site such that $C = {grid, pv, wind, storage, inverter, charge_{dc}, discharge_{dc}, charge_{ac}, discharge_{ac}}$. These constraints below are listed for co-located resources without any capacity reserve margin, long-duration energy storage inter-temporal linkages, or operating reserves. For more information on how the formulation may change with these additions, see \cite{genx_github}.

    \begin{enumerate}
    
        \item \textbf{Total Capacity:} The total capacity for each component $c \in C$ of each resource $y \in Y$ is the sum of the existing capacity plus the new installed capacity minus any retired capacity.
        \begin{equation}
        \Delta^{total, c}_{y, z} = \overline{\Delta}^{c}_{y, z} + \Omega^{c}_{y, z} - \Delta^{c}_{y, z}, \quad \forall y \in \mathcal{VS}, z \in Z, c \in C
        \end{equation}
        
        \item \textbf{Retired Capacity:} The retired capacity for each component $c \in C$ of each resource $y \in Y$ cannot exceed existing capacity for each component.
        \begin{equation}
            \Delta^{c}_{y, z} \leq \overline{\Delta}^{c}_{y, z},
            \quad \forall y \in \mathcal{VS}, z \in Z, c \in C
        \end{equation}
            
        \item \textbf{Maximum Capacity:} A maximum capacity can be defined for each component $c \in C$ of a resource $y \in Y$.
        \begin{equation}
            \Delta^{total, c}_{y, z} \leq \overline{\Omega}^{c}_{y, z},
           \quad  \forall y \in \mathcal{VS}, z \in Z, c \in C
        \end{equation}
            
        \item \textbf{Minimum Capacity:} A minimum capacity can be defined for each component $c \in C$ of a resource $y \in Y$.
        \begin{equation}
            \Delta^{total, c}_{y, z} \geq \underline{\Omega}^{c}_{y, z},
           \quad  \forall y \in \mathcal{VS}, z \in Z, c \in C
        \end{equation}
        
        \item \textbf{Ratio of VRE to Inverter or Grid Connection Sizing:} The ILR between the solar PV component $pv \in C$ and inverter capacity $inv \in C$ and the sizing of the wind component $wind \in C$ to grid connection capacity $grid \in C$ can be constrained to a fixed ratio to mimic behavior of conventional models where this ratio is not endogenously optimized. If this ratio is set to -1 in the input parameters for a given VRE cluster, this constraint is omitted and the module can optimize the capacity of interconnection to the capacity of VRE resources within that cluster.
        \begin{subequations}
        \begin{equation}
            \Delta^{total, pv}_{y, z} = \eta^{ILR,pv}_{y, z} \times \Delta^{total, inv}_{y, z},
            \quad \forall y \in \mathcal{VS}^{pv}, z \in Z
        \end{equation}
        \begin{equation}
            \Delta^{total, pv}_{y, z} = \eta^{ILR,pv}_{y, z} \times \Delta^{total, grid}_{y, z},
            \quad \forall y \in \mathcal{VS}^{pv}, z \in Z
        \end{equation}
        \begin{equation}
            \Delta^{total, wind}_{y, z} = \eta^{ILR,wind}_{y, z} \times \Delta^{total}_{y, z},
            \quad \forall y \in \mathcal{VS}^{wind}, z \in Z
        \end{equation}
        \end{subequations}
            
        \item \textbf{Energy Balance:} Energy balance ensuring net DC power (discharge of battery, PV generation, and charge of battery) and net AC power (discharge of battery, wind generation, and charge of battery) is equal to the technology's total discharging to and charging from the grid.
    
        \begin{equation}
        \begin{split}
            \Theta_{y, z, t}^{grid} - \Pi_{y, z, t}^{grid} = \Theta_{y, z, t}^{wind} + \Theta_{y, z, t}^{ac} - \Pi_{y, z, t}^{ac}\\ + \eta^{inverter}_{y, z} \times (\Theta_{y, z, t}^{pv} + \Theta_{y, z, t}^{dc}) - \frac{\Pi^{dc}_{y, z, t}}{\eta^{inverter}_{y, z}}, \\ \quad \forall y \in \mathcal{VS}, z \in Z, t \in T
        \end{split}
        \end{equation} 
        
        \item \textbf{Grid Export Maximum:} The maximum grid exports and imports must be less than the grid capacity.
        \begin{equation}
            \Theta_{y, z, t}^{grid} + \Pi_{y, z, t}^{grid} \leq \Delta^{total, grid}_{y, z}, \quad \forall y \in \mathcal{VS}, z \in Z, t \in T
        \end{equation}

        \item \textbf{Inverter Export Maximum:} The maximum DC grid exports and imports must be less than the inverter capacity.
        \begin{equation}
        \begin{split}
            \eta^{inverter}_{y, z} \times (\Theta^{pv}_{y, z, t} + \Theta^{dc}_{y, z, t}) +\frac{\Pi_{y, z, t}^{dc}}{\eta^{inverter}_{y, z}} \\ \leq \Delta^{total, inv}_{y, z}, \quad \forall y \in \mathcal{VS}^{inv}, z \in Z, t \in T
        \end{split}
        \end{equation}

        \item \textbf{Generation Maximum:} The maximum power generated per hour by the solar PV and wind components must be less than the hourly capacity factor times the total installed capacity of each component.
        \begin{subequations}
            \begin{equation}
            \Theta^{pv}_{y, z, t} \leq \rho^{max, pv}_{y, z, t} \times \Delta^{total,pv}_{y, z}, 
            \quad \forall y \in \mathcal{VS}^{pv}, z \in Z, t \in T
        \end{equation}
        \begin{equation}
            \Theta^{wind}_{y, z, t} \leq \rho^{max, wind}_{y, z, t} \times \Delta^{total,wind}_{y, z}, 
            \quad \forall y \in \mathcal{VS}^{wind}, z \in Z, t \in T
        \end{equation}
        \end{subequations}
    
        \item \textbf{State of Charge (SOC):} This constraint calculates the battery component's state of charge during each time period. 
        \begin{equation}
        \begin{split}
            \Gamma_{y, z, t} = \Gamma_{y, z, t-1} + (\eta^{dc, cha}_{y, z} \times \Pi^{dc}_{y, z, t} - \frac{\Theta^{dc}_{y, z, t}}{\eta^{dc, dis}_{y, z}})\\ + (\eta^{ac, cha}_{y, z} \times \Pi^{ac}_{y, z, t} - \frac{\Theta^{ac}_{y, z, t}}{\eta^{ac, dis}_{y, z}}), 
            \\\forall y \in \mathcal{VS}^{stor}, \forall z \in Z, \forall t \in T
        \end{split}
        \end{equation}
        
        \item \textbf{SOC Maximum:} The state of charge must be constrained by the maximum energy capacity.
        \begin{equation}
            \Gamma_{y, z, t} \leq \Delta^{total, energy}_{y, z}, \forall y \in \mathcal{VS}^{stor}, \forall z \in Z, \forall t \in T
        \end{equation}
            
        \item \textbf{Symmetric Storage Constraints:} For storage resources with symmetric charge and discharge capacity and a fixed ratio of storage power-to-energy capacity (e.g. $\forall y \in \mathcal{VS}^{sym}$), the maximum storage DC and AC charge and maximum storage DC and AC discharge must be less than the power-to-energy ratio of the battery multiplied by the energy storage capacity.
        \begin{subequations}
        \begin{equation}
        \begin{split}
            \Theta^{dc}_{y, z, t} + \Pi^{dc}_{y, z, t} \leq 
            \mu^{dc, stor}_{y, z} \times \Delta^{total, energy}_{y, z}, \\ \forall y \in \mathcal{VS}^{sym,dc}, \forall z \in Z, \forall t \in T
        \end{split}
        \end{equation}
        \begin{equation}
        \begin{split}
            \Theta^{ac}_{y, z, t} + \Pi^{ac}_{y, z, t} \leq 
            \mu^{ac, stor}_{y, z} \times \Delta^{total, energy}_{y, z}, \\ \forall y \in \mathcal{VS}^{sym,ac}, \forall z \in Z, \forall t \in T
        \end{split}
        \end{equation}
        \end{subequations}
    \end{enumerate}
    
    Note that this formulation enables any study in any region to utilize these modeling capabilities in GenX. For the scope of this case study, this module is applied to one test system: the Western Interconnection.

    \subsection{Overview of Case Study}

        \begin{figure}[H]
            \makebox[\textwidth]{\includegraphics[scale=0.7]{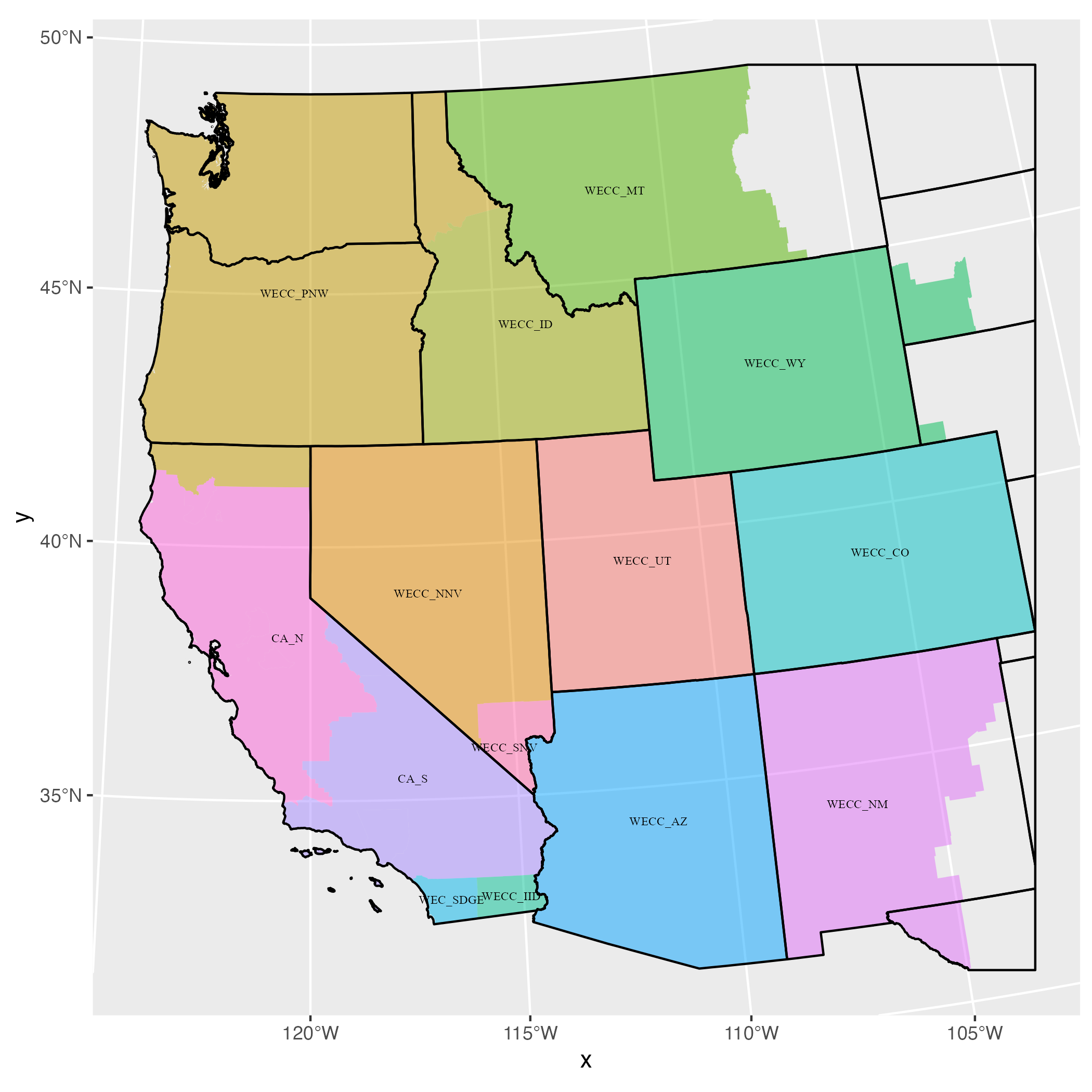}}
            \caption{\textbf{Fourteen-Region Model of the Western Interconnection.}\\ The region CA\_N includes WEC\_CALN and WEC\_BANC in northern California. The region CA\_S includes WECC\_SCE and WEC\_LADW in southern California.}
            \label{fig::wecc_map}
        \end{figure}
    
        This case study analyzes a fourteen-region model of WECC in the U.S. for 2030. The zones are aggregated based on the Environmental Protection Agency's (EPA) Integrated Planning Model (IPM) mapping. The zones are showcased in Fig. \ref{fig::wecc_map} and represent the following aggregations of IPM regions: the region CA\_N includes WEC\_CALN and WEC\_BANC in northern California; CA\_S is composed of the sub-regions WECC\_SCE and WEC\_LADW; WEC\_SDGE, WECC\_IID, WECC\_AZ, WECC\_NM, WECC\_SNV, WECC\_PNW, WECC\_ID, WECC\_MT, WECC\_NNV, WECC\_UT, WECC\_WY, and WECC\_CO are all treated as independent zones \cite{epa}. The model utilizes inter-zonal transport power flow constraints, hourly resolution of resource capacity factors and demand, and linearized unit commitment decisions for thermal generators \cite{genx_github, jenkins_enhanced_2017}. Due to the significant capacity decision differences that time domain reduction methods can cause on the buildout of co-located solar, wind, and storage resources, a full-year model with hourly resolution is run without any time domain reduction methods for all scenarios \cite{li}. Linearized unit commitment relaxes discrete unit start-up and shut-down decisions to continuous decisions, introducing limited abstraction errors. However, it enables the use of more computationally efficient linear (rather than mixed integer) programming solution algorithms, preserving the computational capacity to enable more granular modeling of other dimensions (i.e. resources, space, and time) \cite{palmintier_bryan_s_incorporating_2013}.
        
        Data and input files are compiled via PowerGenome, an open-source scenario generation tool for electricity system capacity expansion models that aggregates data from numerous sources (including the Energy Information Administration (EIA), Federal Energy Regulatory Commission (FERC), National Renewable Energy Laboratory (NREL), and EPA) to format GenX input files \cite{pg_github}. More information on PowerGenome can be found in \ref{app_pg}. The following assumptions about the data are made:
        
        \begin{itemize}    
            \item Batteries have a power-to-energy ratio equivalent to 0.25 (e.g. approximately four-hour battery charge and discharge duration from an empty/full state of charge), as is used in NREL's bottom-up analysis of a DC-coupled, co-located solar PV and lithium-ion battery site \cite{nrel_pv_battery_2021}.
            
            \item Inverter efficiency is considered to be 96\% \cite{nrel_pv_battery_capacity_expansion_2020, pg_github}. Single-trip battery charging and discharging efficiencies for four-hour batteries are assumed to be 95\% \cite{nrel_pv_battery_2021, nrel_pv_battery_capacity_expansion_2020}. The self-discharge rate of batteries is assumed to be 5\% per hour.
            
            \item The real weighted average cost of capital (WACC) for annuitizing overnight capital costs is averaged during the model planning period of 2022 to 2030 from NREL's 2022 Annual Technology Baseline (ATB), equaling 2.5\% for solar and batteries and 3.2\% for wind \cite{nrel_atb_2022}. 

            \item Existing generation and storage capacity includes all operational plants at the end of 2022 to enable GenX to determine the optimal capacity mix for 2030 without consideration of proposed projects that are under construction or undergoing review \cite{pg_github, eia860m}.
            
            \item For simplicity, the following technology resources cannot have new capacity built in the model for 2030 in the case study: geothermal, biomass, or biopower. There are options to expand wind, solar PV, battery storage, natural gas (NG) with carbon capture and sequestration (CCS) (with a 90\% carbon dioxide (CO\textsubscript{2}) capture rate), and nuclear resources, though no NG with 90\% CCS or nuclear capacity is selected in the modeling.  Capital and operating costs for new resources are based on NREL's ATB 2022 \cite{nrel_atb_2022}.
            
            \item The following technologies are eligible to meet state Renewable Portfolio Standards (RPS): onshore wind, offshore wind, solar, biomass, small hydro, and geothermal resources. The following technologies are eligible to meet the Clean Energy Standards (CES): all technologies listed above, conventional hydro, nuclear, and NG with 90\% CCS (with partial credit). Partial crediting for each gigawatt-hour (GWh) of NG with 90\% CCS is calculated as the ratio of the emission rates between new NG technologies built with and without the 90\% CCS component (88\% for partial crediting). Alternative policy rules may apply in practice and in different jurisdictions to account for NG with 90\% CCS in the CES.
            
            \item Demand profiles and projections are adapted from the NREL Electrification Futures Study with the assumption of a reference electrification scenario for 2030 and adjusted to reflect the electrification of other sectors accounting for the impact of the Inflation Reduction Act (IRA) as analyzed in the REPEAT Project mid-range current policies scenario \cite{repeat, pg_github, load, flexible}.
            
            \item We assume 20\% of all EV charging load is time shiftable and can be delayed by up to five hours  \cite{pg_github, flexible}.
            
            \item Capacity reserve margins in GenX are implemented by aggregating the fourteen zones into three regions to ensure 2022 North American Electric Reliability Corporation reliability planning standards are met \cite{cap_2022}. California has a 16.9\% reserve margin requirement. Oregon, Washington, Idaho, Montana, Nevada, Wyoming, and Utah have a 16.1\% reserve margin requirement. New Mexico and Arizona have a 10.2\% reserve margin requirement.  
        \end{itemize}

    \subsection{Scenarios}
    \label{scenarios}
        There are 24 finalized runs, varying the scenario, forced system-wide battery capacity, and cost trajectories of VREs in 2030. We run three scenarios to compare how modeling VREs with optimized interconnection and co-location of storage impacts results. The first scenario, the fixed interconnection scenario, assumes that there is 1.3 MW of PV built for every 1 MW of inverter and grid connection built and 1 MW of wind built for every 1 MW of grid connection built \citep{nrel_pv_battery_2018, nrel_pv_battery_2021, nrel_pv_battery_capacity_expansion_2020}. Additionally, the fixed interconnection scenario models only standalone options to build batteries, solar PV, and wind. This scenario represents standard capacity expansion modeling practices. The second scenario, the optimized interconnection scenario, endogenously optimizes the capacity of solar PV or wind capacity and the associated grid capacity. This scenario only allows the buildout of standalone options for batteries, PV, and wind to compare how co-optimizing grid connection would change the capacity expansion of VREs. The third scenario, the new co-located storage model scenario, assumes the optimized interconnection scenario with the ability to co-locate solar PV and storage or wind and storage resources. Both standalone resources and co-located VREs and storage can be built in this scenario. 
        
        We run all three scenarios over different levels of new battery deployment in the system. Only lithium-ion batteries are modeled in this case study for both standalone and co-located storage options. To capture the substitutability of transmission and storage resources and analyze how increasing the discharge capacities of storage impacts investments and siting of resources, we run several cases requiring total deployment of new storage capacity equal to 2.5\% (3.75 GW), 3.3\% (5 GW), 5\% (7.5 GW), and 10\% (15 GW) of system-wide peak demand. This new storage capacity is in addition to the existing 5.6 GW in WECC as of the end of 2022. By forcing increasing amounts of storage into the system, we can investigate changes in the siting and sizing of resources and transmission upgrade decisions as storage penetration increases. Using NREL's ATB scenarios, two different VRE and energy storage cost projections are utilized: moderate and low-cost scenarios \cite{nrel_atb_2022}.
        
        We run all 24 cases under the current policy environment. First, existing state RPS are aggregated per model region: 49\% for CA\_N, CA\_S, WECC\_IID, and WEC\_SDGE; 6\% for WECC\_AZ; 21\% for WECC\_CO; 41\% for WECC\_NM, WECC\_NNV, and WECC\_SNV; and 16\% for WECC\_PNW. We calculate the actual RPS requirements based on the state nominal RPS value for 2030, RPS coverage over various utilities and retail sales, and distributed generation carve-outs. Washington has an 80\% CES for 2030 in addition to the RPS requirements, so the CES for the Pacific Northwest is calculated to be 58\% \cite{state_rps}. Second, these scenarios also include the production and investment tax credits (PTC and ITC) extended and modified by the IRA, which are further elaborated upon in \ref{app_costs_analysis}. A list of all of the runs can be found in Table \ref{tab:scenarios}. 

        \begin{table}[H]
            \centering
            \small
            \caption{\textbf{Case Study Scenarios.}}
            \label{tab:scenarios}
            {\begin{tabular}{|c|c|c|c|}
                \hline
               Run& \begin{tabular}{@{}l@{}}Scenario\end{tabular} 
               & \begin{tabular}{@{}l@{}}VRE\\Costs\end{tabular} 
               & \begin{tabular}{@{}l@{}}Forced Battery\\Capacity (GW)\end{tabular}\\
               
               \hline
                1 & Fixed Interconnection & Low & 3.75 \\
                2 & Fixed Interconnection & Low & 5 \\
                3 & Fixed Interconnection & Low & 7.5 \\
                4 & Fixed Interconnection & Low & 15 \\
                5 & Optimized Interconnection & Low & 3.75 \\
                6 & Optimized Interconnection & Low & 5 \\
                7 & Optimized Interconnection & Low & 7.5 \\
                8 & Optimized Interconnection & Low & 15 \\
                9 & Co-Located Storage & Low & 3.75 \\
                10 & Co-Located Storage & Low & 5 \\
                11 & Co-Located Storage & Low & 7.5 \\
                12 & Co-Located Storage & Low & 15 \\
                13 & Fixed Interconnection & Mid & 3.75 \\
                14 & Fixed Interconnection & Mid & 5 \\
                15 & Fixed Interconnection & Mid & 7.5 \\
                16 & Fixed Interconnection & Mid & 15 \\
                17 & Optimized Interconnection & Mid & 3.75 \\
                18 & Optimized Interconnection & Mid & 5 \\
                19 & Optimized Interconnection & Mid & 7.5 \\
                20 & Optimized Interconnection & Mid & 15 \\
                21 & Co-Located Storage & Mid & 3.75 \\
                22 & Co-Located Storage & Mid & 5 \\
                23 & Co-Located Storage & Mid & 7.5 \\
                24 & Co-Located Storage & Mid & 15 \\
                \hline
            \end{tabular}}
        \end{table}

    \subsection{Cost Analysis}
    \label{cost_analysis}
        \begin{table}[H]
            \centering
            \resizebox{\textwidth}{!}{\begin{tabular}{l|c|c}
                \hline
                
                & \begin{tabular}{@{}l@{}}Low VRE Costs\end{tabular} 
                & \begin{tabular}{@{}l@{}}Mid VRE Costs \end{tabular} \\ \hline
                Solar Capital Costs (\$/kW$_{DC}$) & 710 & 771 \\
                Solar Fixed O\&M Costs (\$/kW$_{DC}$/yr) & 16.2 & 17.3 \\
                \hline
                Wind Capital Costs (\$/kW$_{AC}$) & 1138 & 1308 \\
                Wind Fixed O\&M Costs (\$/kW$_{AC}$/yr) & 43 & 46 \\
                \hline
                Battery Pack Capital Costs (\$/kWh) & 261 & 290 \\
                Battery Pack Fixed O\&M Costs (\$/kWh/yr) & 6.5 & 7.3 \\
                \hline
                Grid Connection Capital Costs (\$/kW-km$_{AC}$) & 2.9-6.8 & 2.9-6.8 \\
                Inverter Capital Costs (\$/kW$_{AC}$) & 60 & 83 \\
                Inverter Fixed O\&M Costs (\$/kW$_{DC}$/yr) & 2.4 & 2.6 \\
                \hline
            \end{tabular}}
            \caption{\textbf{2030 Projected Capital Cost Breakdown of Co-Located VRE and Storage Systems and Standalone Battery Systems}\\
            The table showcases the solar, wind, battery, grid connection, and inverter overnight capital and fixed O\&M costs for standalone and co-located solar and storage resources. Inverter costs are included in the model whenever storage or solar PV resources are built. The variation in grid connection costs is due to the difference in spur line costs across zones in the Western Interconnection. Note that grid connection overnight capital costs displayed in the table only incorporate substation costs, as additional modeled interconnection costs (e.g. for transmission-to-demand centers) are site-dependent. It is assumed that all of the fixed O\&M costs are allocated across the solar, wind, battery, and inverter components and there are no O\&M costs for the grid connection. Two cost sensitivities are also displayed: 1) low VRE \& battery cost projections, and 2) mid VRE \& battery cost projections according to NREL's ATB \cite{nrel_atb_2022}. Regional multipliers, asset lifespan, and WACC are used to convert capital costs into annuitized costs, which are included in the model objective function. Information on parameter assumptions to find annuitized costs for different technologies can be found in Table \ref{tab:pvcosts}.}
            \label{tab:costbreakdown}
        \end{table}
    
        Cost breakdowns for co-located (or `hybrid') wind or solar PV and battery storage projects are calculated using various NREL reports \cite{nrel_pv_battery_2018, nrel_pv_battery_2021, nrel_atb_2022, dataset_nrel_2021}. The analysis concludes with the following costs inputted into the model: investment and fixed operations and maintenance (O\&M) costs for solar PV resources (\$/MW DC/year (yr)), investment and fixed O\&M costs for wind technologies (\$/MW AC/year), investment and fixed O\&M costs for batteries (\$/megawatt-hour (MWh)/year), investment and fixed O\&M costs for the inverter (\$/MW AC/year), and investment and fixed O\&M costs the for grid connection (\$/MW AC/year). Table \ref{tab:costbreakdown} showcases the finalized 2030 overnight capital cost and fixed O\&M cost breakdown for the two scenarios: low and mid VRE \& battery costs.
        
        \cite{nrel_pv_battery_2021, dataset_nrel_2021} developed a bottom-up analysis of DC-coupled co-located solar PV and battery resources (100 MW PV DC/60 MW storage DC/240 MWh). From the bottom-up analysis, the AC cost breakdown of 2021 costs for standalone and co-located solar PV, batteries, and grid connection are split into 
        1) solar PV modules, 2) battery modules, and 3) inverters \cite{nrel_pv_battery_2021, dataset_nrel_2021}. The DC:AC cost ratios for solar PV and storage resources are calculated and a 2021 baseline overnight capital cost for the inverter is isolated from this analysis. Costs are projected to 2030 estimates based upon cost decline assumptions from NREL's ATB and annuitized with the IRA tax credits applied (see Table \ref{tab:pvcosts} for the assumptions on annuitizing overnight capital costs) \cite{nrel_atb_2022, IRA2022}. For wind and additional co-located storage costs, costs are projected according to estimates from NREL's ATB, annuitized with the IRA tax credits, and outputted by PowerGenome \cite{pg_github, IRA2022, nrel_atb_2022}. Grid connection costs are assumed to be the interconnection capacity and spur line costs annuitized that are outputted by PowerGenome \cite{pg_github}. Fixed O\&M costs for solar, storage, wind, and grid connection are calculated based upon previous NREL reports and NREL's ATB \cite{nrel_pv_battery_2021, nrel_atb_2022}. A full description of how the solar PV, wind, storage, inverter, and grid connection investment and fixed O\&M costs are calculated step-by-step can be found in \ref{app_costs_analysis}. 
            
        \begin{table}
            \centering
            \resizebox{\textwidth}{!}{\begin{tabular}{l|cccc|ccc|ccc}
                
                & \begin{tabular}{@{}l@{}}Solar\end{tabular} 
                & \begin{tabular}{@{}l@{}}Battery \end{tabular} 
                & \begin{tabular}{@{}l@{}}Inverter \end{tabular} 
                & \begin{tabular}{@{}l@{}}Grid  \end{tabular}
                & \begin{tabular}{@{}l@{}}Standalone\\Battery \end{tabular}
                & \begin{tabular}{@{}l@{}}Inverter \end{tabular}
                & \begin{tabular}{@{}l@{}}Grid \end{tabular}
                & \begin{tabular}{@{}l@{}}Wind\end{tabular} 
                & \begin{tabular}{@{}l@{}}Inverter \\(if co-located) \end{tabular} 
                & \begin{tabular}{@{}l@{}}Grid \end{tabular} \\ \hline
                WACC (\%) & 2.5 & 2.5 & 2.5 & 4.4 & 2.5 & 2.5 & 2.5 & 3.2 & 2.5 & 4.4 \\
                Lifespan (years) & 30 & 15 & 15 & 60 & 15 & 30 & 60 & 30 & 30 & 60 \\
                Regional & PV & Battery & Battery & PV & Battery & Battery & PV & Wind & Battery & Wind\\
        
                %19.10 & 18.30 [-1.6] & 50.40 [164.9\%] \\
                %Transmission & 5265 & 5956 [13.1\%] & 2617 [-50.3\%] &
                %6502 & 6101 [-6.2\%] & 6459 [-0.7\%]\\
                
                \hline
            \end{tabular}}
            \caption{\textbf{Parameters in Calculations of Co-Located VRE and Storage Systems and Standalone Battery Systems}\\
            The left side of the table showcases solar, battery, inverter, and grid connection cost breakdown assumptions for standalone solar and co-located solar and storage resources. The center of the table showcases storage, inverter, and grid connection cost breakdown assumptions for standalone storage resources. The right side of the table showcases wind, inverter, and grid connection cost breakdown assumptions for standalone wind and co-located wind and storage resources \cite{nrel_atb_2022}. `Regional' stands for the regional multipliers from NREL used since costs can vary for technologies across WECC \cite{Nrel_regional_multiplier}.}
            \label{tab:pvcosts}
        \end{table}
        
        There are no fuel costs for VRE or battery resources. Nuclear fuel costs are generated in PowerGenome \cite{pg_github}. Other NG fuel costs are derived from the 2021 EIA's Annual Energy Outlook, projected to 2030 costs, and aggregated into two categories: Pacific and Mountain \cite{pg_github}. For natural gas with 90\% CCS resources, the value of the 45Q tax credit net of CO\textsubscript{2} disposal cost is estimated at -\$57.28/ton \cite{IRA2022}.
        
    \subsection{Downscaling and Post-Processing Results}
        For potential solar and wind CPAs selected in this project, candidate projects were identified and post-processed, routed to interconnection lines, and downscaled according to techniques from the appendix in Patankar et al. 2022 \cite{downscaling}.
                
        %We additionally downscale the aggregated results, processing the GenX outputs to 1) 4x4 km\textsuperscript{2} candidate project areas (CPAs) based on the levelized cost of electricity (LCOE) inclusive of interconnection costs, and 2) intra-regional spur line (short distance site-to-substation and intra-regional transmission-to-metro lines) capacity and inter-regional long-distance transmission capacity \cite{repeat}. 

\section{Results \& Discussion}
\label{results}
    \subsection{Co-optimized Renewable Energy and Interconnection Capacity}
    \label{coop_results}
        
\begin{figure}[htbp]
            \centering
            \begin{subfigure}[b]{0.45\textwidth}
                \centering
                \includegraphics[width=\linewidth]{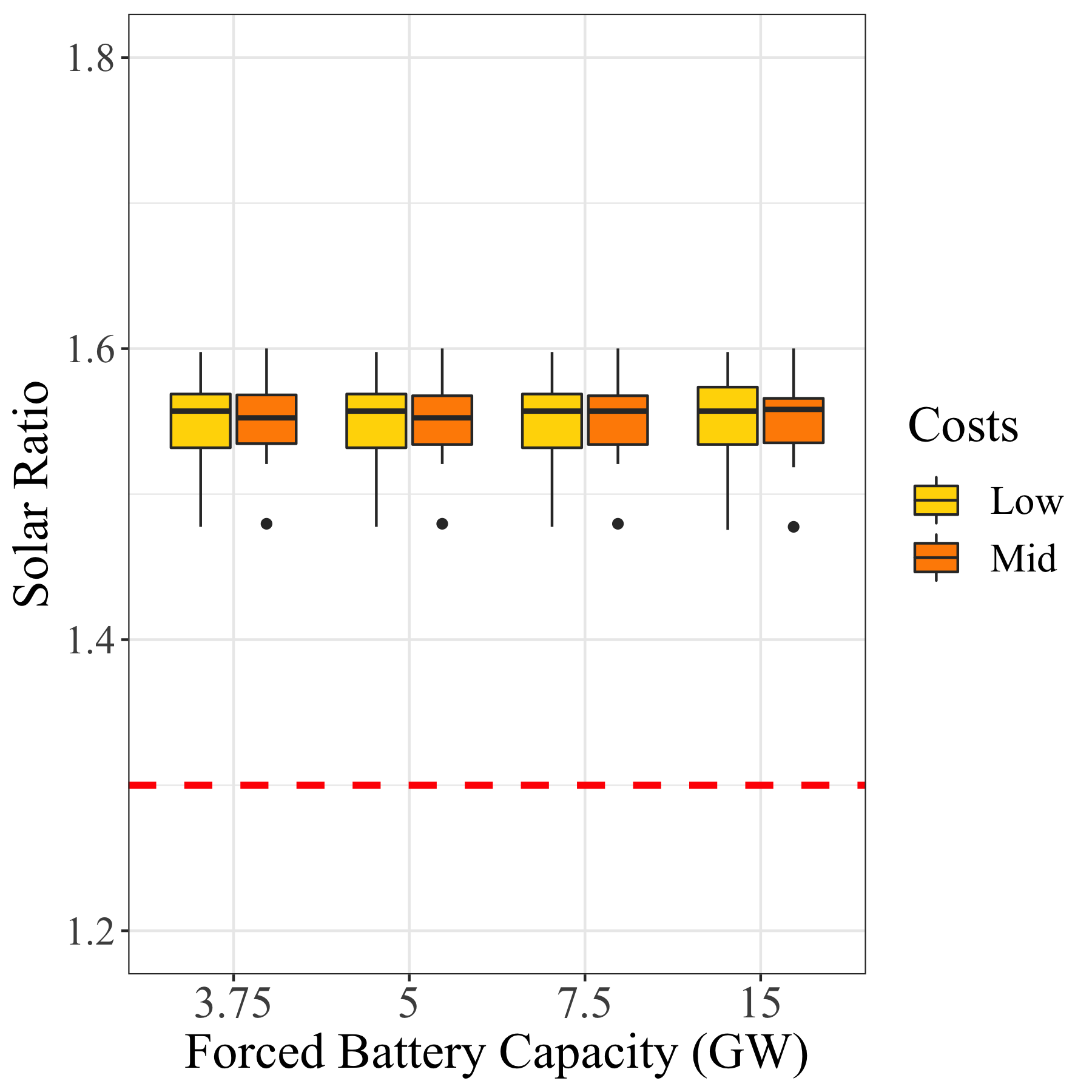}
                \caption{}
                \label{fig:solarboxplot}
            \end{subfigure}
            \hspace{0\textwidth}  % Horizontal space between subfigures
            \begin{subfigure}[b]{0.45\textwidth}
                \centering
                \includegraphics[width=\linewidth]{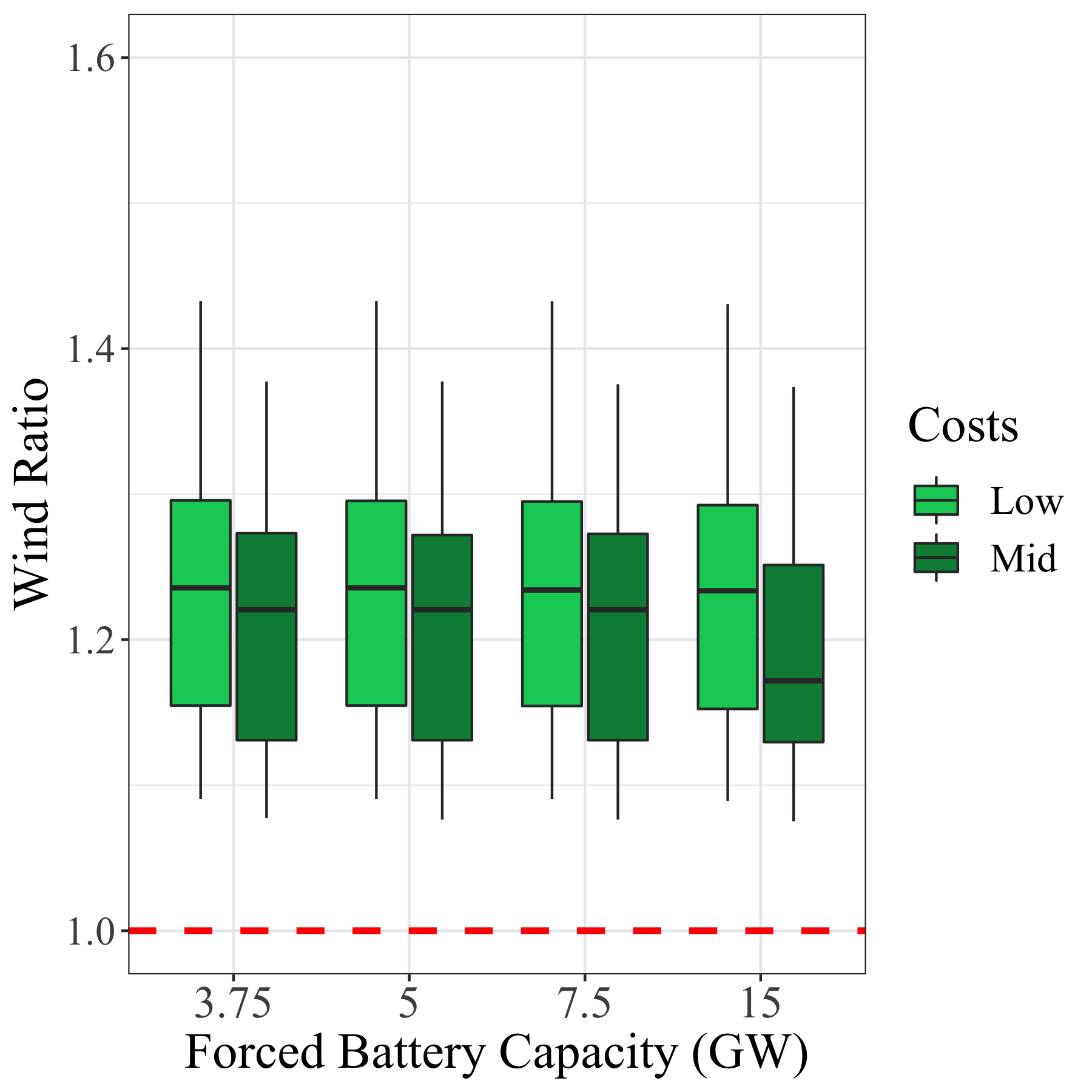}
                \caption{}
                \label{fig:windboxplot}
            \end{subfigure}
            
            \vspace{0\textwidth}  % Vertical space between rows
            
            \begin{subfigure}[b]{0.45\textwidth}
                \centering
                \includegraphics[width=\linewidth]{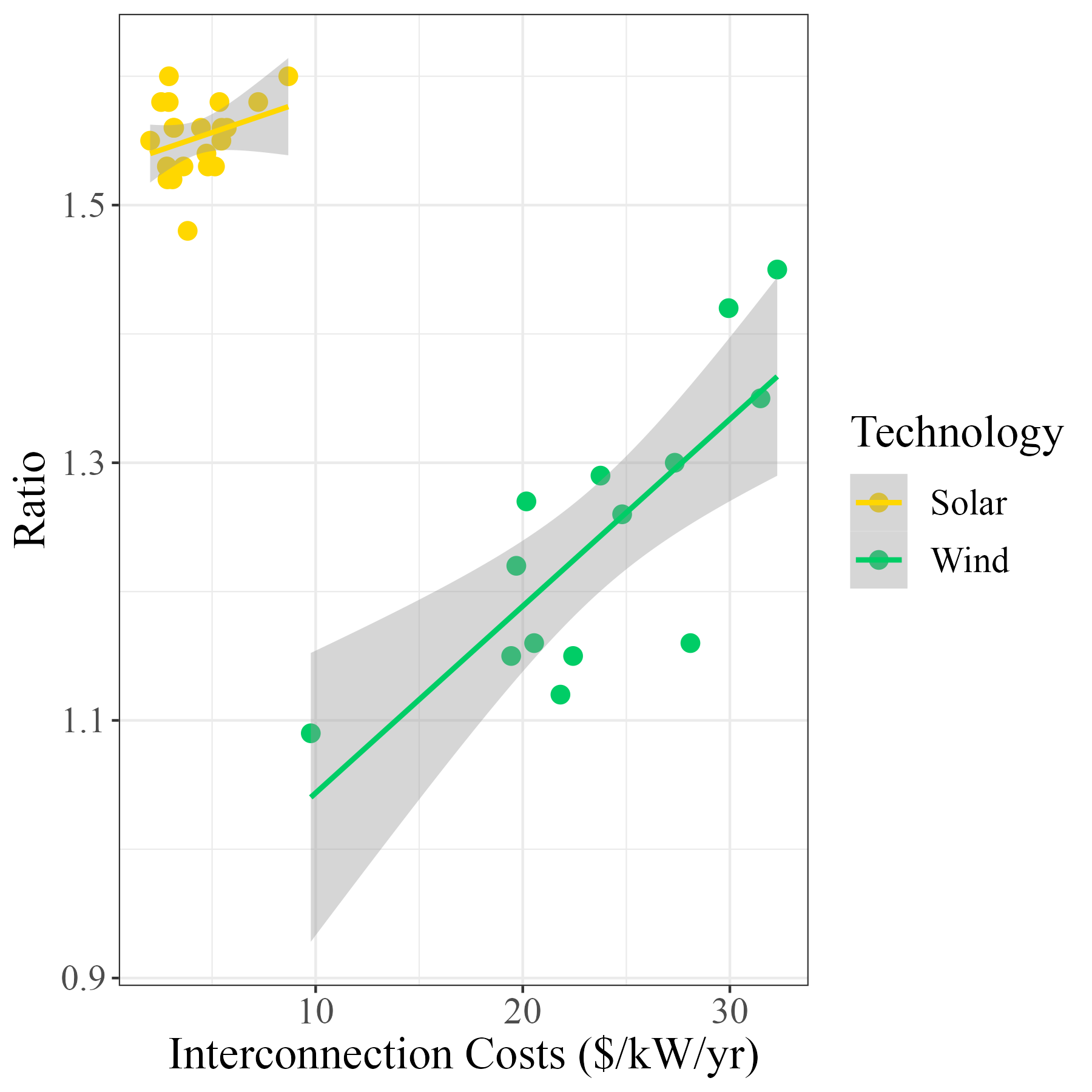}
                \caption{}
                \label{fig:scatterlow}
            \end{subfigure}
            \hspace{0\textwidth}  % Horizontal space between subfigures
            \begin{subfigure}[b]{0.45\textwidth}
                \centering
                \includegraphics[width=\linewidth]{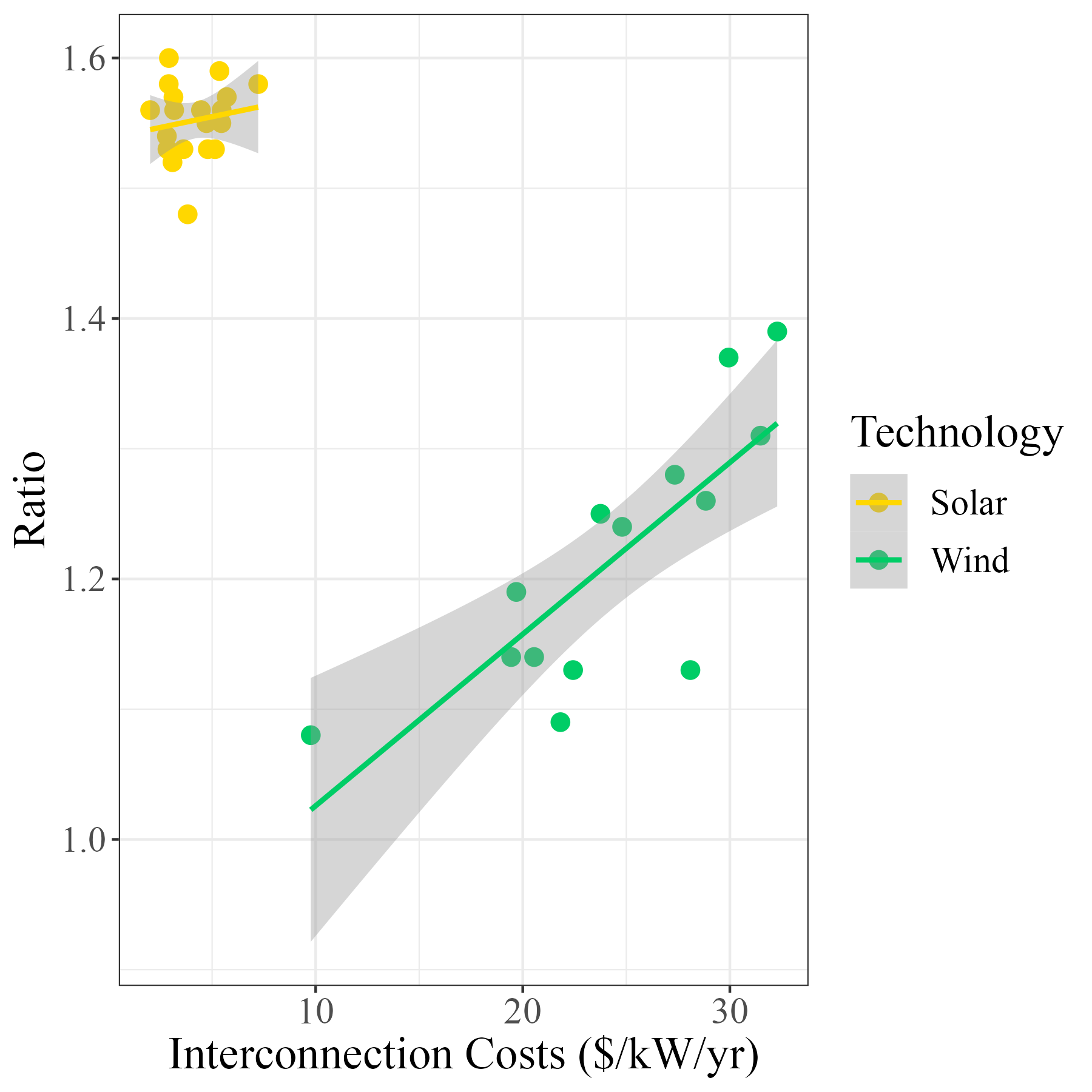}
                \caption{}
                \label{fig:scattermid}
            \end{subfigure}
            
            \caption{\textbf{Box Plots of Wind and Solar PV Clusters' Average Ratio of VRE to Inverter/Grid Connection Built and Scatter Plots of Ratio of VRE to Inverter/Grid Connection Built vs. Interconnection Costs with 3.75 GW of System-Wide Battery Capacity for Optimized Interconnection Scenario.} (a) Average Ratio of PV to Inverter Capacity (red dashed line represents the fixed assumed ratio of 1.3 MW of solar PV to 1 MW of inverter capacity), (b) Average Ratio of Wind to Grid Connection Capacity (red dashed line represents the fixed assumed ratio of 1 MW of wind to 1 MW of grid connection capacity), (c) Scatter Plot of the Ratio of Solar PV and Wind vs. Interconnection Costs (\$/kW/yr) for the Low VRE-Cost Scenario with 3.75 GW of Forced System-Wide Battery Capacity, and (d) Scatter Plot of the Ratio of Solar PV and Wind vs. Interconnection Costs (\$/kW/yr) for the Mid VRE-Cost Scenario with 3.75 GW of Forced System-Wide Battery Capacity.}
            \label{fig:boxplot}
        \end{figure}

        When permitted to co-optimize VRE and transmission interconnection capacity, GenX selects an average of 1.55 MW PV: 1 MW inverter and interconnection capacity and 1.21-1.24 MW wind: 1 MW interconnection capacity across the cost sensitivities and regardless of battery penetration, with significant variation in the optimal ratio across solar and wind sites (Fig. \ref{fig:boxplot}(a-b)). These sites are effectively self-curtailing output (at times when available production exceeds grid connection capacity) to reduce the amount of transmission connection built. Because VRE resources do not consistently generate 100\% of potential output and the hours of VRE curtailment are correlated with peak generation from other renewable resources and low electricity prices, interconnection capacity costs can be reduced to a greater extent than the generation revenue. For example, a wind project sized at a 1.16:1 ratio reduces transmission costs by 14\% relative to a project sized at a 1:1 ratio. This site's total annual generation is decreased by 2.1\% but energy revenues may only decline by the order of 1.1\% due to curtailed output. This lost revenue can be offset by reducing transmission interconnection costs.

        Furthermore, the ratio of VRE to interconnection capacity increases as both the inverter and interconnection costs rise and VRE technologies' costs decline. We observe that the optimal ratio of solar PV or wind to grid connection capacity is positively correlated with the annuitized cost of transmission interconnection for each cluster, as illustrated in Fig. \ref{fig:boxplot}(c-d). Other scenarios with varying forced system-wide storage capacities can be found in Fig. \ref{fig::scatter_interconnection_ratios}. There is a weaker correlation between the solar PV to inverter capacity ratio and grid connection costs relative to the wind to interconnection capacity ratio and grid connection costs. The cheaper interconnection costs for solar PV sites contribute less to the optimized sizing of each solar PV cluster.

        \newgeometry{left=0.7in, right=0.7in, top=0.7in, bottom=1in}

        \restoregeometry

\begin{figure}[htbp]
            \centering
            \begin{subfigure}[b]{0.49\textwidth}
                \centering
                \includegraphics[width=\linewidth]{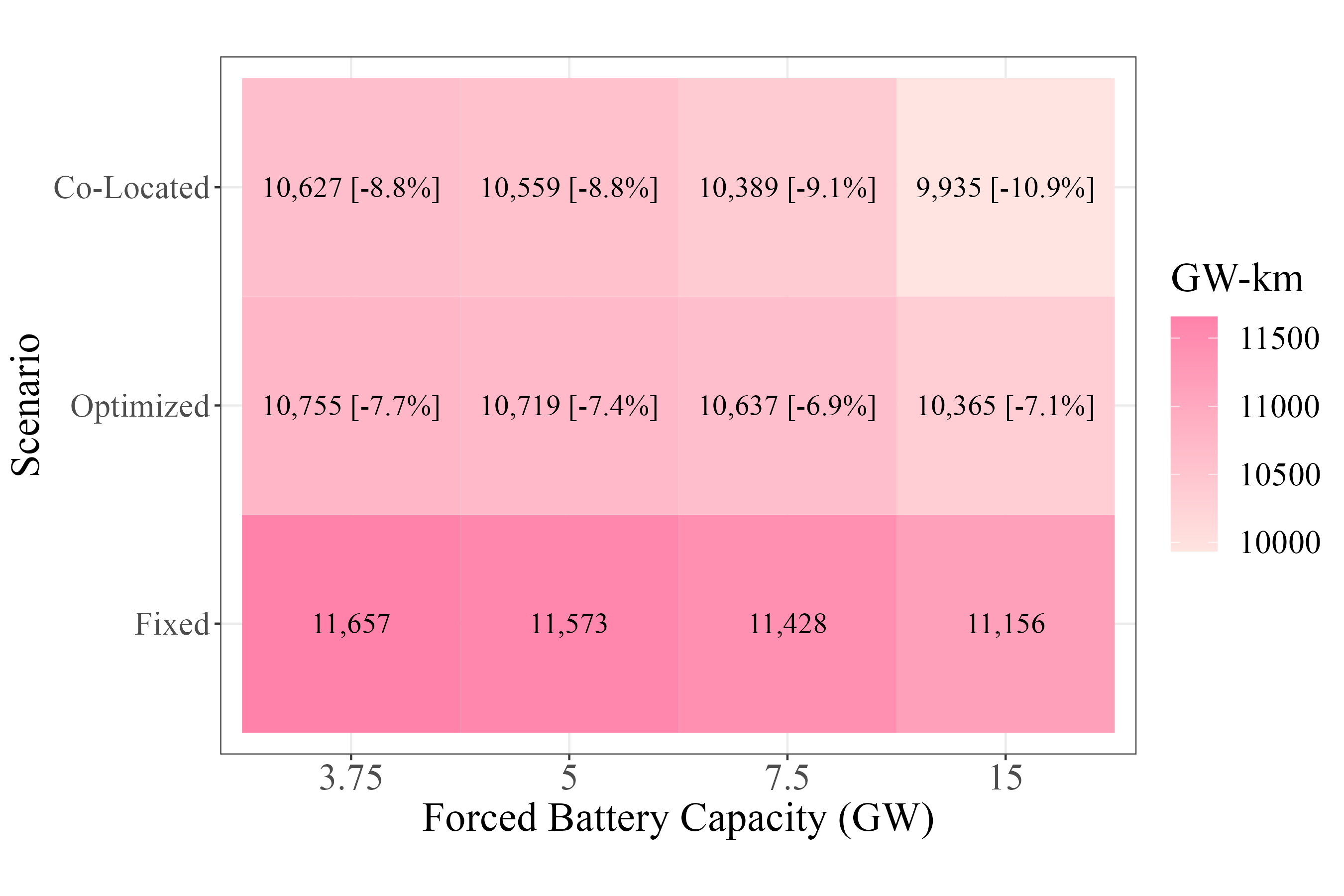}
                \caption{}
                \label{fig:intralow}
            \end{subfigure}
            \hspace{0\textwidth}  % Horizontal space between subfigures
            \begin{subfigure}[b]{0.49\textwidth}
                \centering
                \includegraphics[width=\linewidth]{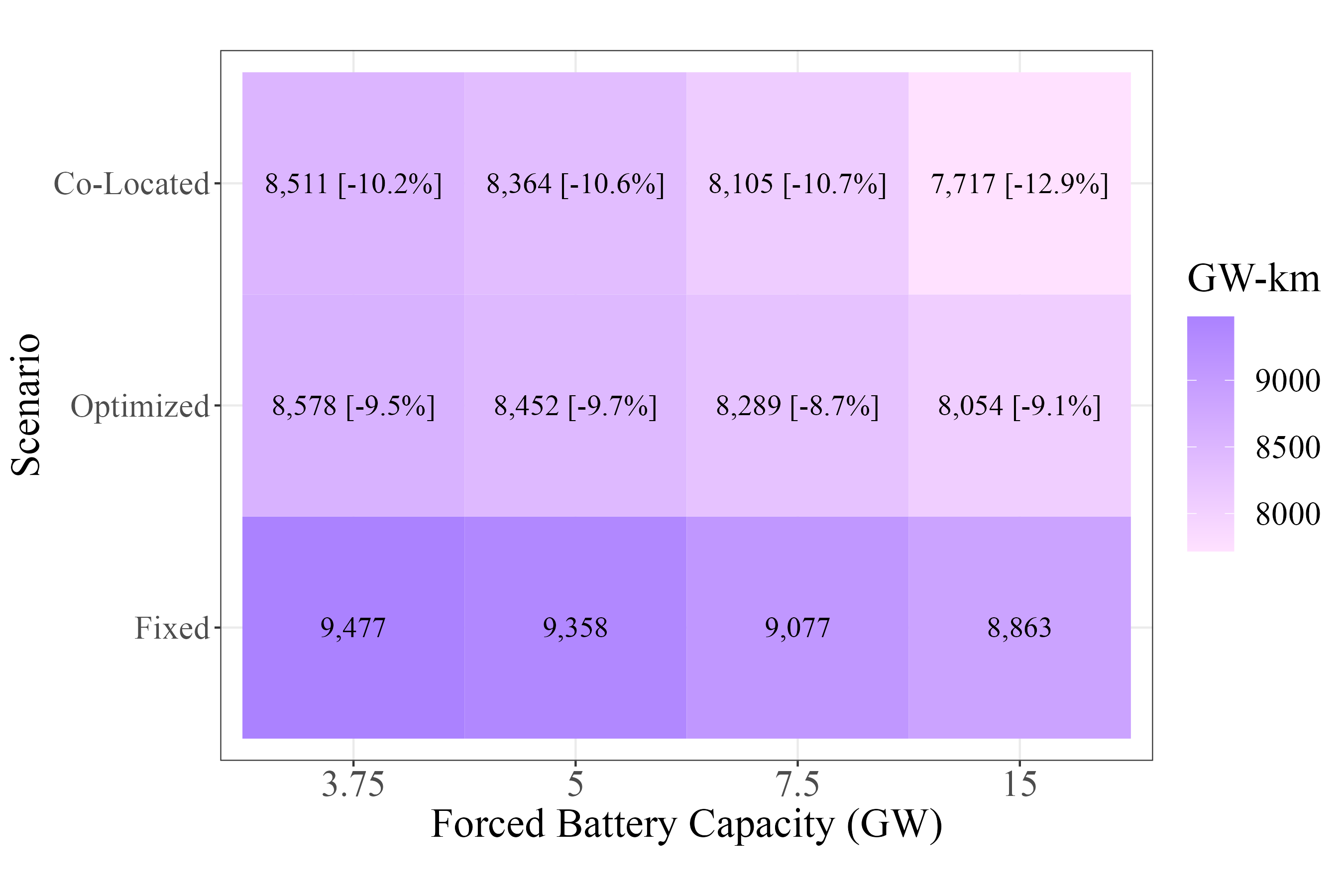}
                \caption{}
                \label{fig:intramid}
            \end{subfigure}
            
            \vspace{0\textwidth}  % Vertical space between rows
            
            \begin{subfigure}[b]{0.49\textwidth}
                \centering
                \includegraphics[width=\linewidth]{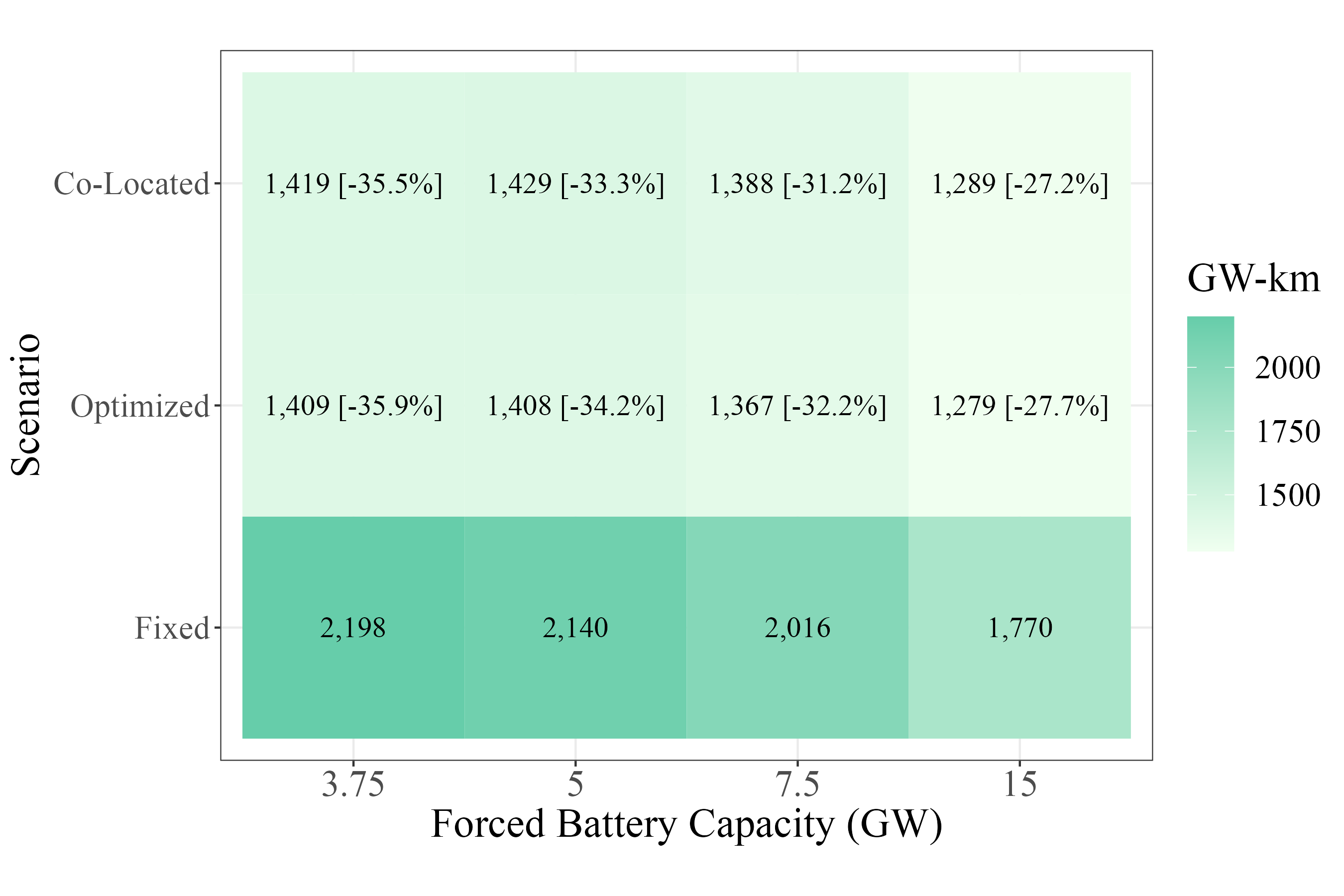}
                \caption{}
                \label{fig:interlow}
            \end{subfigure}
            \hspace{0\textwidth}  % Horizontal space between subfigures
            \begin{subfigure}[b]{0.49\textwidth}
                \centering
                \includegraphics[width=\linewidth]{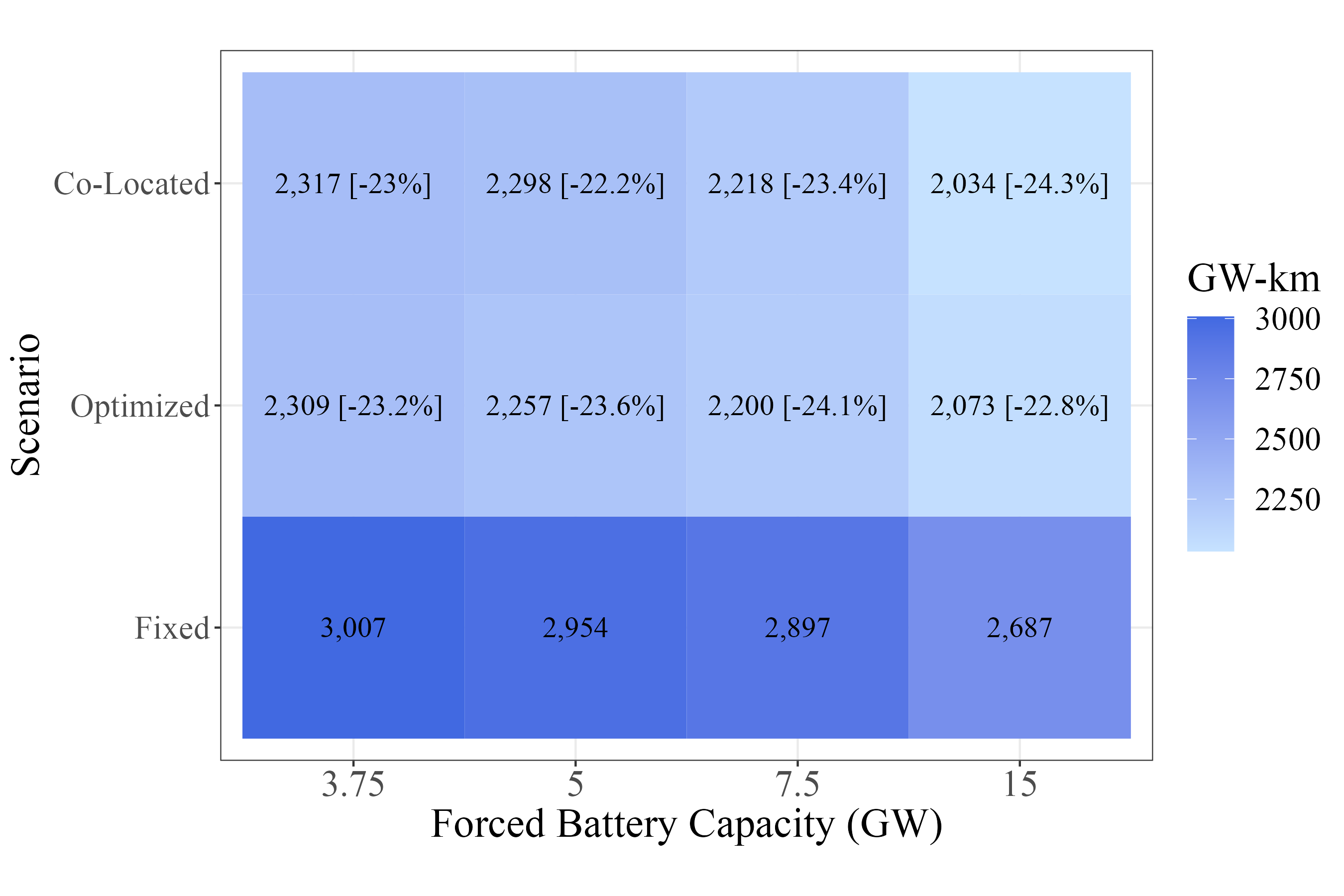}
                \caption{}
                \label{fig:intermid}
            \end{subfigure}
            
            \caption{\textbf{Interconnection Capacity and Inter-Zonal Transmission Buildout.} \\ (a) Interconnection Capacity in Low VRE-Cost Scenario, (b) Interconnection Capacity in Mid VRE-Cost Scenario, (c) Inter-Zonal Transmission Capacity in Low VRE-Cost Scenario, (d) Inter-Zonal Transmission Capacity in Mid VRE-Cost Scenario. Brackets indicate the percent decline of the optimized interconnection or co-located storage scenario relative to the fixed interconnection scenario.}
            \label{fig:transmission_heatmap}
        \end{figure}

        This further oversizing of VREs relative to the assumed ratios in the fixed interconnection scenario corresponds to a 7.7-9.5\% reduction in interconnection capacity (in GW-km) with 3.75 GW of forced system-wide battery capacity across both cost sensitivities (Fig. \ref{fig:transmission_heatmap}(a-b)). As battery penetration in the system increases from 3.75 to 15 GW, there is a 3.6-6.1\% decline in interconnection capacity in the optimized interconnection scenario, which indicates the substitutability of interconnection and storage (similar to the fixed interconnection scenario reduction in interconnection of 4.3-6.5\% as battery penetration in the system rises). The increasing battery discharge capacity results in a shift of wind buildout (6.1-8.1\% decrease) with longer intra-zonal transmission to solar PV (16.7-19.6\% increase) with shorter intra-zonal transmission in the optimized interconnection scenario.
        
        We also find that endogenously optimizing VRE and grid connection capacity sizing affects the optimal generation portfolio. In the 3.75 GW of forced battery capacity scenario, the optimized interconnection scenario builds 8.5-14.6\% more wind capacity, which offsets solar PV inverter and grid connection (16.8-17.1\%) and wind grid connection (8.6-10.1\%) capacity compared to the fixed interconnection scenario. System-wide new solar PV capacity has negligible changes (Fig. \ref{fig::capacity_difference_op}(a)). We see a preference for resource buildout of wind relative to solar PV because wind sites require longer average transmission distances and higher interconnection costs. Wind deployment thus benefits more from the ability to optimally size grid connection capacity.

        \newgeometry{left=0.6in, right=0.6in, top=1in, bottom=1in}

        \restoregeometry
        
\begin{figure}[H]
            \noindent
            \makebox[\textwidth]{\includegraphics[scale=0.6]{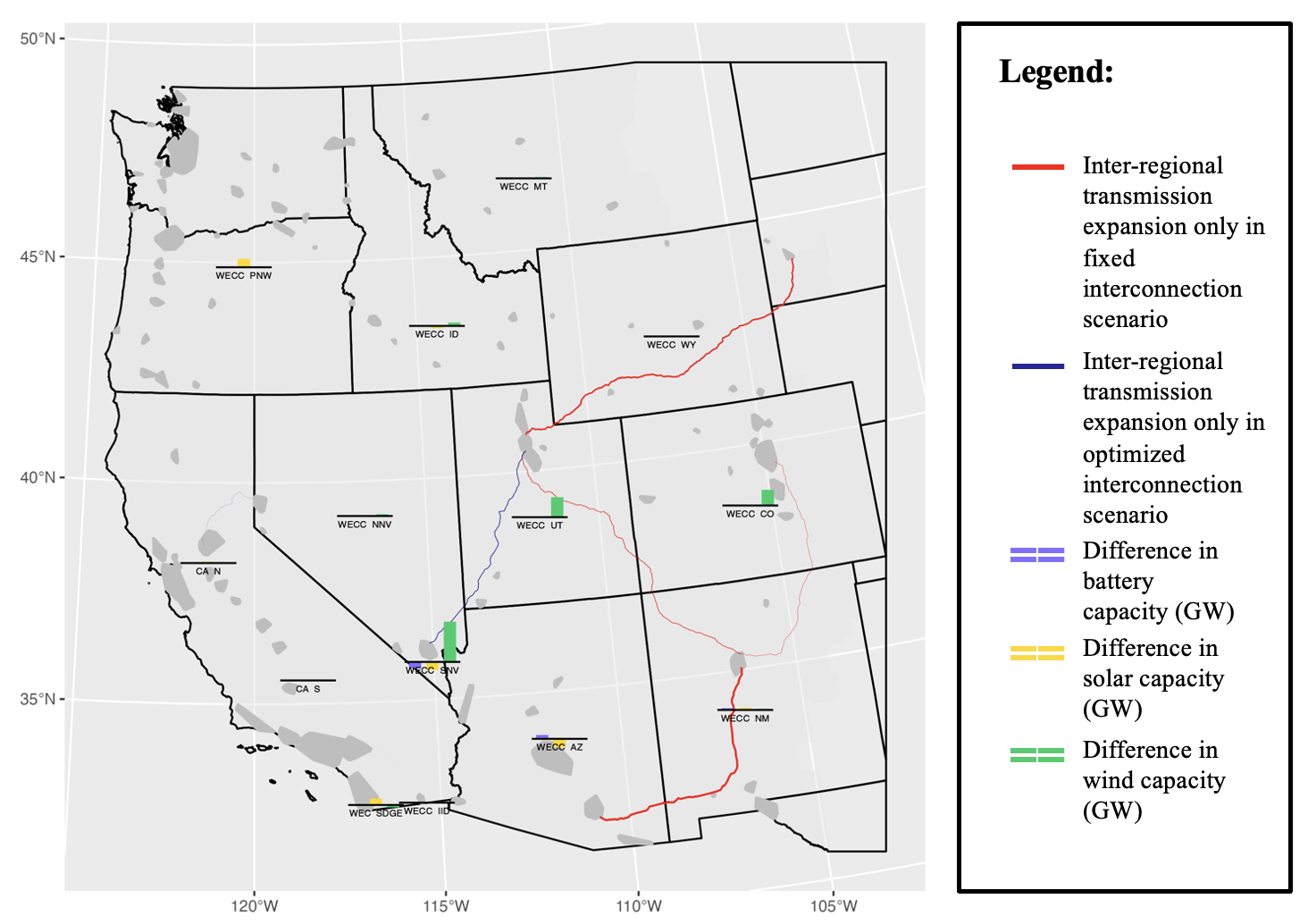}}
            \caption{\textbf{Zonal Solar, Wind, and Battery Capacity and Inter-Zonal Transmission Capacity Differences between the Optimized Interconnection and Fixed Interconnection Scenario.} \\ This scenario represents 3.75 GW of new battery forced into the system. The grey areas represent metropolitan statistical areas treated as demand centers.}
            \label{fig:downscaled_plot}
        \end{figure}

        The shift in system-wide resource deployment from oversizing VRE resources also impacts the inter-zonal transmission buildout. As showcased by Fig. \ref{fig:transmission_heatmap}(c-d), inter-zonal transmission (in GW-km) declines by 23.2-35.9\% in the 3.75 GW of forced system-wide battery capacity scenario from enabling the oversizing of resources compared to the fixed interconnection scenarios. These results are, however, not generalizable for two reasons. First, as indicated by Fig. \ref{fig:downscaled_plot}, these declines in inter-zonal transmission occur due to the reshuffling of battery, solar PV, and wind resources across the fourteen zones. Regardless of the system-wide battery penetration and cost sensitivities, there are inter-zonal transmission upgrade reductions across lines between New Mexico and Utah and Utah and Wyoming, which align with the reshuffling of battery resources from Southern Nevada to Arizona and New Mexico. The optimized interconnection scenario enables more degrees of freedom in lowering overall system costs by moving resources within and across zones. These results of inter-zonal transmission deployment may vary depending on the existing solar, wind, and battery resources and their respective locations. In this case study, inter-zonal transmission declines due to the optimized placement of VRE and battery technologies in different zones with the ability to oversize these resources. Second, the utilization rate of each line is increasing in the optimized interconnection scenario relative to the fixed interconnection scenario in the low VRE-cost scenario (Fig. \ref{fig::utilratelow}). This increase in the utilization rate showcases the model's decision to further optimize inter-zonal transmission capacity and correlates with the reduced deployment. We expect this effect (on long-distance transmission utilization) to be more generalizable than the shift in generation siting, and prior research has demonstrated partial substitutability between storage and transmission expansion \cite{BROWN2018720, VICTORIA2019111977, LIU20191308, CONLON20191085}.

        These changes in interconnection capacity, resource buildout, and inter-zonal transmission from optimizing renewable energy and battery interconnection can result in a 1.4-1.7\% (\$299-354 million per year) reduction in total annual system costs in the 3.75 GW battery scenario. Though these cost savings include reductions in transmission interconnection costs, as indicated by Fig. \ref{fig::cost_optimized_difference}, the majority of the cost declines can be attributed to a decrease of variable O\&M costs. This savings results from a reduction of annual generation from NG combined cycle and coal resources due to increased wind generation enabled by optimally sizing interconnection capacity (Fig. \ref{fig::power_optimized_difference}). 

    \subsection{Co-located Renewable Energy and Battery Capacity}
        In this section, we expand upon the optimized interconnection capabilities, adding and analyzing the impact of permitting lithium-ion battery packs to be sited at the same physical interconnection point as solar PV and wind resources. We compare the co-located storage scenario to traditional modeling practices (the fixed interconnection scenario) to understand how co-location changes model outcomes. Furthermore, we also compare the co-located storage scenario to the optimized interconnection scenario to isolate the impacts of co-location from optimally sizing VRE capacity alone. We investigate changes in interconnection capacity and the optimal sizing of co-located technologies, resource buildout and trade-offs in how solar PV, wind, and storage are sited, the impacts of co-location on inter-zonal transmission buildout, system-wide cost differentials, and the value of co-located storage.

        \subsubsection{System-Wide Impacts on the Inverter Loading Ratio, Resource Buildout, Inter-Zonal Transmission, and Costs}

        We find that co-locating VRE and storage resources can reduce the interconnection capacity reinforcements needed (in GW-km) by 8.8-10.2\% with respect to the fixed interconnection scenario when 3.75 GW of battery are forced into the system (Fig. \ref{fig:transmission_heatmap}(a-b)). Co-locating storage further reduces interconnection capacity buildout compared to only oversizing VREs by 0.8-1.2\%. In co-located facilities, the maximum generation of solar PV and wind resources can charge batteries rather than generate electricity that is directly sent to the grid or curtailed, lowering the necessary grid connection. These results demonstrate the strong substitutability of storage for grid interconnection capacity. This important dynamic is not captured in models that do not permit co-location of storage and VREs.
        
        Furthermore, adding more system-wide storage capacity results in a greater decline in interconnection capacity. In the co-located storage scenario and across cost sensitivities, grid connection decreases by 6.5\%-9.3\% as battery penetration increases from 3.75 to 15 GW. Meanwhile, adding storage capacity results in smaller reductions in interconnection in the optimized interconnection (3.6-6.1\%) and fixed interconnection (4.3-6.5\%) scenarios. The larger decline in the co-located storage scenario indicates that models can only fully capture how storage can substitute for transmission when including the co-location functionality.

        This decline in grid connection from co-locating VRE and storage resources results in greater oversizing of solar PV and wind to grid connection but a decrease in the solar PV ILR compared to the optimized interconnection scenario. In the 3.75 GW battery scenario and across cost sensitivities, co-located solar PV sites average a 1.57-1.59 solar PV to grid connection ratio and 1.50-1.51 solar PV ILR, while standalone clusters average a 1.55-1.56 solar PV to grid connection and solar PV ILR. Co-located wind sites average a 1.25-1.3 wind to grid connection ratio, while standalone wind clusters have a mean of 1.17-1.21 wind to grid connection ratio. The ratio of solar PV or wind to grid connection increases because grid connection for each site is further reduced since the VRE can charge the battery during periods of peak VRE production and optimize interconnection accordingly. As showcased in Table \ref{tab:appcolocatestats} for the 3.75 GW, co-located storage scenario, solar PV interconnection (in GW-km) declines by 16.8-18.4\% and wind interconnection decreases by 7.7-8.8\% compared to the fixed interconnection scenario (1.1-1.4\% and 0-0.6\% reduction respectively compared to the optimized interconnection scenario). Meanwhile, solar and co-located storage resources now share inverter capacity when modeling co-location, which reduces overall system-wide inverter capacity by 2.4-3.2\% and marginally increases inverter capacity for co-located solar PV and storage sites in the 3.75 GW, co-located storage scenario compared to the optimized interconnection scenario (Table \ref{tab:appcolocatestats}). As indicated by Fig. \ref{fig:boxplot_colocated_solar} and \ref{fig:boxplot_colocated_wind}, there is also greater variation in how each solar PV and wind resource is sited because adding the model option to co-locate storage resources may alter the optimized sizing since the grid connection and inverter capacity must be sized for both the VRE and storage components.

        Co-locating VRE and storage technologies results in similar resource and inter-zonal transmission deployment trends as the optimized interconnection scenario. With 3.75 GW of battery forced into the system, the co-located storage scenario results in a 1.1\% reduction in PV capacity relative to the fixed interconnection scenario and 0.7-0.9\% decline compared to the optimized interconnection scenario. The decrease in solar PV infrastructure buildout can be attributed to both the ability to optimize the sizing of solar PV to inverter and interconnection capacity and to the strategic siting of batteries with solar resources to increase the utilization of existing infrastructure. Wind deployment increases in the 3.75 GW scenario by 9.7-15.4\% compared to the fixed interconnection scenario and 0.7-1.1\% relative to the optimized interconnection scenario. Similar to the optimized interconnection scenario, the model takes advantage of the longer interconnection distances and greater grid connection costs to benefit wind deployment. Furthermore, as system-wide battery penetration increases for the co-located storage scenario, more solar PV is built. As showcased in Fig. \ref{fig:marginalsubstitution}, an increase of a GW of battery capacity is associated with increased solar PV in substitution of wind, NG combustion turbines, NG  combined cycle, and coal across scenarios. The magnitude of capacity displacement declines as battery capacity increases (e.g. less solar is built, while wind and other capacity are displaced less per unit of storage built). %While there is this displacement, wind still dominates generation across sensitivities when comparing the co-located storage scenario to either the fixed interconnection or optimized interconnection because the magnitude increase in wind across scenarios is greater than the solar PV capacity substitutability from increasing battery penetration.
        
        Inter-zonal transmission deployment declines in the 3.75 GW battery, co-located storage scenario by 23-35.5\% relative to the fixed interconnection scenario due to the reshuffling of battery, wind, and solar PV resources across zones, as further elaborated upon in Section \ref{coop_results}. These changes in interconnection capacity, resource shuffling, and inter-zonal transmission upgrades result in a 1.6-2\% (\$342-404 million per year) decline in overall system costs compared to the fixed interconnection scenario with 3.75 GW of forced system-wide battery. Cost changes between the co-located storage and fixed interconnection scenarios are not particularly sensitive to increasing battery penetration in the system. As in the optimized interconnection scenario, storage co-location reduces transmission costs but results in larger savings by enabling increased generation from co-located wind and solar resources and reductions in variable O\&M costs associated with fossil-based generation (see Fig. \ref{fig::cost_colocated_difference} and Fig. \ref{fig::power_optimized_difference}).

        \subsubsection{The Value of Co-located Storage and Siting Decisions of Co-Located Storage}

        \begin{figure}[H]
            \noindent
            \makebox[\textwidth]{\includegraphics[scale=0.5]{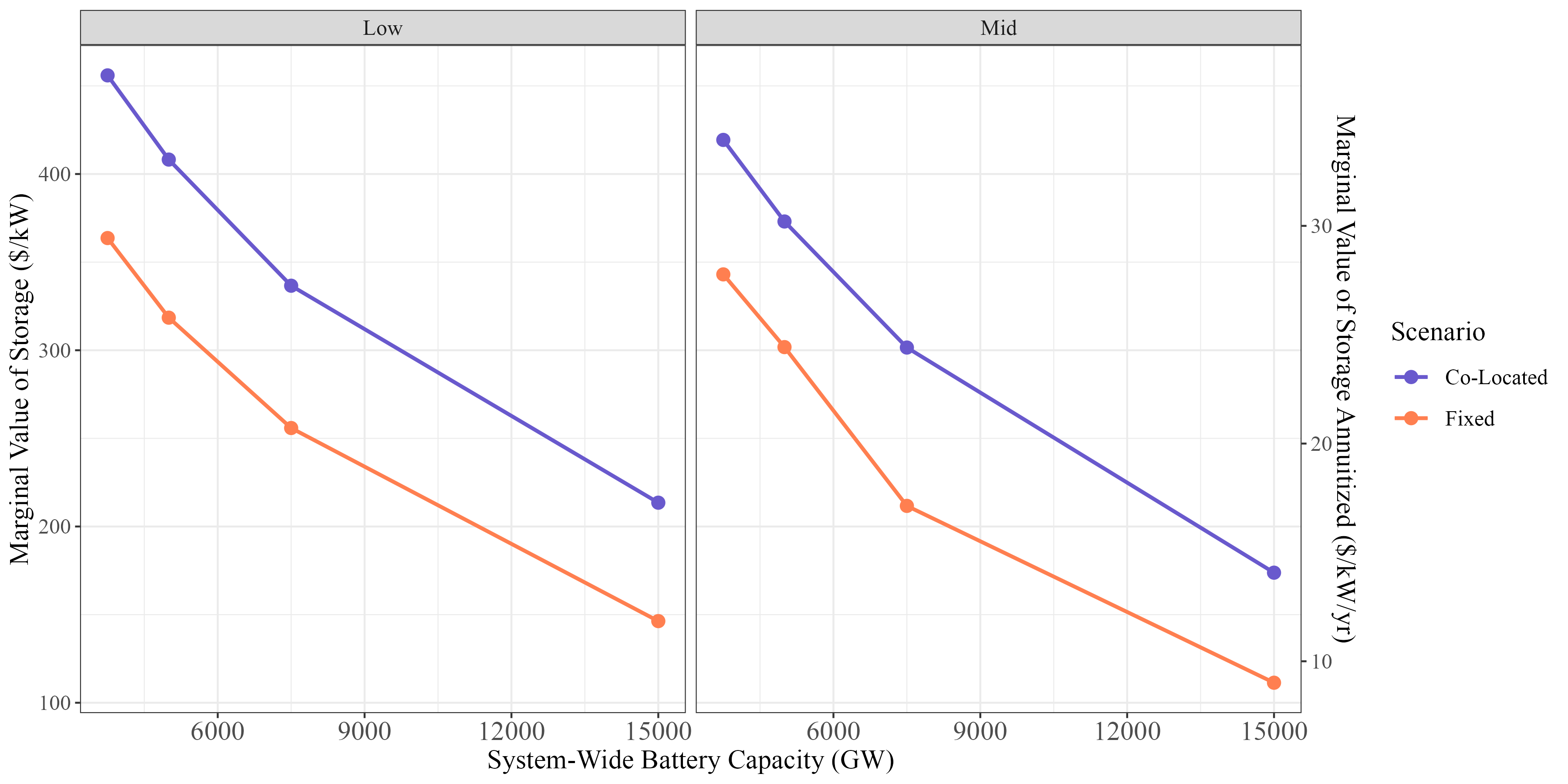}}
            \caption{\textbf{Marginal Value of Storage for the Co-Located Storage and Fixed Interconnection Scenarios.} }
            \label{fig:valuestorage}
        \end{figure}

        When co-location of storage is permitted, our model sites all battery capacity built across cost sensitivities and regardless of battery penetration in the system alongside either solar PV or wind resources. Co-location is preferred because the value of co-located storage is greater than standalone storage options due to the reduction in required inverter and interconnection buildout, as showcased in Fig. \ref{fig:valuestorage}. The value of co-located storage is calculated from the shadow price. In the lower battery penetration systems (3.75 GW), the marginal value of co-located storage is 22.2-25.4\% greater than when modeling standalone storage at the same penetration level in the fixed interconnection scenario. With 15 GW of system-wide battery capacity, the value of co-located storage is 45.9-56.1\% higher than the fixed interconnection scenario.

        We note, however, that this study employs aggregated inter-zonal power flow constraints and does not capture all transmission congestion within a region. By permitting co-location with VRE, we allow storage to substitute for transmission interconnection costs, but not other potential intra-zonal transmission constraints. In reality, there are likely other constrained locations within the grid where storage could be sited independently of VRE resources and deliver similar (or greater) value, and thus we encourage readers not to over-interpret our results as implying standalone storage resources are never justified. Rather, this finding indicates that the most cost-effective storage locations are likely to enable some reduction in transmission network capacity, in addition to other sources of value.
        
        \begin{figure}[H]
            \noindent
            \makebox[\textwidth]{\includegraphics[scale=0.5]{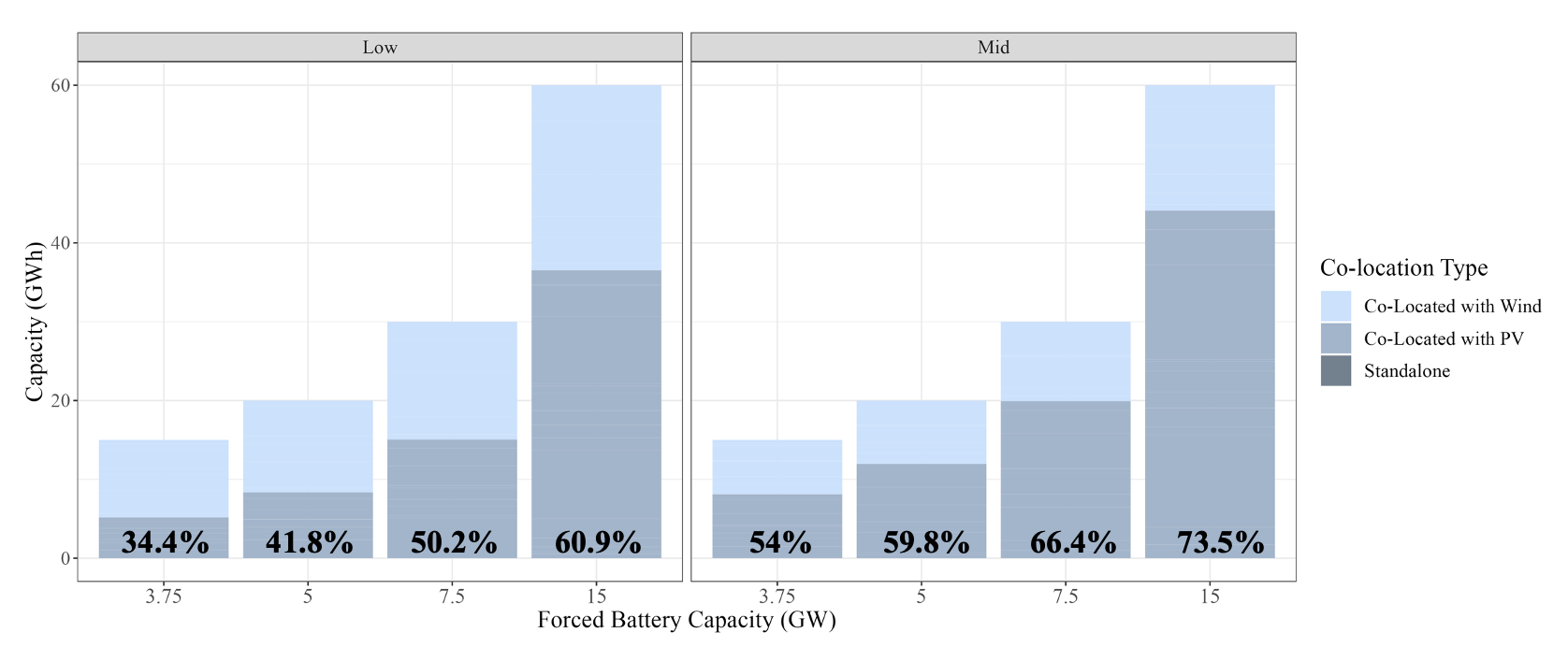}}
            \caption{\textbf{Siting Breakdown of Batteries Built across System-Wide Forced Battery Capacity and Cost Sensitivities.} \\ Bolded percentages in the figure represent the percentage of storage co-located with solar PV. The remaining percentage is co-located with wind.}
            \label{fig:battery_breakdown}
        \end{figure}

        Furthermore, storage projects are sited at a mix of wind and solar PV sites. The share of battery capacity co-located with each resource varies depending upon the scenario. As showcased in Fig. \ref{fig:battery_breakdown}, in the 3.75 GW, low VRE-cost scenario, $\sim$34\% of batteries are co-located with solar PV resources. Meanwhile, in the 3.75 GW, mid VRE-cost scenario, $\sim$54\% of batteries are co-located with solar PV resources. Co-location is thus sensitive to cost projections. In the model setup, solar PV cost projections in the mid VRE-cost scenario represent an 8.6\% increase relative to the low VRE-cost scenario, while wind cost projections are 14.9\% higher \cite{nrel_atb_2022}. The greater increase in wind relative to solar PV costs between the two scenarios is responsible for the decline in co-location with wind in the mid VRE-cost scenario, as fewer wind project areas are cost-competitive at this higher cost. 

        \begin{figure}[htbp]
            \centering
            \begin{subfigure}[b]{0.6\textwidth}
                \centering
                \includegraphics[width=\linewidth]{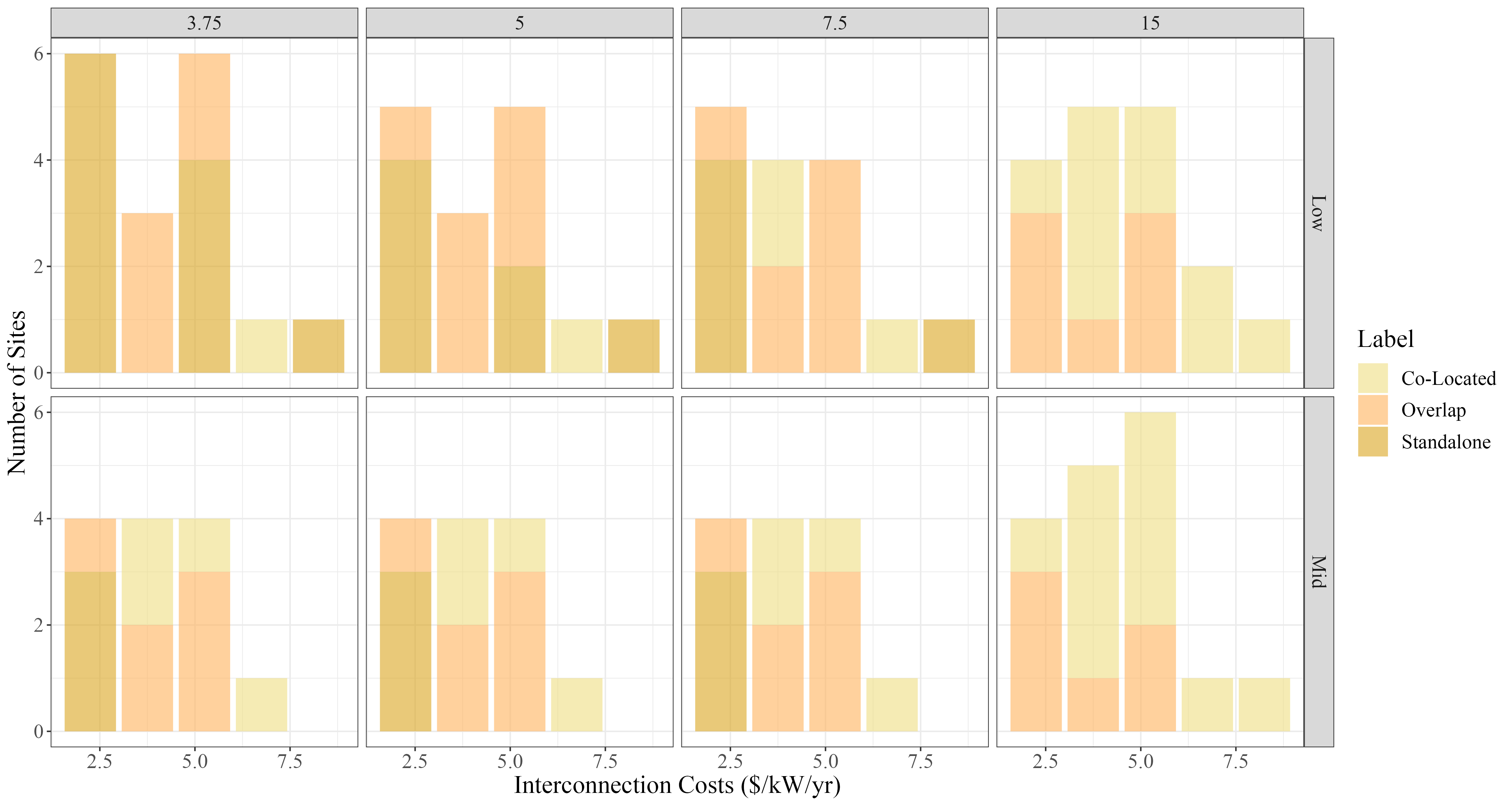}
                \caption{}
                \label{fig:pvhisto}
            \end{subfigure}
            
            \vspace{0\textwidth}  % Vertical space between rows
            
            \begin{subfigure}[b]{0.6\textwidth}
                \centering
                \includegraphics[width=\linewidth]{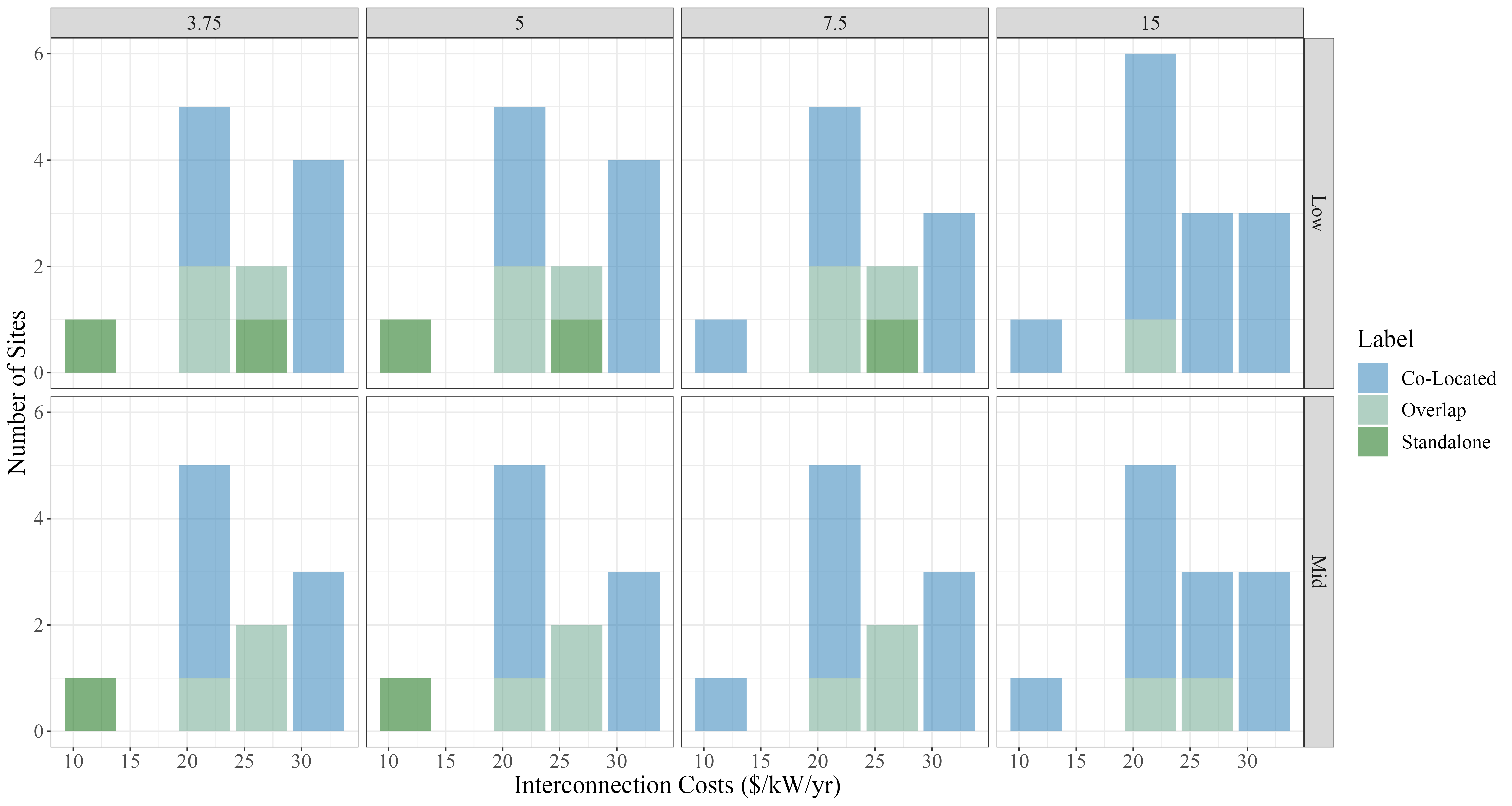}
                \caption{}
                \label{fig:windhisto}
            \end{subfigure}
            
            \caption{\textbf{Histogram of the Interconnection Costs for Sites that are Co-located with Solar PV and Wind Resources.} \\ (a) Solar PV, (b) Wind. Overlap in the legend describes the number of both standalone and co-located sites that fall within the respective interconnection cost range.}
            \label{fig:histointerconnection}
        \end{figure}
        
        Meanwhile, we see an increasing share of storage co-located with solar PV as the penetration of storage in the system increases. In the low-cost scenario, our model sites $\sim$34\% of batteries with solar PV at 3.75 GW battery penetration, while $\sim$61\% of batteries are co-located with solar PV at 15 GW battery penetration. On the one hand, wind sites tend to depend on longer transmission and higher interconnection costs, which makes siting storage at wind farms more valuable in substituting for transmission costs. In our scenarios, there are a few wind sites with long interconnection distances (mainly in Arizona, Colorado, New Mexico, Nevada, Utah, and Wyoming) that are ideal candidates for storage co-location to reduce transmission buildout. As showcased in Fig. \ref{fig:histointerconnection}, the wind sites with the highest interconnection costs tend to be co-located with storage. However, because wind farms do not regularly produce zero output, wind generation can frequently compete with storage discharge for the use of shared interconnection capacity. In contrast, solar PV exhibits consistent diurnal generation patterns. As a result, batteries co-located with solar can reliably charge the battery when prices are low or negative during mid-day and discharge in the evening or at night when there is ample shared interconnection capacity available (Fig. \ref{fig:wecc_az}). Thus, we observe that after a few key wind sites are saturated with storage, the model takes advantage of the complementarity of solar PV and storage, increasing the share of battery capacity co-located with solar as storage penetration increases.
        
        The initial conditions of the system can also alter co-location results. The modeled Western Interconnection system already has 5.6 GW/20.1 GWh of lithium-ion battery capacity in the system. Over 90\% of this storage exists in California where there is high buildout of solar PV in the system. Hence, we hypothesize that storage is initially deployed elsewhere in the American West due to the existing battery infrastructure in California. We test this by removing all initial batteries in the system across increasing forced battery penetration scenarios to determine how existing battery capacity changes siting decisions. With 3.75 GW of forced battery with low VRE-costs, in the system with no initial battery capacity, $\sim$56\% of batteries are co-located with solar PV (compared to the $\sim$34\% co-located with solar PV in the existing system of 5.6 GW of battery). These trends indicate that the current storage buildout in California for this case study is already utilized to balance the state's solar PV generation, which skews the additional storage capacity forced in through our experiment to co-locate more frequently with wind sites across the rest of the West.

\section{Implications \& Conclusion}
\label{conclusion}
    Wind and solar PV are low-cost resources that are likely to provide the cornerstone for power sector decarbonization. However, modeling of a low-carbon future showcases how large-scale wind and solar expansion requires significant increases in transmission capacity, which in practice faces development, permitting, and social license challenges that may constrain wind and solar deployment. We thus investigate how the optimal sizing of wind or solar resources relative to transmission interconnection capacity and the co-location of `hybrid' VRE and storage capacities can reduce overall transmission expansion and alter resource portfolio decisions in the American West. 

    We investigate two potential strategies to reduce transmission buildout. First, across the Western Interconnection, there are opportunities to decrease interconnection capacity upgrades through co-location of storage with wind and solar resources and optimal sizing of grid connection capacity. Capacity expansion models that do not include these capabilities (e.g. \cite{andlinger_center_for_energy_and_the_environment_acee_net-zero_2020, BROWN2018720, MALLAPRAGADA2020115390, SEPULVEDA20182403}) can potentially overestimate interconnection reinforcements on the order of 10\% or more with battery capacity at 2.5-10\% of regional peak demand. The declines in interconnection capacity correspond with an increase in the ratio of VRE capacity to interconnection capacity (1.5-1.6 MW PV: 1 MW inverter, 1.2-1.3 MW wind: 1 MW grid connection) and further wind technology buildout (+10-15\% in scenarios with battery capacity at 2.5\% of peak demand). Second, while modeling co-location and optimized interconnection capacity can reduce modeled grid connection requirements, we also find that increasing new battery penetration from 2.5\% to 10\% of regional peak demand can result in a further $\sim$7-9\% decline in interconnection capacity when the co-location modeling functionality is activated (compared to a 4-7\% decrease if co-location of VREs and storage is not modeled).

    Given the substantial differences in modeled capacity decisions and system costs observed in this study from enabling the features of co-optimized interconnection and co-location of renewables and storage, we recommend that the capacity expansion modeling community more broadly implements methods to represent co-location of storage and VRE resources to capture critical trade-offs relevant to policymakers, investors, and planners. We note, however, that this increase in model fidelity requires additional decision variables and constraints which come at a computational cost. By increasing the dimensionality of the capacity expansion model, modeling co-located storage can roughly double or triple the required runtime (+115-277\%) vs. modeling fixed interconnection sizing and standalone storage options only. Researchers should thus continue developing best practices for modeling these resources and explore improvements in software or formulation to reduce the computational burden. We also note that the wide variation in optimal interconnection sizing observed across modeled sites indicates that VRE developers should carefully optimize the sizing of each site. The methods developed in this paper can also provide decision support for such planning decisions.

    We find that no standalone storage assets are built when co-location with VREs is modeled. However, since our model represents the value of storage at a bulk transmission system level with zonal resolution, we interpret this result not to mean that standalone storage resources are never competitive, but rather that the most cost-effective storage locations are likely to enable some reduction in transmission network capacity, in addition to other sources of value. In this study, we do not incorporate optimal power flow and the full suite of intra-zonal transmission constraints, nor do we consider distribution network constraints, which could create other opportunities for storage to capture locational value aside from co-location with renewables. Future efforts could improve the resolution of capacity expansion models to represent these opportunities and more accurately incorporate the optimal siting of storage across the system. Furthermore, future capacity expansion modeling could include the co-location of long-duration energy storage solutions with VREs, as various emerging long-duration energy storage technologies have distinct cost and performance parameters from lithium-ion batteries and may demonstrate different substitution and behavioral dynamics \cite{JENKINS20212241}. Finally, this case study was run for the American West circa 2030, but quantitative results and trade-offs may also vary depending on long-term decarbonization strategy and geographic context. 

    Despite these caveats, our modeling illustrates the ability of batteries to reduce required interconnection expansion and demonstrates the importance of modeling the optimal sizing of grid connection and co-location of storage and VRE resources to accurately capture the opportunity in saving on inverter and interconnection costs and estimate the value of energy storage.

\section{Data Availability}
Code related to this article can be found at:\\ \url{https://github.com/GenXProject/GenX}, an open-source online data repository hosted at Github \cite{genx_github}. The dataset for this article can be found at: \url{https://zenodo.org/records/13340214}.

\section{Acknowledgements}
%This section is not included in the double-anonymous peer review.
Funding for this manuscript was provided by Princeton University’s Carbon Mitigation Initiative (supported by BP), Princeton University's High Meadows Environmental Institute (HMEI) and the associated HMEI Environmental Scholars Program, and the Princeton Zero-carbon Technology Consortium (supported by unrestricted gifts from Google, General Electric, ClearPath, and Breakthrough Energy).

\section{Disclosure of Interests}
%This section is not included in the double-anonymous peer review.

J.D.J is part owner of DeSolve, LLC, which provides techno-economic analysis and decision support for clean energy technology ventures and investors. A list of clients can be found at \url{https://www.linkedin.com/in/jessedjenkins}. He serves on the advisory boards of Eavor Technologies Inc., a closed-loop geothermal technology company, Rondo Energy, a provider of high-temperature thermal energy storage and industrial decarbonization solutions, and Dig Energy, a developer of low-cost drilling solutions for geothermal heating and cooling, and he has an equity interest in each company. He also serves as a technical advisor to MUUS Climate Partners and Energy Impact Partners, both investors in early-stage climate technology companies, including energy storage companies.

\newpage
\appendix

\section{Model Notation}
\label{model_notation}
    \subsection{Model Indices \& Sets}
    \label{model_sets}

        \begin{longtable}[c]{|c|p{9.8cm}|}%[tbp]
            \caption{GenX VRE-Storage Module Model Indices \& Sets}
            \label{indices} \\
            
            \hline
            \multicolumn{1}{|c|}{\textbf{Notation}} & \multicolumn{1}{c|}{\textbf{Description}}  \\ \hline
            $t \in T$ & $t$ denotes the indexed time step over the set of $T$ time steps \\
            
            \hline
            $z \in Z$ & $z$ denotes the zone over the set of $Z$ zones in the model \\

            \hline
            $c \in C$ & $c$ denotes each component at a co-located site: the grid connection, solar PV, wind, storage, inverter, DC charge capacity, DC discharge capacity, AC charge capacity, and AC discharge capacity \\
            
            \hline
            $y \in G$ & $y$ denotes the indexed generator over the set of $G$ available generators \\
            
            \hline
            $\mathcal{VS} \subseteq \mathcal{G}$ & $\mathcal{VS}$ denotes the subset of co-located VRE and storage resources \\
            
            \hline
            $\mathcal{VS}^{pv} \subseteq \mathcal{VS}$ & $\mathcal{VS}^{pv}$ denotes the subset of co-located VRE and storage resources with a solar PV component\\
            
            \hline
            $\mathcal{VS}^{wind} \subseteq \mathcal{VS}$ & $\mathcal{VS}^{wind}$ denotes the subset of co-located VRE and storage resources with a wind component\\
            
            \hline
            $\mathcal{VS}^{inv} \subseteq \mathcal{VS}$ & $\mathcal{VS}^{inv}$ denotes the subset of co-located VRE and storage resources with an inverter component\\
            
            \hline
            $\mathcal{VS}^{stor} \subseteq \mathcal{VS}$ & $\mathcal{VS}^{stor}$ denotes the subset of co-located VRE and storage resources with a storage component\\
            
            \hline
            $\mathcal{VS}^{sym,dc} \subseteq \mathcal{VS}$ & $\mathcal{VS}^{sym,dc}$ denotes the subset of co-located VRE and storage resources with a storage DC component with equal (or symmetric) charge and discharge power capacities \\
            
            \hline
            $\mathcal{VS}^{sym,ac} \subseteq \mathcal{VS}$ & $\mathcal{VS}^{sym,ac}$ denotes the subset of co-located VRE and storage resources with a storage AC component with equal (or symmetric) charge and discharge power capacities \\
            
            \hline
            $\mathcal{VS}^{asym,dc,dis} \subseteq \mathcal{VS}$ & $\mathcal{VS}^{asym,dc,dis}$ denotes the subset of co-located VRE and storage resources with a storage DC component with independently sized (or asymmetric) discharge power capabilities \\
            
            \hline
            $\mathcal{VS}^{asym,dc,cha} \subseteq \mathcal{VS}$ & $\mathcal{VS}^{asym,dc,cha}$ denotes the subset of co-located VRE and storage resources with a storage DC component with independently sized (or asymmetric) charge power capabilities \\
            
            \hline
            $\mathcal{VS}^{asym,ac,dis} \subseteq \mathcal{VS}$ & $\mathcal{VS}^{asym,ac,dis}$ denotes the subset of co-located VRE and storage resources with a storage AC component with independently sized (or asymmetric) discharge power capabilities \\
            
            \hline
            $\mathcal{VS}^{asym,ac,cha} \subseteq \mathcal{VS}$ & $\mathcal{VS}^{asym,ac,cha}$ denotes the subset of co-located VRE and storage resources with a storage AC component with independently sized (or asymmetric) charge power capabilities \\

            \hline
            $\mathcal{VS}^{dc,dis} \subseteq \mathcal{VS}$ & $\mathcal{VS}^{dc,dis}$ denotes the subset of co-located VRE and storage resources with a storage DC discharge component \\
            
            \hline
            $\mathcal{VS}^{dc,cha} \subseteq \mathcal{VS}$ & $\mathcal{VS}^{dc,cha}$ denotes the subset of co-located VRE and storage resources with a storage DC charge component \\
            
            \hline
            $\mathcal{VS}^{ac,dis} \subseteq \mathcal{VS}$ & $\mathcal{VS}^{ac,dis}$ denotes the subset of co-located VRE and storage resources with a storage AC discharge component \\
            
            \hline
            $\mathcal{VS}^{ac,cha} \subseteq \mathcal{VS}$ & $\mathcal{VS}^{ac,cha}$ denotes the subset of co-located VRE and storage resources with a storage AC charge component \\
        
            \hline
        \end{longtable}

    \subsection{Model Constants}
    \label{model_constants}
        \begin{longtable}[c]{|c|p{9.8cm}|}    
             \caption{GenX VRE-Storage Module Constants} 
             \label{constants}\\
            
            \hline
            \textbf{Notation} & \textbf{Description} \\
        
            \hline
            $\overline{\Delta}_{y, z}$ &  Existing installed capacity of the grid connection component of technology \textit{y} in zone \textit{z} [MW AC] \\
            
            \hline
            $\overline{\Delta}^{pv}_{y, z}$ & Existing installed capacity of the solar PV component of technology \textit{y} in zone \textit{z} [MW DC] \\
        
            \hline
            $\overline{\Delta}^{wind}_{y, z}$ & Existing installed capacity of the wind component of technology \textit{y} in zone \textit{z} [MW AC] \\
            
            \hline
            $\overline{\Delta}^{energy}_{y, z}$ & Existing installed storage energy capacity of technology \textit{y} in zone \textit{z} [MWh] \\
        
            \hline
            $\overline{\Delta}^{inv}_{y, z}$ & Existing installed inverter capacity of technology \textit{y} in zone \textit{z} [MW AC] \\
        
            \hline
            $\overline{\Delta}^{dc, dis}_{y, z}$ & Existing installed storage DC discharge capacity of technology \textit{y} in zone \textit{z} [MW DC] \\
        
            \hline
            $\overline{\Delta}^{dc, cha}_{y, z}$ & Existing installed storage DC charge capacity of technology \textit{y} in zone \textit{z} [MW DC] \\
        
            \hline
            $\overline{\Delta}^{ac, dis}_{y, z}$ & Existing installed storage AC discharge capacity of technology \textit{y} in zone \textit{z} [MW AC] \\
        
            \hline
            $\overline{\Delta}^{ac, cha}_{y, z}$ & Existing installed storage AC charge capacity of technology \textit{y} in zone \textit{z} [MW AC] \\
            
            \hline
            $\eta^{ILR, pv}_{y, z}$ & Inverter loading ratio (the solar PV capacity sizing to inverter capacity built) of technology \textit{y} in zone \textit{z} [\%]\\
        
            \hline
            $\eta^{ILR, wind}_{y, z}$ & Ratio of the wind capacity sizing to grid connection capacity built of technology \textit{y} in zone \textit{z} [\%]\\
            
            \hline
            $\eta^{dc, cha}_{y, z}$ & Single-trip efficiency of storage DC charging for technology \textit{y} in zone \textit{z} [\%]\\
            
            \hline
            $\eta^{dc, dis}_{y, z}$ & Single-trip efficiency of storage DC discharging for technology \textit{y} in zone \textit{z} [\%] \\
        
            \hline
            $\eta^{ac, cha}_{y, z}$ & Single-trip efficiency of storage AC charging for technology \textit{y} in zone \textit{z} [\%]\\
            
            \hline
            $\eta^{ac, dis}_{y, z}$ & Single-trip efficiency of storage AC discharging for technology \textit{y} in zone \textit{z} [\%] \\
            
            \hline
            $\eta^{inverter}_{y, z}$ & Inverter efficiency representing losses from converting DC to AC power and vice versa for technology \textit{y} in zone \textit{z} [\%] \\
            
            \hline
            $\rho^{max, pv}_{y, t, z}$ & Maximum available generation per unit of solar PV installed capacity during time step \textit{t} for the solar PV component of technology \textit{y} in zone \textit{z} [\%] \\
        
            \hline
            $\rho^{max, wind}_{y, t, z}$ & Maximum available generation per unit of wind installed capacity during time step \textit{t} for the wind component of technology \textit{y} in zone \textit{z} [\%] \\
            
            \hline
            $\underline{\Omega}_{y, z}$ & Minimum grid connection capacity requirement for technology \textit{y} in zone \textit{z} [MW AC]\\
        
            \hline
            $\underline{\Omega}^{pv}_{y, z}$ & Minimum solar PV capacity requirement for technology \textit{y} in zone \textit{z} [MW DC]\\
        
            \hline
            $\underline{\Omega}^{wind}_{y, z}$ & Minimum wind capacity requirement for technology \textit{y} in zone \textit{z} [MW AC]\\
            
            \hline
            $\underline{\Omega}^{energy}_{y, z}$ & Minimum storage energy capacity requirement for technology \textit{y} in zone \textit{z}  [MWh]\\
            
            \hline
            $\underline{\Omega}^{inv}_{y, z}$ & Minimum inverter capacity requirement for technology \textit{y} in zone \textit{z} [MW AC]\\
        
            \hline
            $\underline{\Omega}^{dc, dis}_{y, z}$ & Minimum storage DC discharge capacity requirement for technology \textit{y} in zone \textit{z} [MW DC]\\
        
            \hline
            $\underline{\Omega}^{dc, cha}_{y, z}$ & Minimum storage DC charge capacity requirement for technology \textit{y} in zone \textit{z} [MW DC]\\
        
            \hline
            $\underline{\Omega}^{ac, dis}_{y, z}$ & Minimum storage AC discharge capacity requirement for technology \textit{y} in zone \textit{z} [MW AC]\\
        
            \hline
            $\underline{\Omega}^{ac, cha}_{y, z}$ & Minimum storage AC charge capacity requirement for technology \textit{y} in zone \textit{z} [MW AC]\\
            
            \hline
            $\overline{\Omega}_{y, z}$ & Maximum grid connection capacity requirement for technology \textit{y} in zone \textit{z} [MW AC]\\
        
            \hline
            $\overline{\Omega}^{pv}_{y, z}$ & Maximum solar PV capacity requirement for technology \textit{y} in zone \textit{z} [MW DC]\\
        
            \hline
            $\overline{\Omega}^{wind}_{y, z}$ & Maximum wind capacity requirement for technology \textit{y} in zone \textit{z} [MW AC]\\
            
            \hline
            $\overline{\Omega}^{energy}_{y, z}$ & Maximum storage energy capacity requirement for technology \textit{y} in zone \textit{z} [MWh]\\
            
            \hline
            $\overline{\Omega}^{inv}_{y, z}$ & Maximum inverter capacity requirement for technology \textit{y} in zone \textit{z} [MW AC]\\
        
            \hline
            $\overline{\Omega}^{dc, dis}_{y, z}$ & Maximum storage DC discharge capacity requirement for technology \textit{y} in zone \textit{z} [MW DC]\\
        
            \hline
            $\overline{\Omega}^{dc, cha}_{y, z}$ & Maximum storage DC charge capacity requirement for technology \textit{y} in zone \textit{z} [MW DC]\\
        
            \hline
            $\overline{\Omega}^{ac, dis}_{y, z}$ & Maximum storage AC discharge capacity requirement for technology \textit{y} in zone \textit{z} [MW AC]\\
        
            \hline
            $\overline{\Omega}^{ac, cha}_{y, z}$ & Maximum storage AC charge capacity requirement for technology \textit{y} in zone \textit{z} [MW AC]\\
            
            \hline
            $\mu^{dc, stor}_{y, z}$ & Power-to-energy ratio for the storage DC component of technology \textit{y} in zone \textit{z} [\%]\\
        
            \hline
            $\mu^{ac, stor}_{y, z}$ & Power-to-energy ratio for the storage AC component of technology \textit{y} in zone \textit{z} [\%]\\
            
            \hline
            $\pi^{INVEST}_{y, z}$ & Investment cost (annual amortization of total construction cost) for the grid connection component power capacity of technology \textit{y} in zone \textit{z} [\$/MW AC-yr]\\
        
            \hline
            $\pi^{INVEST, pv}_{y, z}$ & Investment cost (annual amortization of total construction cost) for the solar PV component power capacity of technology \textit{y} in zone \textit{z} [\$/MW DC-yr]\\
        
            \hline
            $\pi^{INVEST, wind}_{y, z}$ & Investment cost (annual amortization of total construction cost) for the wind component power capacity of technology \textit{y} in zone \textit{z} [\$/MW AC-yr]\\
            
            \hline
            $\pi^{INVEST, energy}_{y, z}$ & Investment cost (annual amortization of total construction cost) for the storage component energy capacity of technology \textit{y} in zone \textit{z} [\$/MWh-yr]\\
            
            \hline
            $\pi^{INVEST, inv}_{y, z}$ & Investment cost (annual amortization of total construction cost) for the inverter component power capacity of technology \textit{y} in zone \textit{z} [\$/MW AC-yr]\\
        
            \hline
            $\pi^{INVEST, dc, dis}_{y, z}$ & Investment cost (annual amortization of total construction cost) for the storage DC discharge component power capacity of technology \textit{y} in zone \textit{z} [\$/MW DC-yr]\\
        
            \hline
            $\pi^{INVEST, dc, cha}_{y, z}$ & Investment cost (annual amortization of total construction cost) for the storage DC charge component power capacity of technology \textit{y} in zone \textit{z} [\$/MW DC-yr]\\
        
            \hline
            $\pi^{INVEST, ac, dis}_{y, z}$ & Investment cost (annual amortization of total construction cost) for the storage AC discharge component power capacity of technology \textit{y} in zone \textit{z} [\$/MW AC-yr]\\
        
            \hline
            $\pi^{INVEST, ac, cha}_{y, z}$ & Investment cost (annual amortization of total construction cost) for the storage AC charge component power capacity of technology \textit{y} in zone \textit{z} [\$/MW AC-yr]\\
            
            \hline
            $\pi^{FOM}_{y, z}$ & Fixed O\&M cost of the grid connection component of technology \textit{y} in zone \textit{z} [\$/MW AC-yr]\\
        
            \hline
            $\pi^{FOM, pv}_{y, z}$ & Fixed O\&M cost of the solar PV component of technology \textit{y} in zone \textit{z} [\$/MW DC-yr]\\
        
            \hline
            $\pi^{FOM, wind}_{y, z}$ & Fixed O\&M cost of the wind component of technology \textit{y} in zone \textit{z} [\$/MW AC-yr]\\
            
            \hline
            $\pi^{FOM, energy}_{y, z}$ & Fixed O\&M cost of the storage energy component of technology \textit{y} in zone \textit{z} [\$/MWh-yr]\\
            
            \hline
            $\pi^{FOM, inv}_{y, z}$ & Fixed O\&M cost of the inverter component of technology \textit{y} in zone \textit{z} [\$/MW AC-yr]\\
        
            \hline
            $\pi^{FOM, dc, dis}_{y, z}$ & Fixed O\&M cost of the storage DC discharge component of technology \textit{y} in zone \textit{z} [\$/MW DC-yr]\\
        
            \hline
            $\pi^{FOM, dc, cha}_{y, z}$ & Fixed O\&M cost of the storage DC charge component of technology \textit{y} in zone \textit{z} [\$/MW DC-yr]\\
        
            \hline
            $\pi^{FOM, ac, dis}_{y, z}$ & Fixed O\&M cost of the storage AC discharge component of technology \textit{y} in zone \textit{z} [\$/MW AC-yr]\\
        
            \hline
            $\pi^{FOM, ac, cha}_{y, z}$ & Fixed O\&M cost of the storage AC charge component of technology \textit{y} in zone \textit{z} [\$/MW AC-yr]\\
            
            \hline
            $\pi^{VOM, pv}_{y, z}$ & Variable O\&M cost of the solar PV component of technology \textit{y} in zone \textit{z} [\$/MWh]\\
        
            \hline
            $\pi^{VOM, wind}_{y, z}$ & Variable O\&M cost of the wind component of technology \textit{y} in zone \textit{z} [\$/MWh]\\
        
            \hline
            $\pi^{VOM, dc, dis}_{y, z}$ & Variable O\&M cost of the storage DC discharge component of technology \textit{y} in zone \textit{z} [\$/MWh]\\
        
            \hline
            $\pi^{VOM, dc, cha}_{y, z}$ & Variable O\&M cost of the storage DC charge component of technology \textit{y} in zone \textit{z} [\$/MWh]\\
        
            \hline
            $\pi^{VOM, ac, dis}_{y, z}$ & Variable O\&M cost of the storage AC discharge component of technology \textit{y} in zone \textit{z} [\$/MWh]\\
        
            \hline
            $\pi^{VOM, ac, cha}_{y, z}$ & Variable O\&M cost of the storage AC charge component of technology \textit{y} in zone \textit{z} [\$/MWh]\\

            \hline
        \end{longtable}

    \subsection{Model Decision Variables}
    \label{model_variables}
        \begin{longtable}[c]{|c|p{9.8cm}|}
            \caption{GenX VRE-Storage Module Variables} \label{variables}\\
        
            \hline
            \textbf{Notation} & \textbf{Description} \\
            
            \hline
            $\Omega_{y, z}$ & Installed capacity of the grid connection component of technology \textit{y} in zone \textit{z} [MW AC]\\
        
            \hline
            $\Omega^{pv}_{y, z}$ & Installed capacity of the solar PV component of technology \textit{y} in zone \textit{z} [MW DC]\\
        
            \hline
            $\Omega^{wind}_{y, z}$ & Installed capacity of the wind component of technology \textit{y} in zone \textit{z} [MW AC]\\
            
            \hline
            $\Omega^{energy}_{y, z}$ & Installed energy capacity of the storage component of technology \textit{y} in zone \textit{z} [MWh]\\
            
            \hline
            $\Omega^{inv}_{y, z}$ & Installed capacity of the inverter component of technology \textit{y} in zone \textit{z} [MW AC]\\
        
            \hline
            $\Omega^{dc, dis}_{y, z}$ & Installed capacity of the storage DC discharge component of technology \textit{y} in zone \textit{z} [MW DC]\\
        
            \hline
            $\Omega^{dc, cha}_{y, z}$ & Installed capacity of the storage DC charge component of technology \textit{y} in zone \textit{z} [MW DC]\\
        
            \hline
            $\Omega^{ac, dis}_{y, z}$ & Installed capacity of the storage AC discharge component of technology \textit{y} in zone \textit{z} [MW AC]\\
        
            \hline
            $\Omega^{ac, cha}_{y, z}$ & Installed capacity of the storage AC charge component of technology \textit{y} in zone \textit{z} [MW AC]\\
            
            \hline
            $\Delta_{y, z}$ & Retired capacity of the grid connection component of technology \textit{y} in zone \textit{z} [MW AC] \\
        
            \hline
            $\Delta^{pv}_{y, z}$ & Retired capacity of the solar PV component of technology \textit{y} in zone \textit{z} [MW DC] \\
        
            \hline
            $\Delta^{wind}_{y, z}$ & Retired capacity of the wind component of technology \textit{y} in zone \textit{z} [MW AC] \\
            
            \hline
            $\Delta^{energy}_{y, z}$ & Retired energy capacity of the storage component of technology \textit{y} in zone \textit{z} [MWh] \\
            
            \hline
            $\Delta^{inv}_{y, z}$ & Retired capacity of the inverter component of technology \textit{y} in zone \textit{z} [MW AC] \\
        
            \hline
            $\Delta^{dc, dis}_{y, z}$ & Retired capacity of the storage DC discharge component of technology \textit{y} in zone \textit{z} [MW DC] \\
        
            \hline
            $\Delta^{dc, cha}_{y, z}$ & Retired capacity of the storage DC charge component of technology \textit{y} in zone \textit{z} [MW DC] \\
        
            \hline
            $\Delta^{ac, dis}_{y, z}$ & Retired capacity of the storage AC discharge component of technology \textit{y} in zone \textit{z} [MW AC] \\
        
            \hline
            $\Delta^{ac, cha}_{y, z}$ & Retired capacity of the storage AC charge component of technology \textit{y} in zone \textit{z} [MW AC] \\
            
            \hline
            $\Delta^{total}_{y, z}$ & Total installed capacity of the grid connection component of technology \textit{y} in zone \textit{z} [MW AC] \\
        
            \hline
            $\Delta^{total, pv}_{y, z}$ & Total installed capacity of the solar PV component of technology \textit{y} in zone \textit{z} [MW DC] \\
        
            \hline
            $\Delta^{total, wind}_{y, z}$ & Total installed capacity of the wind component of technology \textit{y} in zone \textit{z} [MW AC] \\
            
            \hline
            $\Delta^{total, energy}_{y, z}$ & Total installed energy capacity of the storage component of technology \textit{y} in zone \textit{z} [MWh] \\
            
            \hline
            $\Delta^{total, inv}_{y, z}$ & Total installed capacity of the inverter component of technology \textit{y} in zone \textit{z} [MW AC] \\
        
            \hline
            $\Delta^{total, dc, dis}_{y, z}$ & Total installed capacity of the storage DC discharge component of technology \textit{y} in zone \textit{z} [MW DC] \\
        
            \hline
            $\Delta^{total, dc, cha}_{y, z}$ & Total installed capacity of the storage DC charge component of technology \textit{y} in zone \textit{z} [MW DC] \\
        
            \hline
            $\Delta^{total, ac, dis}_{y, z}$ & Total installed capacity of the storage AC discharge component of technology \textit{y} in zone \textit{z} [MW AC] \\
        
            \hline
            $\Delta^{total, ac, cha}_{y, z}$ & Total installed capacity of the storage AC charge component of technology \textit{y} in zone \textit{z} [MW AC] \\
            
            \hline
            $\Theta^{pv}_{y, z, t}$ & Energy generated by the solar PV DC component of technology \textit{y} at time step \textit{t} in zone \textit{z} [MWh]\\
        
            \hline
            $\Theta^{wind}_{y, z, t}$ & Energy generated by the wind AC component of technology \textit{y} at time step \textit{t} in zone \textit{z} [MWh]\\
            
            \hline
            $\Theta^{dc}_{y, z, t}$ & Energy discharged by the storage DC component of technology \textit{y} at time step \textit{t} in zone \textit{z} [MWh]\\
        
            \hline
            $\Theta^{ac}_{y, z, t}$ & Energy generated by the storage AC component of technology \textit{y} at time step \textit{t} in zone \textit{z} [MWh]\\
            
            \hline
            $\Theta_{y, z, t}$ & Energy injected into the grid by technology \textit{y} at time step \textit{t} in zone \textit{z} [MWh]\\
            
            \hline
            $\Pi^{dc}_{y, z, t}$ & Energy withdrawn from the VRE resources and grid to charge the storage DC component for technology \textit{y} at time step \textit{t} in zone \textit{z} [MWh]  \\
        
            \hline
            $\Pi^{ac}_{y, z, t}$ & Energy withdrawn from the VRE resources and grid to charge the storage AC component for technology \textit{y} at time step \textit{t} in zone \textit{z} [MWh]  \\
            
            \hline
            $\Pi_{y, z, t}$ & Energy withdrawn from the grid for technology \textit{y} (the storage component) at time step \textit{t} in zone \textit{z} [MWh]  \\
            
            \hline
            $\Gamma_{y, z, t}$ & Stored energy level of the storage component of technology \textit{y} at the end of time step \textit{t} in zone \textit{z} [MWh] \\

            \hline 
            $\Theta^{CRM,dc}_{y,z,t}$ & ``Virtual" energy discharged by the storage DC component that contributes to the capacity reserve margin for technology $y$ at time step $t$ in zone $z$ - only applicable for co-located VRE and storage resources with activated capacity reserve margin policies, $y \in \mathcal{VS}^{dc,dis}$ [MWh] \\

            \hline
            $\Pi^{CRM,dc}_{y,z,t}$ & ``Virtual" energy withdrawn by the storage DC component from the grid by technology $y$ at time step $t$ in zone $z$ - only applicable for co-located VRE and storage resources with activated capacity reserve margin policies, $y \in \mathcal{VS}^{dc,cha}$ [MWh] \\

            \hline
            $\Theta^{CRM,ac}_{y,z,t}$ & ``Virtual" energy discharged by the storage AC component that contributes to the capacity reserve margin for technology $y$ at time step $t$ in zone $z$ - only applicable for co-located VRE and storage resources with activated capacity reserve margin policies, $y \in \mathcal{VS}^{ac,dis}$ [MWh] \\
            
            \hline
            $\Pi^{CRM,ac}_{y,z,t}$ &``Virtual" energy withdrawn by the storage AC component from the grid by technology $y$ at time step $t$ in zone $z$ - only applicable for co-located VRE and storage resources with activated capacity reserve margin policies, $y \in \mathcal{VS}^{ac,cha}$ [MWh] \\

            \hline
            $\Gamma^{CRM}_{y,z,t}$ & Total ``virtual" state of charge being held in reserves for technology $y$ at time step $t$ in zone $z$ - only applicable for co-located VRE and storage resources with activated capacity reserve margin policies, $y \in \mathcal{VS}^{stor}$ [MWh]\\
            
            \hline
        \end{longtable}

    \subsection{Model Objective Function Additions}
    There are fixed O\&M and overnight capital costs associated with the power capacity decisions of solar PV ($\Omega^{pv}_{y, z}$), wind ($\Omega^{wind}_{y, z}$), inverter ($\Omega^{inv}_{y, z}$), storage DC discharge ($\Omega^{dc, dis}_{y, z}$), storage DC charge ($\Omega^{dc, cha}_{y, z}$), storage AC discharge ($\Omega^{ac, dis}_{y, z}$), storage AC charge ($\Omega^{ac, cha}_{y, z}$), and storage energy capacity ($\Omega^{energy}_{y, z}$) decisions. Grid connection and inverter decision variables added in this new formulation ensure that there is a more detailed representation and separation of the DC and AC connection. There are six operational decision variables for a co-located resource, enabling variable O\&M costs to also exist for the solar PV generation ($\Theta^{pv}_{y, z, t}$), wind generation ($\Theta^{wind}_{y, z, t}$), storage DC discharge ($\Theta^{dc}_{y, z, t}$), storage DC charge ($\Pi^{dc}_{y, z, t}$), storage AC discharge ($\Theta^{ac}_{y, z, t}$), and storage AC charge ($\Pi^{ac}_{y, z, t}$) components of the resource. The following mathematical formulation is added to the least-cost objective function for the costs of co-located VRE and storage resources:
    \label{obj_function}
        
        \begin{equation}
        \begin{split}
                    \sum_{y \in \mathcal{VS}} \sum_{z \in Z}((\pi^{INVEST}_{y, z} \times \Omega_{y, z}) + (\pi^{FOM}_{y, z} \times \Delta^{total}_{y, z})) +\\
                    \sum_{y \in \mathcal{VS}^{pv}} \sum_{z \in Z}((\pi^{INVEST, pv}_{y, z} \times \Omega^{pv}_{y, z}) + (\pi^{FOM, pv}_{y, z} \times \Delta^{total,pv}_{y, z})) +\\
                    \sum_{y \in \mathcal{VS}^{wind}} \sum_{z \in Z}((\pi^{INVEST, wind}_{y, z} \times \Omega^{wind}_{y, z}) + (\pi^{FOM, wind}_{y, z} \times \Delta^{total,wind}_{y, z})) +\\
                    \sum_{y \in \mathcal{VS}^{stor}} \sum_{z \in Z} (( \pi^{INVEST, energy}_{y, z} \times \Omega^{energy}_{y, z}) +  (\pi^{FOM, energy}_{y, z} \times \Delta^{total, energy}_{y, z})) +\\
                    \sum_{y \in \mathcal{VS}^{inv}} \sum_{z \in Z} ((\pi^{INVEST, inv}_{y, z} \times \Omega^{inv}_{y, z}) + (\pi^{FOM, inv}_{y, z} \times \Delta^{total, inv}_{y, z})) + \\
                    \sum_{y \in \mathcal{VS}^{asym,dc,dis}} \sum_{z \in Z} ((\pi^{INVEST,dc,dis}_{y, z} \times \Omega^{dc, dis}_{y, z}) + (\pi^{FOM, dc, dis}_{y, z} \times \Delta^{total, dc, dis}_{y, z})) + \\
                    \sum_{y \in \mathcal{VS}^{asym,dc,cha}} \sum_{z \in Z} ((\pi^{INVEST,dc,cha}_{y, z} \times \Omega^{dc, cha}_{y, z}) + (\pi^{FOM, dc, cha}_{y, z} \times \Delta^{total, dc, cha}_{y, z})) + \\
                    \sum_{y \in \mathcal{VS}^{asym,ac,dis}} \sum_{z \in Z} ((\pi^{INVEST,ac,dis}_{y, z} \times \Omega^{ac, dis}_{y, z}) + (\pi^{FOM, ac, dis}_{y, z} \times \Delta^{total, ac, dis}_{y, z})) + \\
                    \sum_{y \in \mathcal{VS}^{asym,ac,cha}} \sum_{z \in Z} ((\pi^{INVEST,ac,cha}_{y, z} \times \Omega^{ac, cha}_{y, z}) + (\pi^{FOM, ac, cha}_{y, z} \times \Delta^{total, ac, cha}_{y, z})) + \\
                    \sum_{y \in \mathcal{VS}^{pv}} \sum_{z \in Z} \sum_{t \in T}
                    (\pi^{VOM, pv}_{y, z} \times \Theta^{pv}_{y, z, t}) + 
                    \sum_{y \in \mathcal{VS}^{wind}} \sum_{z \in Z} \sum_{t \in T}
                    (\pi^{VOM, wind}_{y, z} \times \Theta^{wind}_{y, z, t}) + \\
                    \sum_{y \in \mathcal{VS}^{dc,dis}} \sum_{z \in Z} \sum_{t \in T}
                    (\pi^{VOM, dc, dis}_{y, z} \times \Theta^{dc}_{y, z, t}) + 
                    \sum_{y \in \mathcal{VS}^{dc,cha}} \sum_{z \in Z} \sum_{t \in T}
                    (\pi^{VOM, dc, cha}_{y, z} \times \Pi^{dc}_{y, z, t}) + \\
                    \sum_{y \in \mathcal{VS}^{ac,dis}} \sum_{z \in Z} \sum_{t \in T}
                    (\pi^{VOM, ac, dis}_{y, z} \times \Theta^{ac}_{y, z, t}) + 
                    \sum_{y \in \mathcal{VS}^{ac,cha}} \sum_{z \in Z} \sum_{t \in T}
                    (\pi^{VOM, ac, cha}_{y, z} \times \Pi^{ac}_{y, z, t})
        \end{split}
        \end{equation} 

\newpage

\section{Model Information}

    \subsection{PowerGenome Overview}
    \label{app_pg}
    
    PowerGenome is an open-source scenario generator for GenX that creates inputs for power system optimization models. We used PowerGenome to gather and synthesize the necessary inputs for the case study on existing generating units, inter-zonal transmission between regions, hourly load profiles, vehicle electrification and flexible demand profiles, hourly capacity factors for wind and solar sites, and cost projections for the planning period. Data for PowerGenome is assembled from a few different resources that are compiled in the Public Utility Data Liberation project database: EIA (for regional cost multipliers), EPA (for IPM regions), and NREL (projected costs of technologies) \cite{pg_github, PUDL}.

    \subsection{Cost Analysis Description}
    \label{app_costs_analysis}
    
    There are two configurations to build co-located VRE and storage resources, DC- and AC-coupled systems, as described further in \cite{nrel_pv_battery_2018, nrel_pv_battery_2021}. For the purposes of this case study and the cost analysis, the DC-coupled configuration is utilized. From co-locating resources, there are cost declines from sharing a single bidirectional inverter, installation labor equipment, engineering, procurement, and construction (EPC) overhead, sales tax, land acquisition, permitting fees, interconnection fees, transmission lines, contingency, developer overhead, and EPC/developer profit. There are additionally some cost increases in the structural and electrical balance of system (BOS), specifically due to the additional DC wiring and foundation for holding the inverters and transformers. A co-located 100 MW PV DC and 60 MW/240 MWh battery DC-coupled system in comparison to the standalone resources saves system costs by a net 6.3\% \cite{nrel_pv_battery_2018, nrel_pv_battery_2021, nrel_pv_bat_2022}.
    
    \textbf{Solar PV, Storage, Inverter, and Grid Connection Investment Costs}: The model enables standalone solar, co-located solar and storage, and standalone storage resources to be built. Because the model can configure each site to be built either as co-located with storage or as a standalone resource, the costs for the solar PV component, inverter, battery, and interconnection must resemble the costs for either a co-located or standalone resource. If all of the system soft cost savings from co-location were captured in this cost breakdown, standalone sites would be cheaper in the model than they are in the real-world, biasing VRE buildout in the model. The full net 6.3\% of cost savings from co-location is thus not captured in the cost analysis since the soft cost savings are not incorporated, but all of the hardware cost savings are captured. The hardware cost savings result in a roughly 2-3\% cost decline from co-location relative to two standalone PV and storage sites. More specifically, the cost savings captured relate to the inverter and the interconnection fees \cite{dataset_nrel_2021, nrel_pv_bat_2022}. The following steps were taken to breakdown each cost category:

    \begin{table}[H]
        \begin{center}
        \caption{\textbf{2021 DC Cost Breakdown (\$ million) of Solar PV Panels (100 MW DC), Battery Pack (60 MW DC/240 MWh), and Inverter (77 MW AC) for a Co-located or Standalone Solar and Battery Resource.} This table was calculated and adapted from NREL's AC cost breakdown with the inverter costs split out of the solar PV standalone and storage standalone sites to isolate the AC and DC components \cite{nrel_pv_battery_2021, dataset_nrel_2021}. Dollars are in 2020 USD.}
        \label{tab:appcostsdc}
        \resizebox{\textwidth}{!}{
        \begin{tabular}{l|c|c|c}
        \hline
        & \textbf{Standalone PV Cost} & 
        \textbf{Standalone Storage Cost} & \textbf{Inverter} \\
        & \textbf{Breakdown (100 MW DC)} & \textbf{Breakdown (60 MW/240 MWh)} & \textbf{(77 MW AC)} \\
        \hline

        PV Module & 33.4 & 0 & 0 \\
        Lithium-Ion Battery & 0 & 53.2 & 0 \\
        Solar Inverter & 0 & 0 & 0 \\
        Bidirectional Inverter & 0 & 0 & 4.5 \\
        Structural BOS & 12.5 & 0.8 & 0 \\
        Electrical BOS & 7.2 & 9.2 & 0 \\
        Installation Labor/Equip & 11.1 & 4.1 & 0 \\
        EPC Overhead & 5.4 & 2.6 & 0 \\
        Sales Tax & 3.4 & 4.1 & 0.3 \\
        Land Acquisition & 0 & 0 & 0 \\
        Permitting Fee & 0.2 & 0.2 & 0 \\
        Interconnection Fee & 1.4 & 0.8 & 0 \\
        Contingency & 2.1 & 2.5 & 0.2 \\
        Developer Overhead & 2.8 & 3.4 & 0.2 \\
        EPC/Developer Net Profit & 3.7 & 4.5 & 0.3 \\
        \hline
        Sum & 83.2 & 85.4 & 5.5 \\
        \hline
        \end{tabular}}
        \end{center}
    \end{table}

    \begin{enumerate}
        \item \textbf{Derive the bottom-up solar PV DC, battery pack, and inverter costs in 2021 (\$ millions) using NREL's AC cost breakdown \cite{nrel_pv_battery_2021, dataset_nrel_2021}}: NREL's bottom-up cost analysis on building a DC-coupled co-located solar and battery system splits each cost contributor as a line item to differentiate the breakdown of costs for standalone solar PV (77 MW AC/100 MW DC), standalone battery (46 MW AC/60 MW DC/240 MWh), and co-located solar and battery resources (100 MW DC PV/60 MW DC storage/240 MWh storage) \cite{nrel_pv_battery_2021, dataset_nrel_2021}. However, while the cost breakdown from NREL captures the savings that occur from co-location, the breakdown does not isolate the costs spent on grid connection or the inverter. Since the goal of our study relies on modeling both the inverter and grid connection capacities separately from VRE and storage capacities, it is critical to find and differentiate DC and AC components and costs, particularly the grid connection and inverter from solar panels and battery packs. Utilizing NREL's AC cost breakdown and line items from \cite{nrel_pv_battery_2021, dataset_nrel_2021}, a further splitting of the costs is conducted to separate them into three categories as indicated in Table \ref{tab:appcostsdc} based upon the hardware cost savings that occur from co-locating resources: 1) PV modules (DC costs), 2) battery modules (DC costs), and 3) inverter costs (AC costs). While there are cost increases from the additional racks and wirings required by co-location for the structural and electrical BOS and soft cost decreases from installation labor, EPC overhead, sales tax, permitting fees, contingency, developer overhead, and EPC/developer net profit, these cost differentials are not captured in this analysis to ensure that the model does not make standalone solar PV or battery sites cheaper to build. For the line items pertaining to the structural BOS, electrical BOS, installation labor and equipment, EPC overhead, and permitting fee, the same costs for standalone PV and storage resources were assumed and allocated to their respective technologies. The cost savings for the inverter and interconnection fee were captured. The costs for the line items involving sales tax, contingency, developer overhead, and EPC/developer net profit were split amongst the PV, storage, and inverter according to the percent that each component contributes to the total technology and hardware costs (see Table \ref{tab:appcostsdc}). Grid connection (or interconnection capacity) costs represent spur line and other grid substation costs, but since PowerGenome calculated these separately, these costs were not included in this part of the analysis \cite{pg_github}.
        
        \item \textbf{Calculate 2021 \$/MW DC, \$/MWh, and \$/MW AC costs for each of the three components (the solar PV, storage, and inverter)}: 2021 \$/MW costs were calculated for the AC-components (in \$ million) as \$1.14/MW AC for PV (divided total costs by 77 MW AC) and \$0.38/MWh for storage (divided total costs by 240 MWh) \cite{nrel_pv_battery_2021}. 2021 \$/MW costs were calculated for the DC-components (in \$ million): \$0.83/MW DC for solar PV (divided total costs by 100 MW DC), \$0.35/MWh for storage (divided total costs by 240 MWh), and \$0.07/MW AC for the inverter (divided total costs by 77 MW AC) (Table \ref{tab:appcostsdc}).

        \item \textbf{Calculate the DC:AC cost ratios for solar PV and storage resources}: The DC:AC cost ratio is 0.73 for solar PV and 0.95 for storage resources. The inverter costs padded with the additional soft costs is estimated to be roughly 120\% of the hardware costs for a total of \$0.07/MW AC (in \$ millions).
                
        \item \textbf{Apply these ratios to NREL's ATB solar and storage 2021 projected costs \cite{nrel_atb_2022}}: The calculated ratios from the NREL studies for DC:AC costs are applied to the NREL's ATB cost values for 2021 to isolate the DC costs for solar PV and storage since NREL's ATB is utilized for technology cost projections in GenX \cite{pg_github,nrel_atb_2022}.
        
        \item \textbf{Calculate 2030 solar and storage cost projections}: With the 2021 DC costs isolated for solar PV and storage resources, the technology cost percent declines from NREL's ATB for both the low-cost and mid-cost scenarios are calculated, applied to the DC cost values, and averaged over the planning period 2022 to 2030. The storage resources follow the cost percent declines that the battery energy capital costs assume \cite{nrel_atb_2022}.

        \item \textbf{Calculate 2030 inverter cost projections}: The inverter is assumed to follow similar cost declines as the battery power capital costs, since the power capital costs for batteries include the inverter for the low- and mid-cost scenarios, and averaged over the 2022 to 2030 planning period \cite{nrel_pv_battery_2021, nrel_atb_2022}. At the end of this step, overnight capital costs for solar PV DC (\$/MW DC), storage (\$/MWh), and the inverter (\$/MW AC) components were calculated for 2030, matching key decision variables of GenX (see Table \ref{tab:costbreakdown}). The grid connection costs are calculated in PowerGenome and are site-dependent.
        
        \item \textbf{Annuitize costs}: The capital costs of the solar PV DC components were multiplied by the solar PV regional multipliers. The inverter and battery components were multiplied by battery regional multipliers from NREL since the costs of technologies may vary based upon the physical location \cite{Nrel_regional_multiplier}. Regional capital costs were annuitized assuming a 30-year lifespan for PV, a 15-year lifespan for the inverter, a 60-year lifespan for the substation, and a 15-year lifespan for batteries with spur line costs added to the grid connection costs \cite{nrel_atb_2022}. Solar PV DC assumed a 2.5\% WACC, the battery DC assumed a 2.5\% WACC, the inverter assumed a 2.5\% WACC, and the substation assumed a 4.4\% WACC \cite{nrel_atb_2022, LARSEN201647}. See Table \ref{tab:pvcosts} for a breakdown of the assumptions made on the WACC, lifespan, and regional multipliers for co-located solar and storage systems, standalone solar resources, standalone battery systems, co-located wind and storage systems, and standalone wind resources.      
        
        \item \textbf{Differentiate standalone storage and co-located storage costs}: While standalone and co-located storage costs are calculated in the same manner,  standalone storage systems have separate grid connection and spur line costs with varying regional multipliers and lifespans (which affect the investment \$/MW AC costs). These assumptions are noted in Table \ref{tab:pvcosts} for how the inverter and substation costs for standalone storage were annuitized differently than co-located storage resources. It is assumed that standalone storage sites have roughly 10 miles of spur line costs, which are calculated in PowerGenome and added to the grid connection costs \cite{spurline}.
    \end{enumerate}

    \textbf{Solar PV, Storage, and Grid Connection Fixed O\&M Costs}: PV fixed O\&M costs were averaged from NREL's ATB over the planning period. PV DC fixed O\&M costs were assumed to be 87\% of the amount assumed when modeling PV AC and the remaining 13\% of fixed O\&M costs were allocated as inverter fixed O\&M costs. For storage, fixed O\&M costs were assumed to be 2.5\% of capital costs \cite{nrel_pv_battery_2021, nrel_atb_2022}. No grid connection fixed O\&M costs are assumed to exist. Table \ref{tab:costbreakdown} showcases the finalized fixed O\&M costs for solar PV, inverter, and battery technologies. 
    
    \textbf{Wind Investment Costs}: The model enables standalone wind, co-located wind and storage, and standalone storage resources to be built, following these steps:
    
    \begin{enumerate}
        \item \textbf{Calculate wind costs by subtracting substation/grid connection AC costs}: Annuitized 2030 investment wind AC costs (\$/MW AC) were outputted by PowerGenome \cite{pg_github}. Grid connection costs were subtracted from wind AC costs for 2030 to isolate the wind costs (\$/MW AC) and grid connection/spur line costs (\$/MW AC). These costs were annuitized in PowerGenome using a 3.3\% WACC, 30-year lifespan, and wind regional multipliers \cite{nrel_atb_2022, Nrel_regional_multiplier, pg_github}. 
        
        \item \textbf{Add co-located battery costs}: For the option to co-locate batteries and wind resources, the same battery (\$/MWh) and inverter (\$/MW AC) costs that were previously calculated in the solar PV and battery analysis were used. 
    \end{enumerate}

    \textbf{Wind Fixed O\&M Costs}: Wind AC fixed O\&M costs were calculated from NREL's projections during the planning period \cite{nrel_atb_2022}. Table \ref{tab:costbreakdown} showcases the finalized fixed O\&M costs for wind resources.

    \textbf{Tax Credits \& Policies}: Various ITCs and PTCs from the Inflation Reduction Act (Section 45Y and 48E) are included and applied to solar PV and wind technologies between 2022-2032 (\$26/MWh PTC and 30\% ITC) \cite{IRA2022}. The IRA has a 30\% ITC on standalone and co-located storage resources (calculated to be 35.1\% after discount rates and corporate tax rates are applied), which is incorporated in the cost analysis. It is assumed that the PTC instead of the ITC is applied to all solar PV and wind resources. The IRA PTC on solar PV was annuitized to be \$12.7/MWh and the IRA PTC on wind was annuitized to be \$13.5/MWh when including discount and corporate tax rates.

\newpage

\section{Supplementary Results}

    \subsection{Solar to Inverter Capacity and Wind to Grid Connection Capacity Optimized Ratio vs. Interconnection Costs}
    \label{scatter_ratio_interconnection}
    
        \begin{figure}[H]
            \noindent
            \makebox[\textwidth]{\includegraphics[scale=0.5]{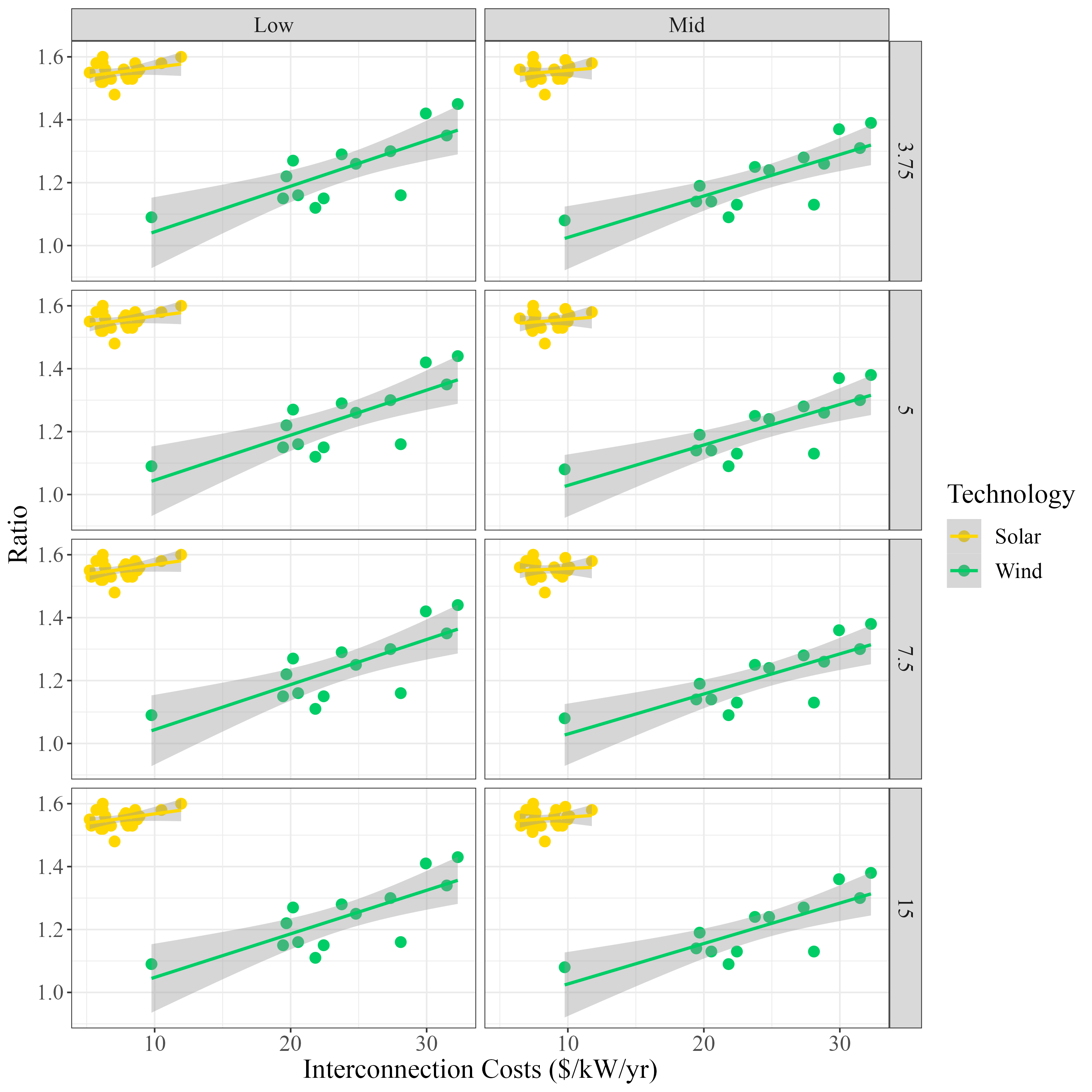}}
            \caption{\textbf{Scatter Plot of Optimized Ratio of Solar PV to Inverter Capacity and Wind to Grid Connection for Each Built Cluster vs. Annuitized Interconnection Costs (\$/kW/yr).\\
            } The annuitized interconnection costs for solar PV resources include both the annuitized inverter and interconnection capacity costs, while the annuitized interconnection costs for wind technologies do not include any inverter component costs.} \label{fig::scatter_interconnection_ratios}
        \end{figure}

    \newpage

    \subsection{Capacity Difference Plots}
    \label{capacity_difference}

        \begin{figure}[H]
            \noindent
            \makebox[\textwidth]{\includegraphics[scale=0.4]{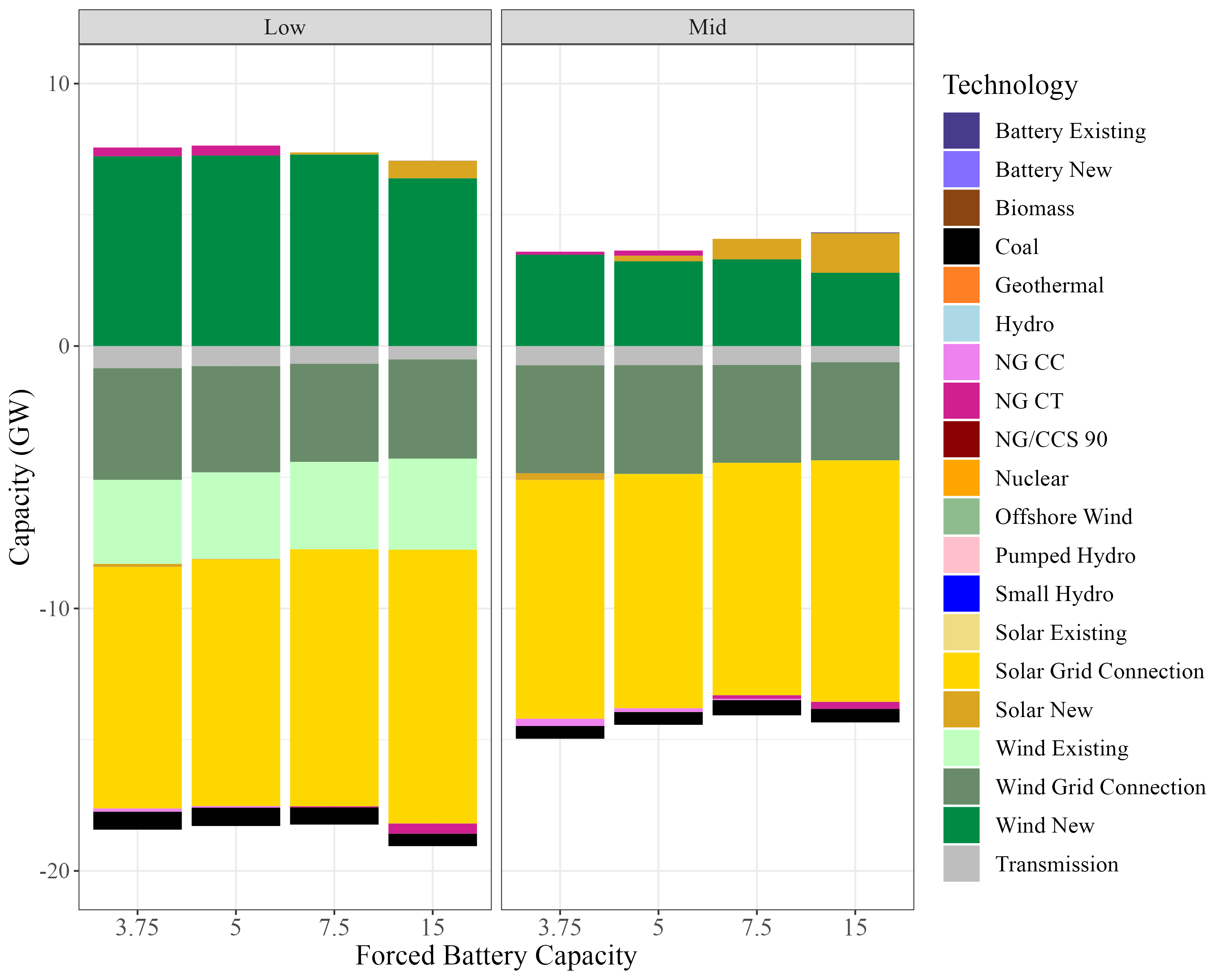}}
            \caption{\textbf{System-Wide Capacity Difference Plots (GW) between the Optimized Interconnection and Fixed Interconnection Scenario.
            }} \label{fig::capacity_difference_op}
        \end{figure}

        \begin{figure}[H]
            \noindent
            \makebox[\textwidth]{\includegraphics[scale=0.4]{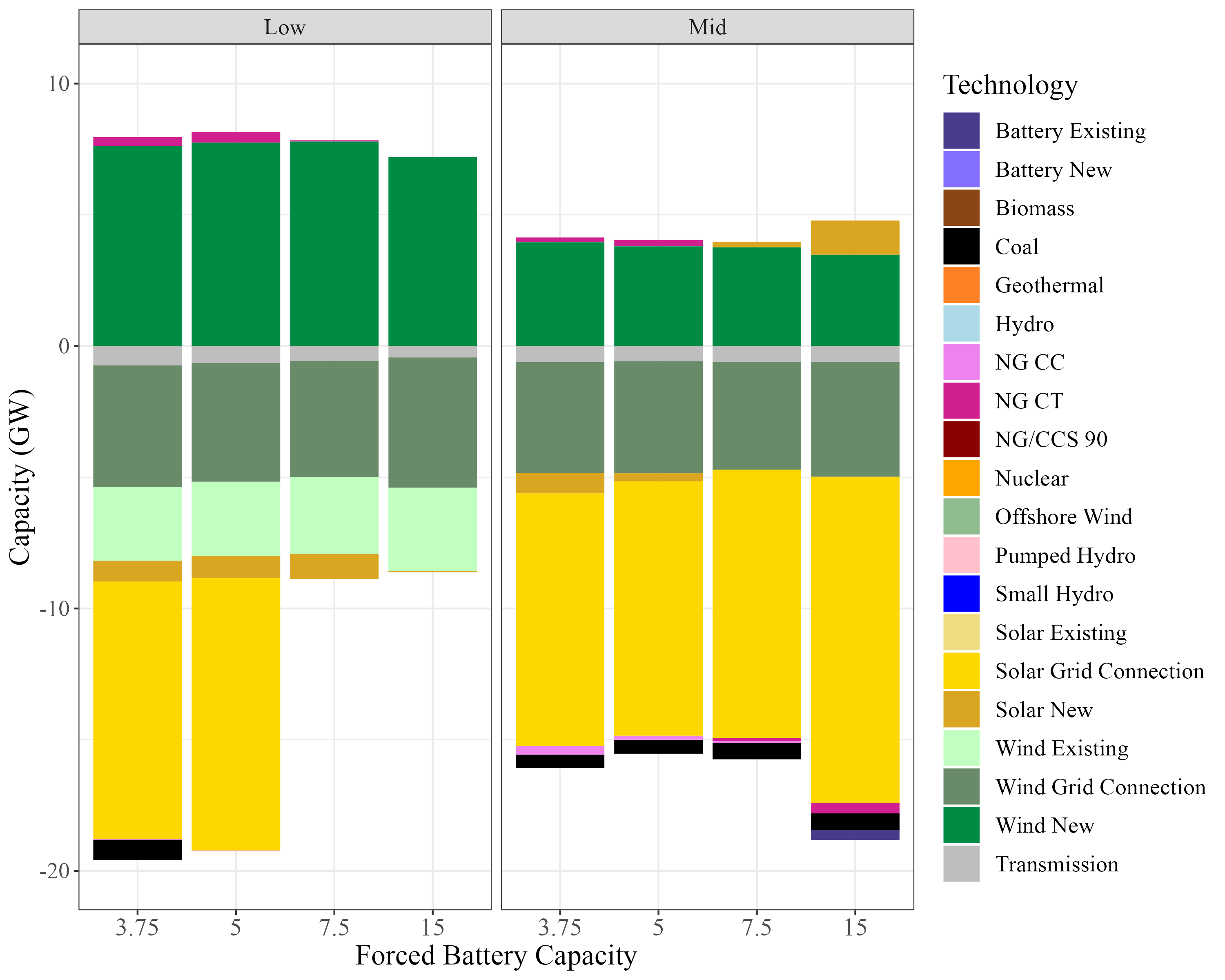}}
            \caption{\textbf{System-Wide Capacity Difference Plots (GW) between the Co-Located Storage and Fixed Interconnection Scenario.
            }} \label{fig::capacity_difference_co}
        \end{figure}

    \newpage

    \subsection{Utilization Rates of Inter-Zonal Transmission Line Heatmaps}
    \label{utilization_rate}

        \begin{figure}[H]
            \noindent
            \makebox[\textwidth]{\includegraphics[scale=0.4]{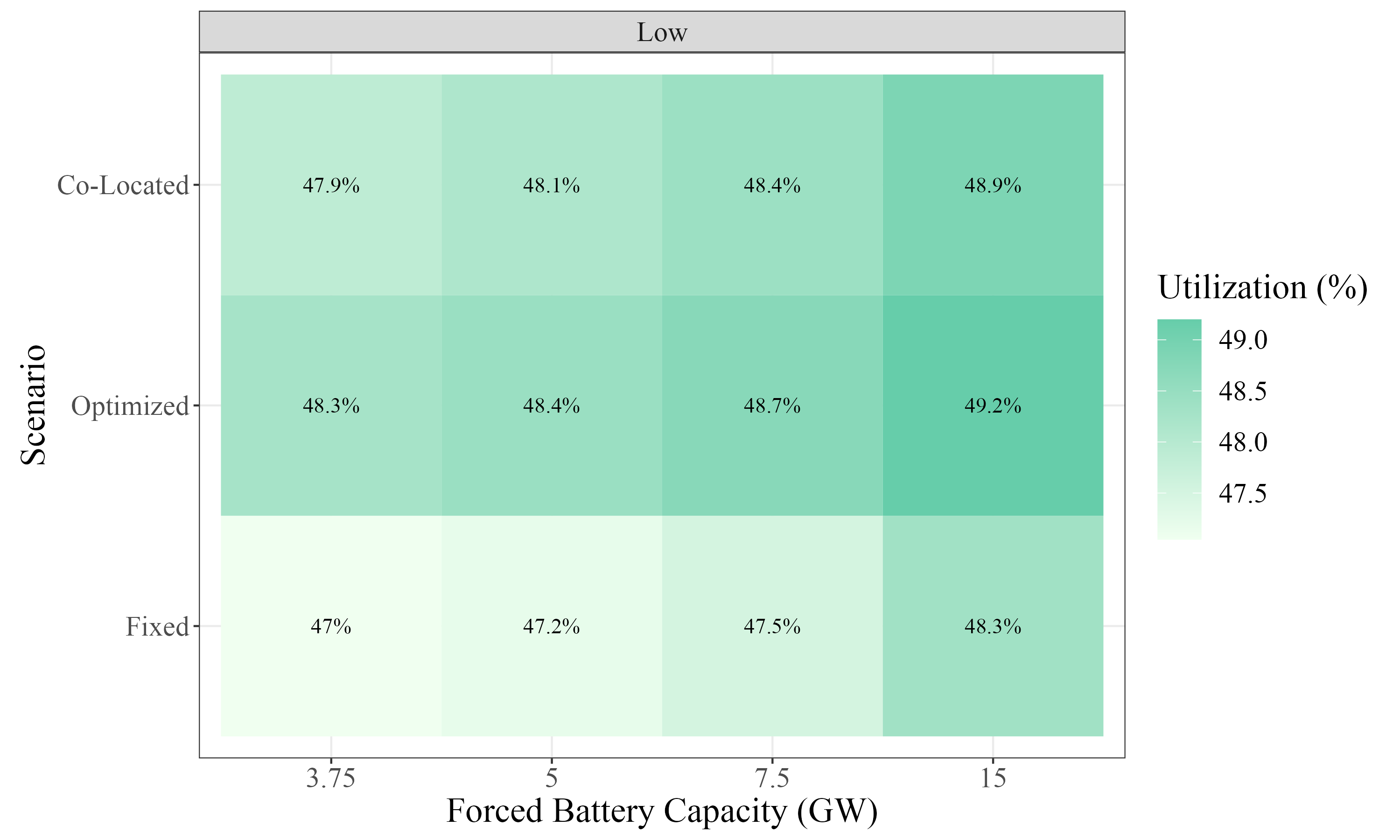}}
            \caption{\textbf{Utilization Rate of Inter-Zonal Transmission Line Heatmap for Low VRE-Cost Scenario.
            }} \label{fig::utilratelow}
        \end{figure}

        \begin{figure}[H]
            \noindent
            \makebox[\textwidth]{\includegraphics[scale=0.4]{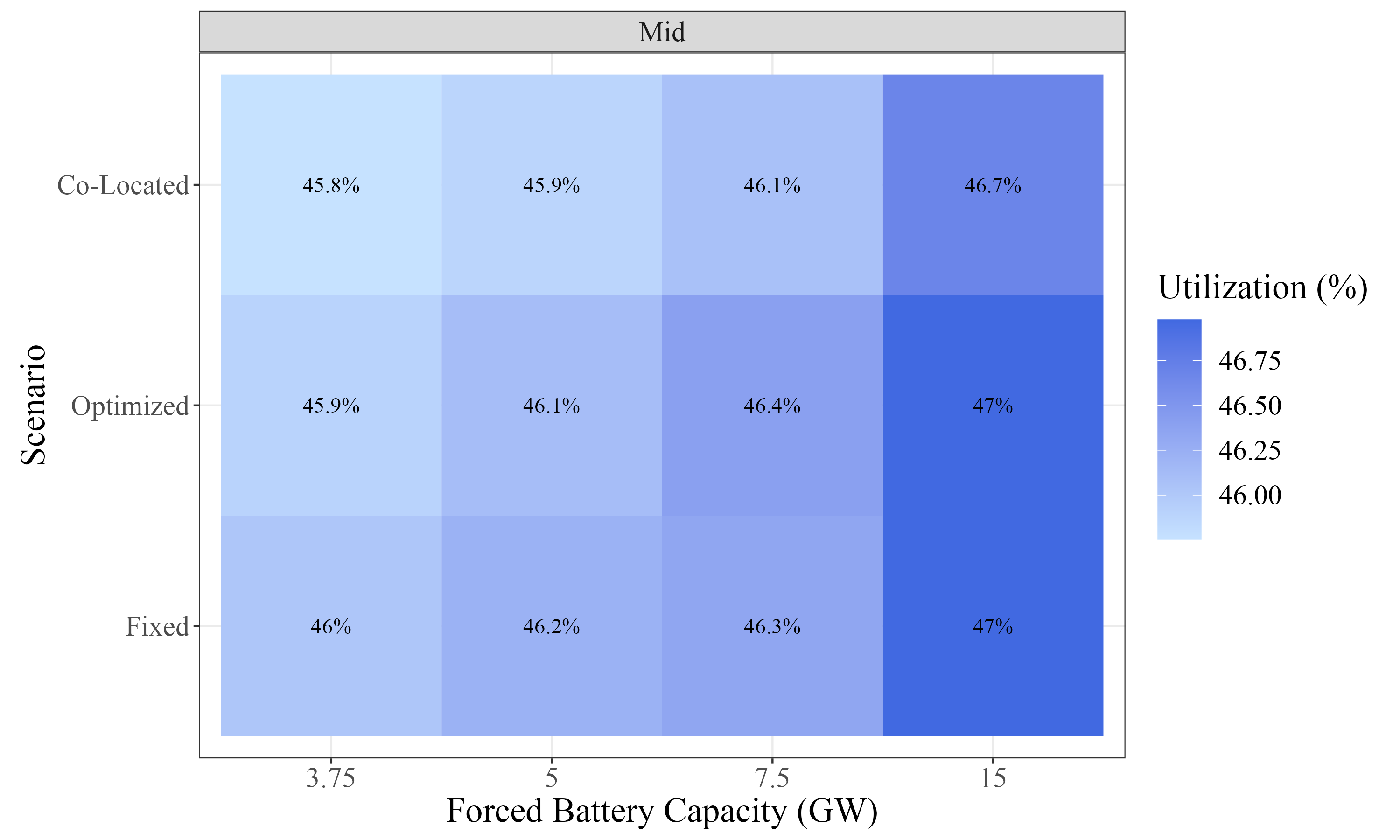}}
            \caption{\textbf{Utilization Rate of Inter-Zonal Transmission Line Heatmap for Mid VRE-Cost Scenario.
            }} \label{fig::utilratemid}
        \end{figure}

    \newpage

    \subsection{Cost Difference Plots}
    \label{cost_difference}
    
        \begin{figure}[H]
            \noindent
            \makebox[\textwidth]{\includegraphics[scale=0.4]{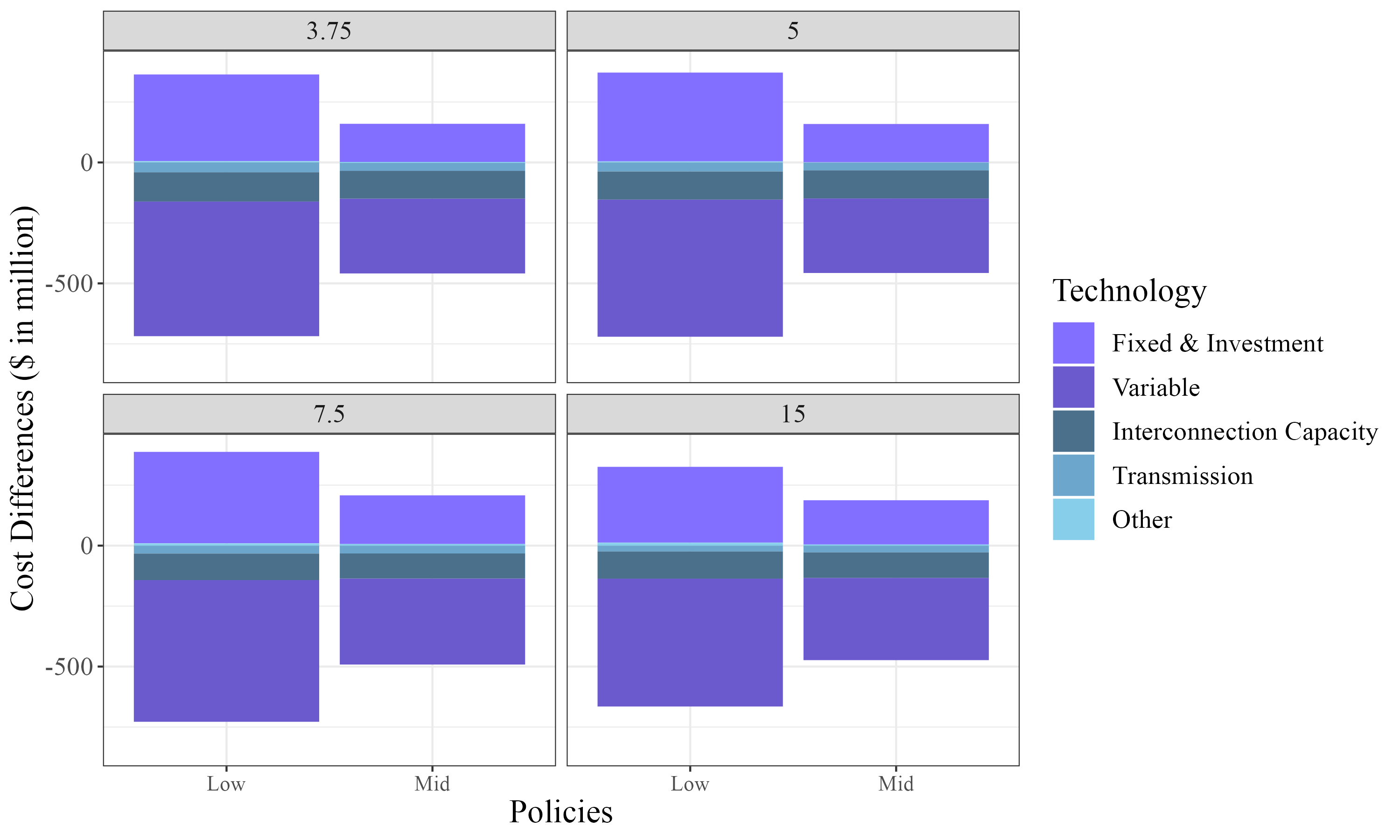}}
            \caption{\textbf{Cost Difference between Optimized Interconnection and Fixed Interconnection Scenario (\$ million).
            }} \label{fig::cost_optimized_difference}
        \end{figure}

        \begin{figure}[H]
            \noindent
            \makebox[\textwidth]{\includegraphics[scale=0.4]{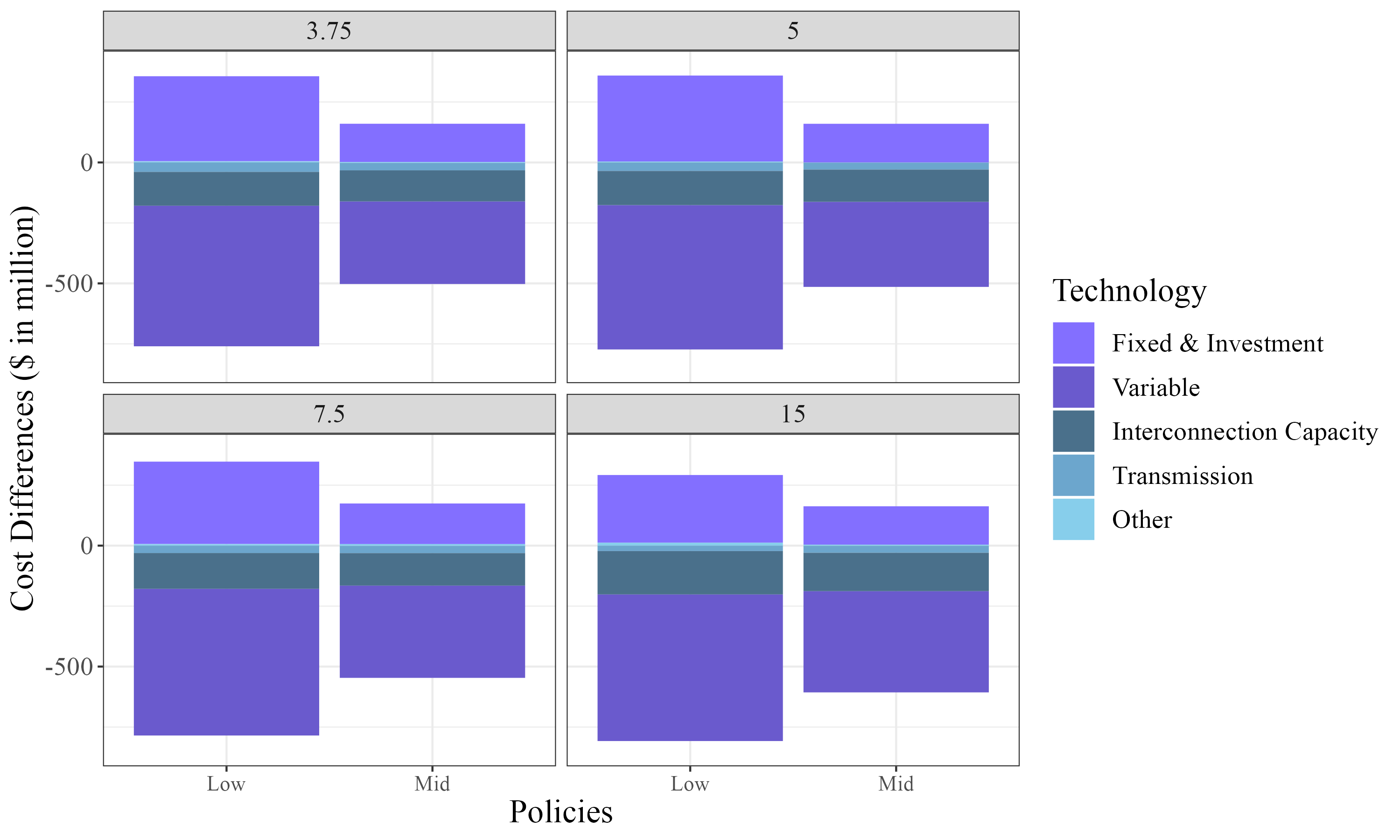}}
            \caption{\textbf{Cost Difference between Co-Located Storage and Fixed Interconnection Scenario (\$ million).
            }} \label{fig::cost_colocated_difference}
        \end{figure}

    \newpage
    
    \subsection{Power Generation Difference Plots}
    \label{power_difference}
    
        \begin{figure}[H]
            \noindent
            \makebox[\textwidth]{\includegraphics[scale=0.4]{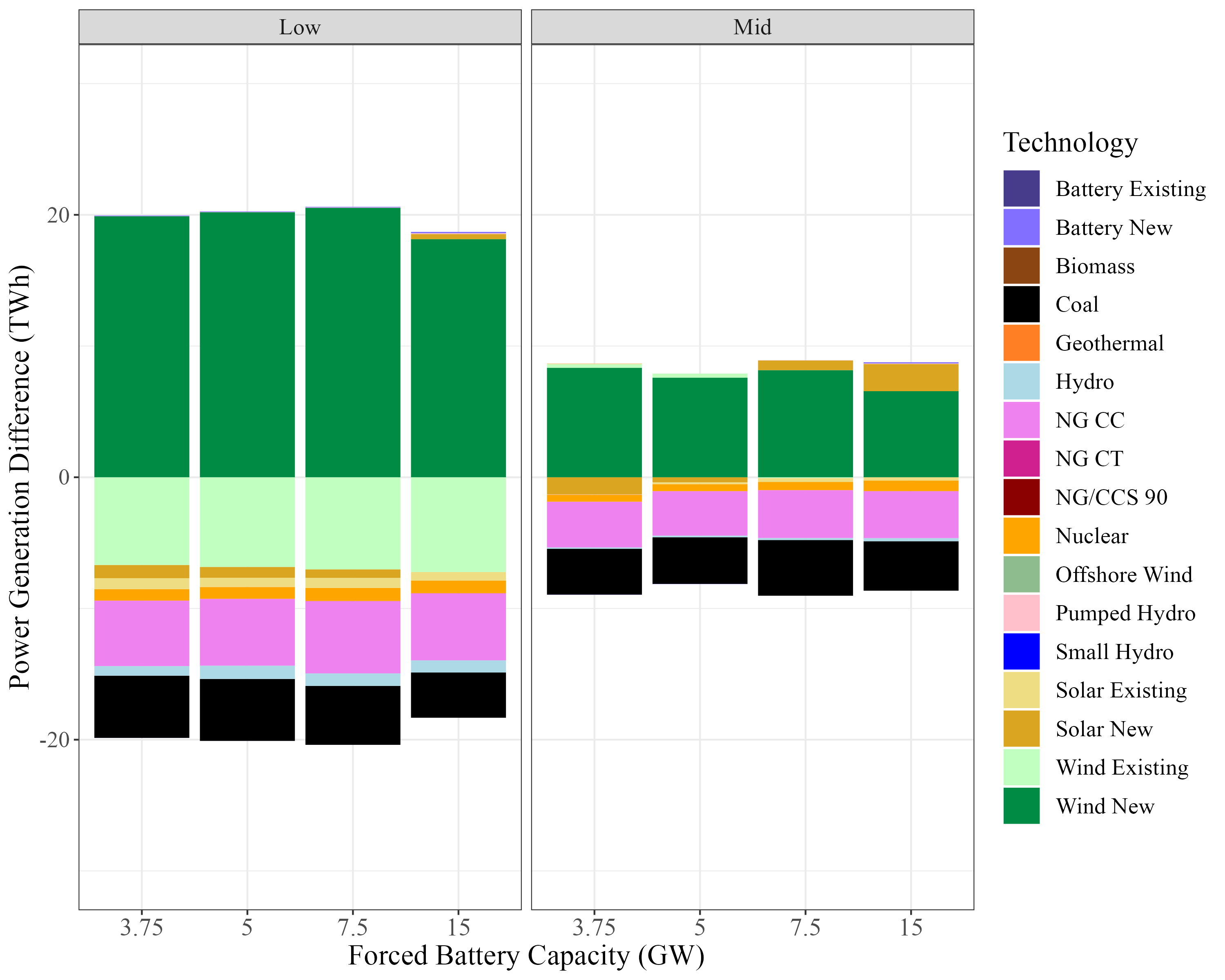}}
            \caption{\textbf{Power Generation Difference between Optimized Interconnection and Fixed Interconnection Scenario (TWh).
            }\\ For battery existing, battery new, and pumped hydro, the differences showcase storage losses (charge of each resource minus discharge of each resource).} \label{fig::power_optimized_difference}
        \end{figure}

        \begin{figure}[H]
            \noindent
            \makebox[\textwidth]{\includegraphics[scale=0.4]{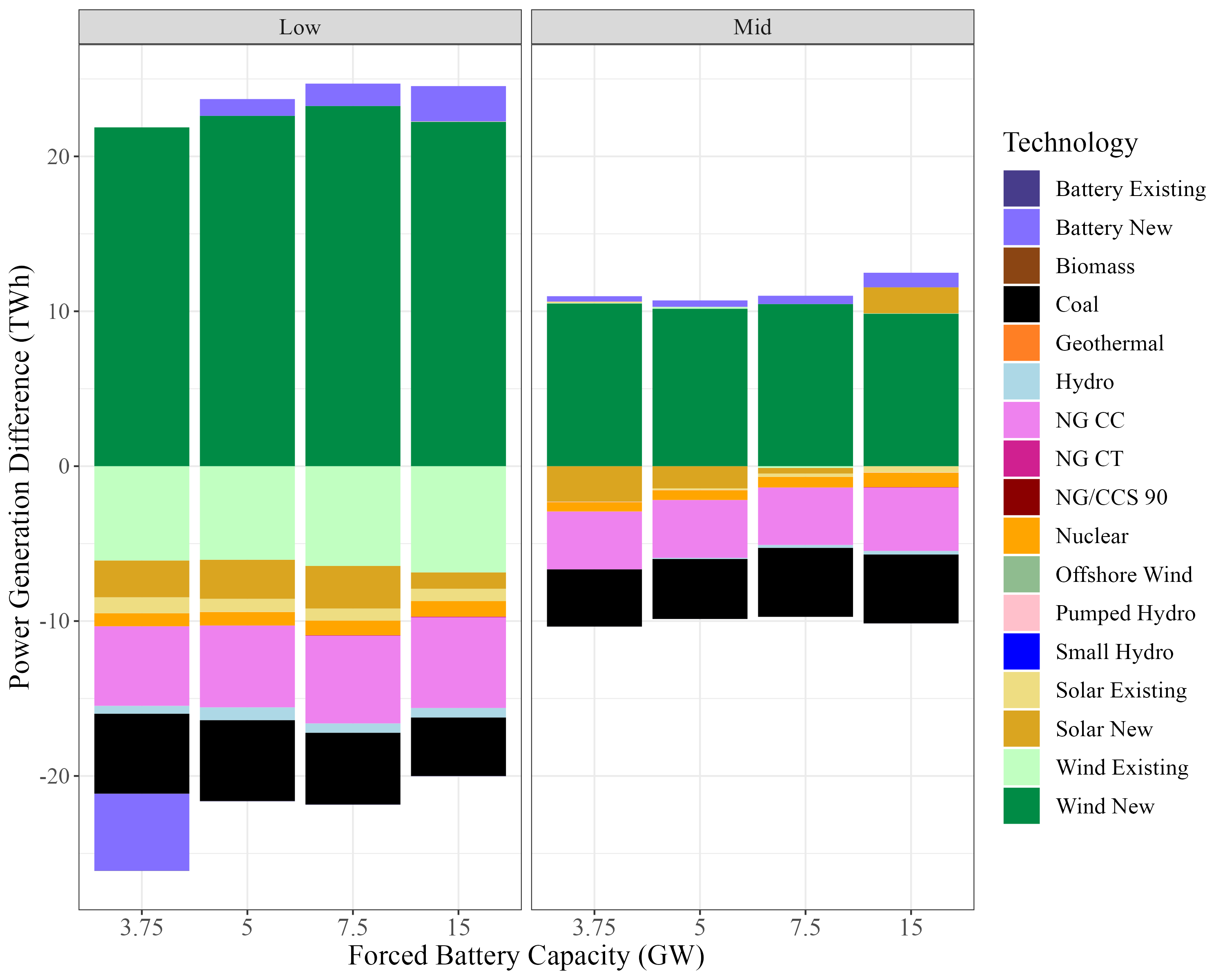}}
            \caption{\textbf{Power Generation Difference between Co-Located Storage and Fixed Interconnection Scenario (TWh).
            }\\For battery existing, battery new, and pumped hydro, the differences showcase storage losses (charge of each resource minus discharge of each resource).} \label{fig::power_colocated_difference}
        \end{figure}

    \newpage

    \subsection{Box Plots of Average Solar and Wind Sizing for the Co-located Storage Scenario}
    \label{box_plot_colocated}

        \begin{figure}[H]
            \noindent
            \makebox[\textwidth]{\includegraphics[scale=0.5]{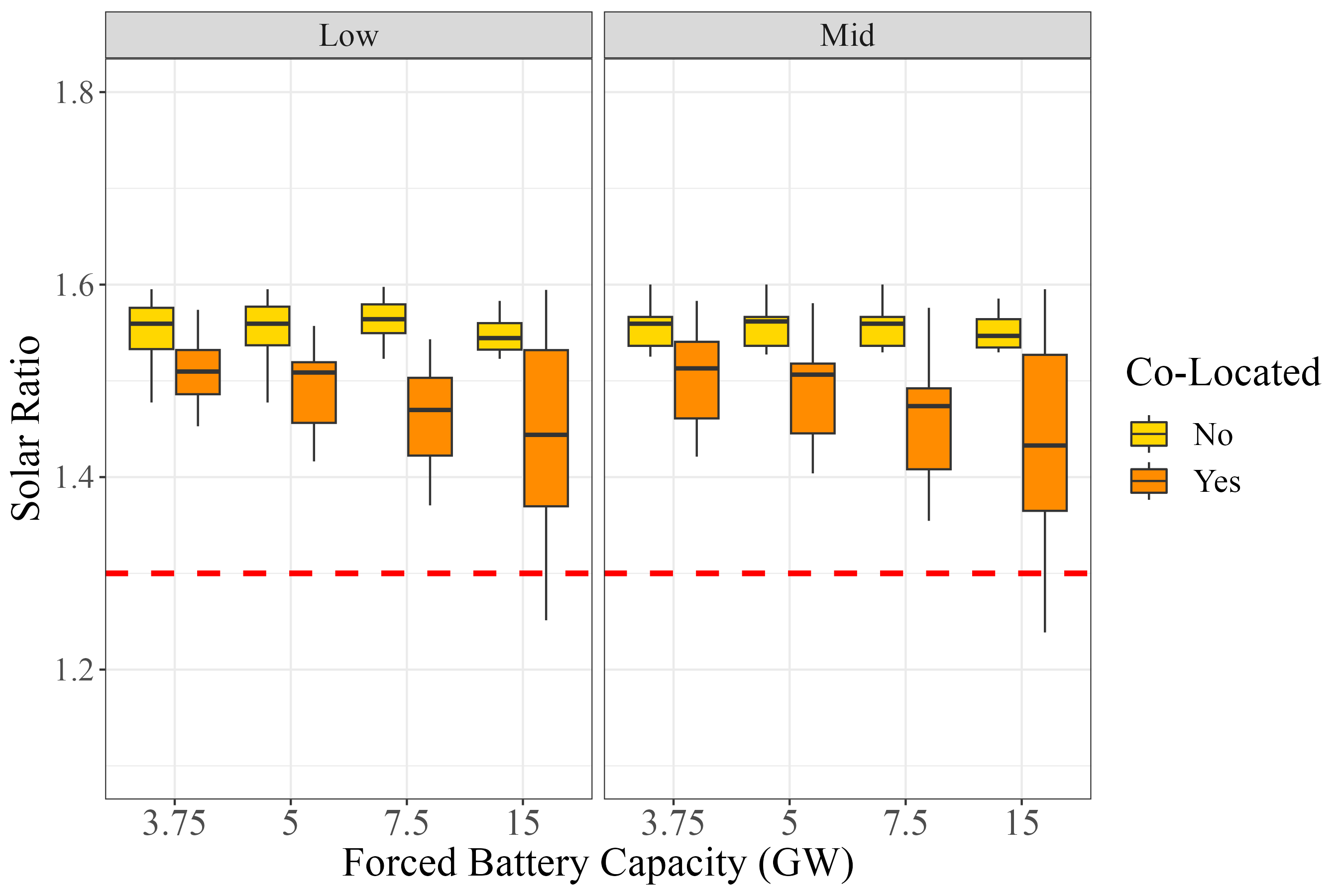}}
            \caption{\textbf{Box Plot of Solar PV Clusters' Average Ratio of VRE to Inverter Built for Co-located Storage Scenario.} \\ The red dashed line represents the fixed assumed ratio of solar PV (1.3) to inverter capacity that most capacity expansion models assume. `No' in the legend means the sites in this scenario were built as standalone solar PV resources, while `Yes' means the sites are co-located with batteries.}
            \label{fig:boxplot_colocated_solar}
        \end{figure}

        \begin{figure}[H]
            \noindent
            \makebox[\textwidth]{\includegraphics[scale=0.5]{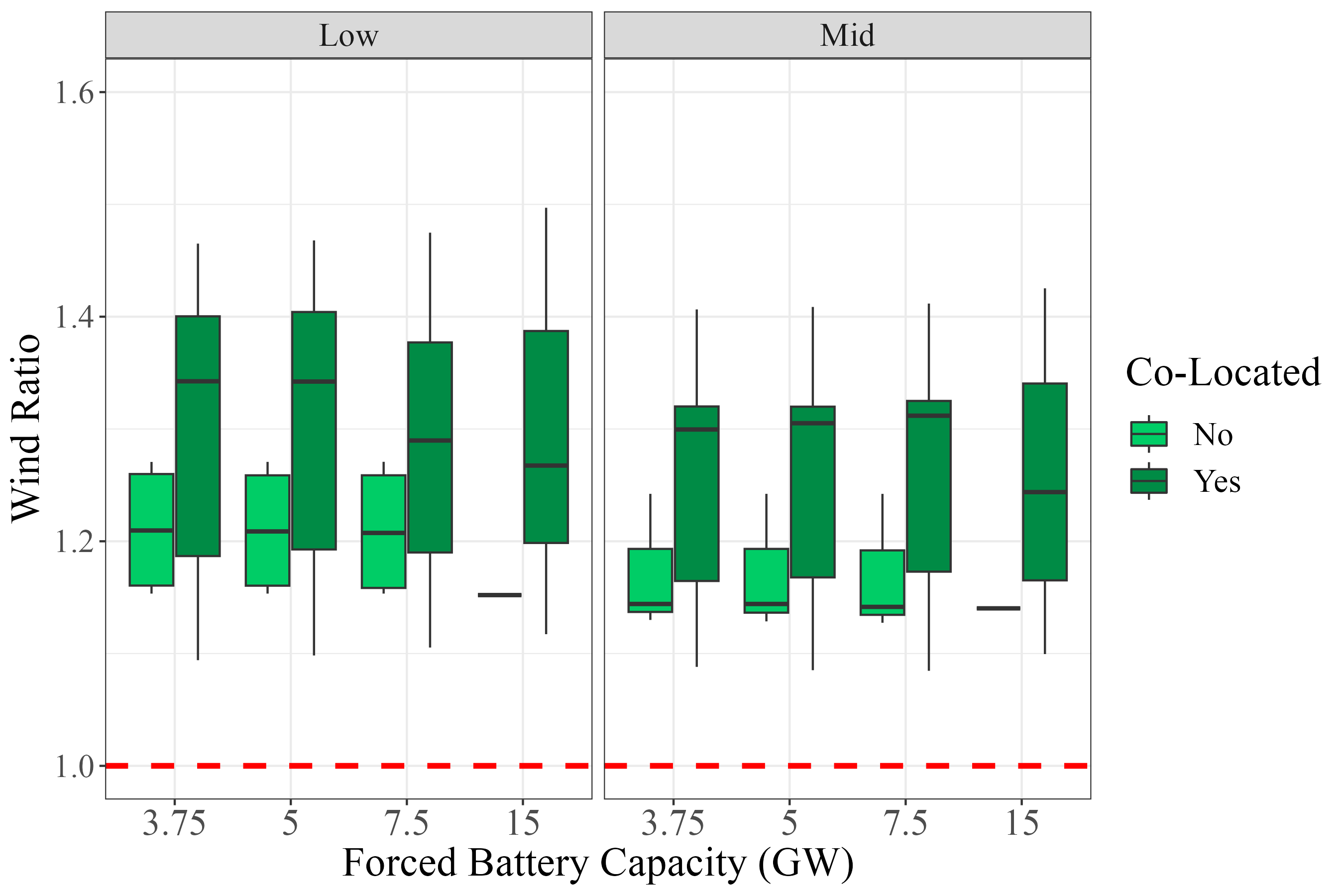}}
            \caption{\textbf{Box Plot of Wind Clusters' Average Ratio of VRE to Grid Connection Capacity Built for Co-located Storage Scenario.} \\ The red dashed line represents the fixed assumed ratio of wind (1.0) to grid connection capacity that most capacity expansion models assume. `No' in the legend means the sites in this scenario were built as standalone wind resources, while `Yes' means the sites are co-located with batteries.}
            \label{fig:boxplot_colocated_wind}
        \end{figure}

    \newpage 

    \subsection{Solar PV, Wind, Grid Connection, and Inverter Capacity Changes between Scenarios}
    \label{vre_stats}

        \begin{table}[H]
            \begin{center}
            \caption{\textbf{Percent Change for Various VRE, Grid Connection, and Inverter Components between the Co-Located Storage and Fixed Interconnection Scenario. }\\The breakdown is conducted for changes to the VRE capacity, VRE inverter, VRE grid connection, total system-wide inverter capacity, and total system-wide interconnection capacity (both GW and GW-km). Brackets indicate the percent change between the co-located storage and optimized interconnection scenario. Columns are labeled by the scenario of forced system-wide battery capacity (3.75, 5, 7.5, or 15 GW) and cost sensitivity (low- or mid-VRE cost).} 
            \label{tab:appcolocatestats}
            \resizebox{\textwidth}{!}{
            \begin{tabular}{l|c|c|c|c|c|c|c|c}
            \hline
            & \textbf{3.75, Low} & \textbf{3.75, Mid} 
            & \textbf{5, Low} & \textbf{5, Mid}
            & \textbf{7.5, Low} & \textbf{7.5, Mid} 
            & \textbf{15, Low} & \textbf{15, Mid}\\
            \hline
    
            PV Module & -1.1\% [-0.9\%] & -1.1\% [-0.7\%] & -1.2\% [-1.1\%] & -0.4\% [-0.7\%] & -1.2\% [-1.4\%] & 0.3\% [-0.8\%] & 0\% [-0.8\%] & 1.7\% [-0.3\%] \\
            PV Inverter & -16.7\% [0.1\%] & -16.3\% [1.0\%] & -16.1\% [0.7\%] & -15.1\% [1.8\%] & -14.8\% [2.1\%] & -12.8\% [3.6\%] & -10.0\% [7.2\%] & -7.3\% [9.3\%] \\
            PV Interconnection (GW) & -17.9\% [-1.3\%] & -18.1\% [-1.2\%] & -18.5\% [-2.1\%] & -17.9\% [-1.7\%] & -19.5\% [-3.5\%] & -18.3\% [-2.9\%] & -21.4\% [-6.4\%] & -20.5\% [-6.3\%] \\
            PV Interconnection (GW-km) & -16.8\% [-1.1\%] & -18.4\% [-1.4\%] & -17.0\% [-1.9\%] & -18.4\% [-2.0\%] & -17.5\% [-2.8\%] & -18.2\% [-2.9\%] & -17.9\% [-4.9\%] & -19.5\% [-5.5\%] \\
            Wind & 15.4\% [0.7\%] & 9.7\% [1.1\%] & 15.8\% [0.9\%] & 9.4\% [1.3\%] & 16.1\% [0.9\%] & 9.6\% [1.1\%] & 15.3\% [1.5\%] & 9.2\% [1.7\%] \\
            Wind Interconnection (GW) & -9.3\% [-0.8\%] & -10.3\% [-0.3\%] & -9.2\% [-1.1\%] & -10.6\% [-0.3\%] & -9.2\% [-1.5\%] & -10.5\% [-1.1\%] & -10.6\% [-2.7\%] & -11.5\% [-1.9\%] \\
            Wind Interconnection (GW-km) & -7.7\% [-0.6\%] & -8.8\% [0\%] & -7.4\% [-0.7\%] & -9.0\% [0.1\%] & -7.3\% [-1.1\%] & -8.6\% [-0.6\%] & -8.4\% [-2.0\%] & -10.0\% [-1.4\%] \\
            Total Inverter & -17.8\% [-2.4\%] & -18.7\% [-3.2\%] & -18.1\% [-3.2\%] & -18.7\% [-4.2\%] & -18.7\% [-4.5\%] & -18.8\% [-5.5\%] & -17.2\% [-4.2\%] & -17.1\% [-5.2\%] \\
            Total Interconnection (GW) & -16.7\% [-4.9\%] & -17.9\% [-5.0\%] & -17.9\% [-6.5\%] & -18.9\% [-6.6\%] & -20.2\% [-9.5\%] & -21.0\% [-9.9\%] & -25.2\% [-15.5\%] & -25.7\% [-16.0\%] \\
            Total Interconnection (GW-km) & -8.8\% [-1.2\%] & -10.2\% [-0.8\%] & -8.8\% [-1.5\%] & -10.6\% [-1.0\%] & -9.1\% [-2.3\%] & -10.7\% [-2.2\%] & -10.9\% [-4.1\%] & -12.9\% [-4.2\%] \\
            
            \hline
            \end{tabular}}
            \end{center}
        \end{table}

    \newpage

    \subsection{Marginal Change of Resource Capacities}
    \label{marginal_cap_value}

        \begin{figure}[H]
                \noindent
                \makebox[\textwidth]{\includegraphics[scale=0.4]{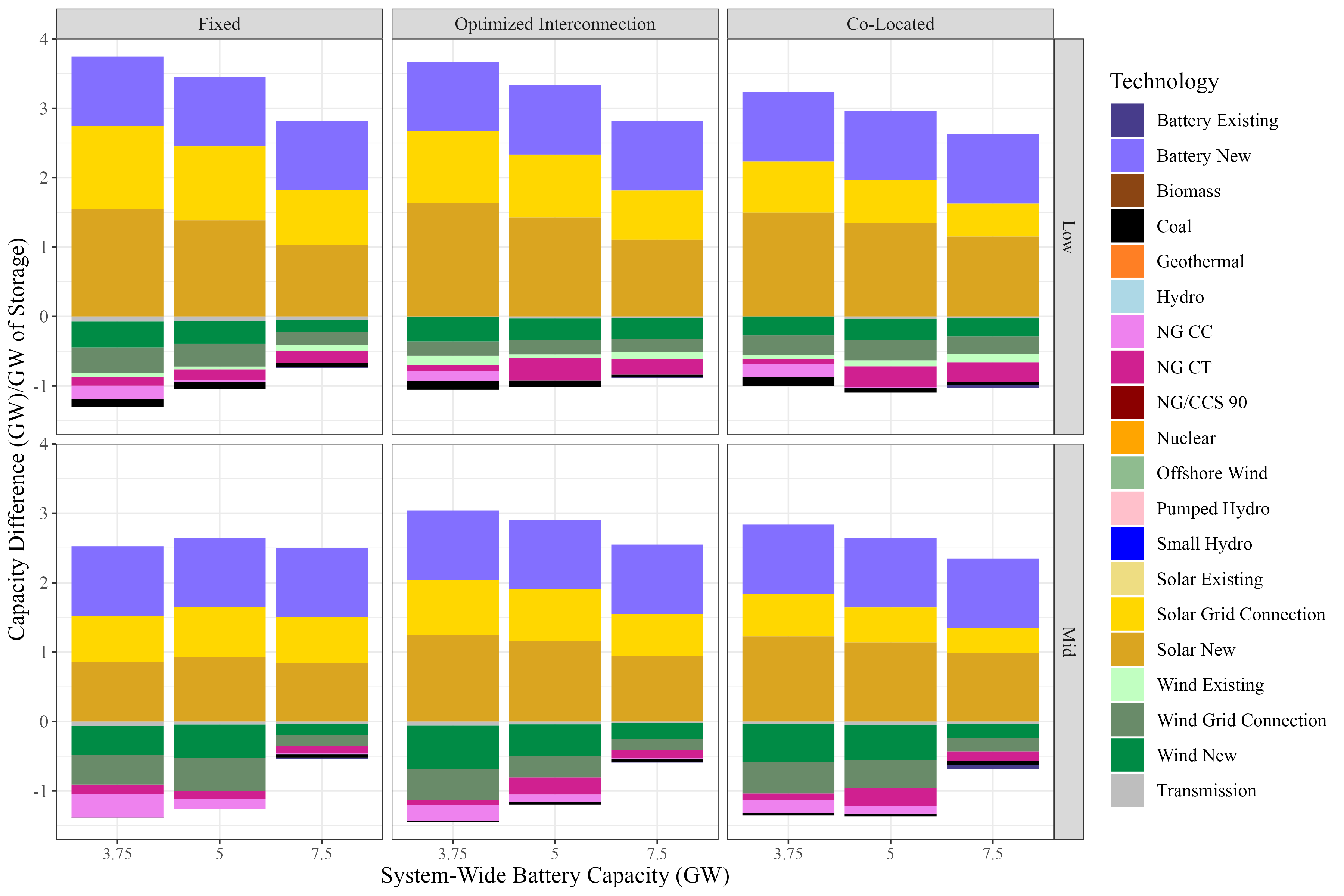}}
                \caption{\textbf{Marginal Change of Resource Capacities between Various Storage Penetrations in the System.} \\ Values are calculated for each column by subtracting the differences in capacities between each scenario and dividing by the GW increase in battery capacity between the two scenarios. For example, the 3.75 column was calculated as the capacity difference for each resource between the 5 and 3.75 GW scenarios, divided by 1.25 GW.}
                \label{fig:marginalsubstitution}
        \end{figure}
    
    \subsection{Scatter Plot of Battery Energy Capacity Built for Each Co-located Wind and Solar Site vs. Interconnection Costs}
    \label{interconnection_colocation}

        \begin{figure}[H]
            \noindent
            \makebox[\textwidth]{\includegraphics[scale=0.4]{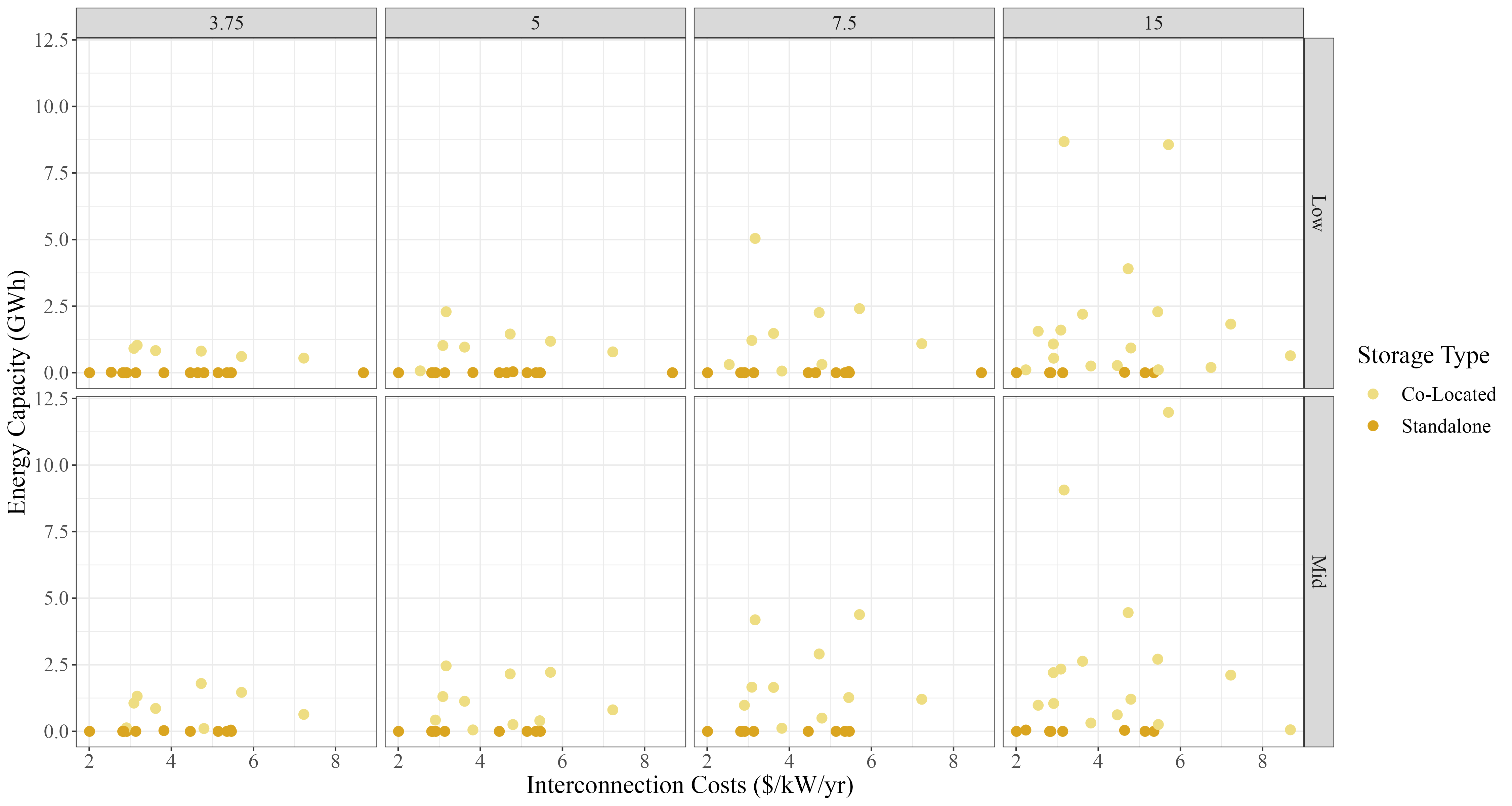}}
            \caption{\textbf{Energy Capacity of Co-located Storage with Solar Resources (GWh) across Interconnection Capacity Costs (\$/kW/yr).}}
            \label{fig:energy_interconnection_solar}
        \end{figure}

        \begin{figure}[H]
            \noindent
            \makebox[\textwidth]{\includegraphics[scale=0.4]{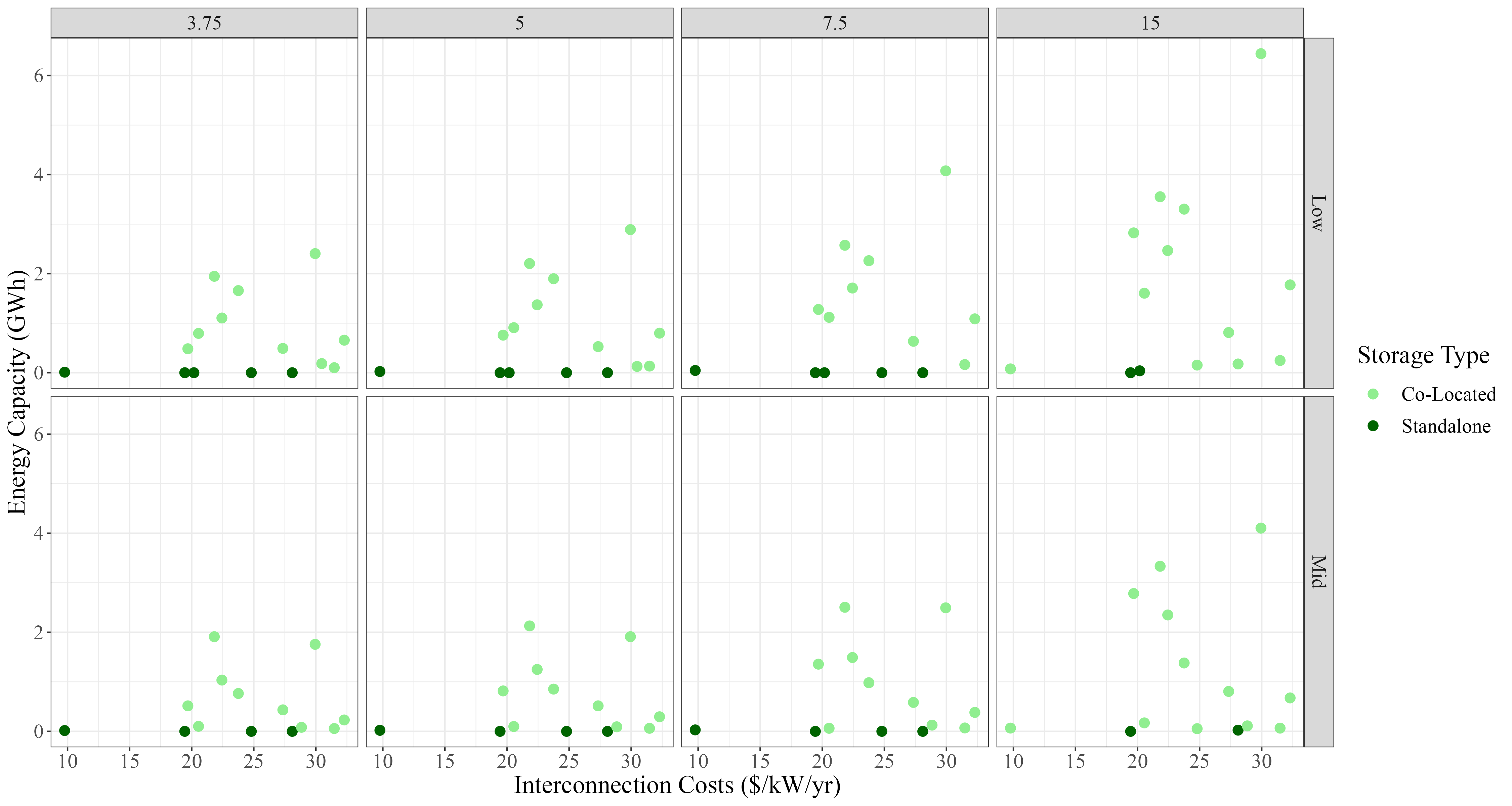}}
            \caption{\textbf{Energy Capacity of Co-located Storage with Wind Resources (GWh) across Interconnection Capacity Costs (\$/kW/yr).}}
            \label{fig:energy_interconnection_wind}
        \end{figure}

    \newpage
    
    \subsection{Patterns of Charging and Discharging of the Largest Co-Located PV and Wind Sites}
    \label{colocated_charging_discharging}

        \begin{figure}[H]
            \noindent
            \makebox[\textwidth]{\includegraphics[scale=0.3]{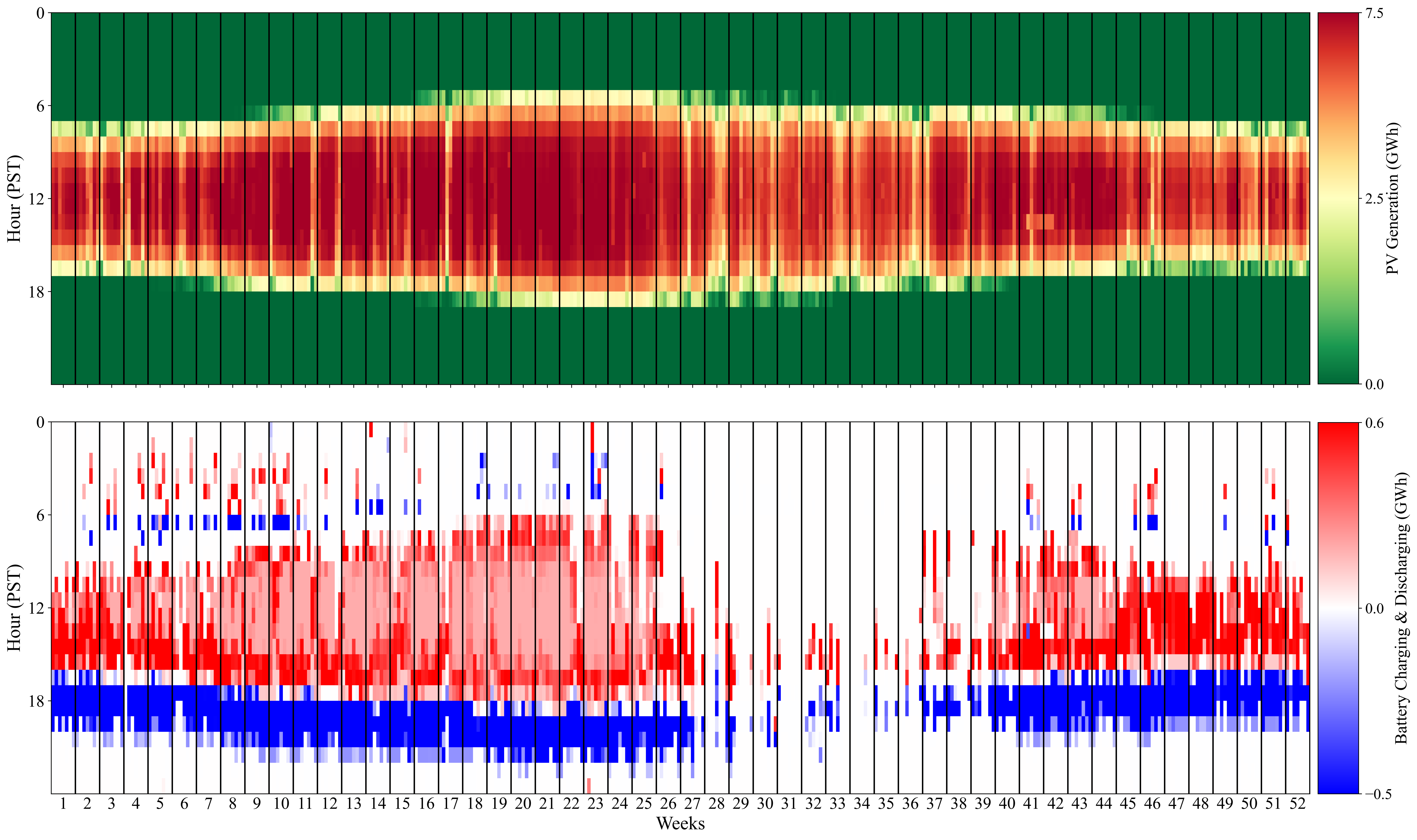}}
            \caption{\textbf{One of the Largest Sites in Western Interconnection (located in Arizona) PV Generation and Battery Charging/Discharging.}\\
            The two plots showcase the heat maps of the 5 GW of forced system-wide battery capacity, low-cost, co-located storage scenario of (a) Solar PV generation of one of the largest solar PV-battery clusters in the Western Interconnection (GWh), and (b) The co-located battery charging (+0.6) and discharging (-0.5) at one of the largest solar PV-battery clusters (GWh). For Arizona's largest cluster of a co-located solar PV-battery resource built, 11.7 GW of PV DC, 7.7 GW of inverter capacity, 7.3 GW of interconnection capacity, and 2.3 GWh of energy storage are built. The bottom plot of battery charging and discharging is normalized between -0.5 to 0.6, where the blue region showcases the discharging of the battery and the red region showcases the charging of the battery.}
            \label{fig:wecc_az}
        \end{figure}

        \begin{figure}[H]
            \noindent
            \makebox[\textwidth]{\includegraphics[scale=0.3]{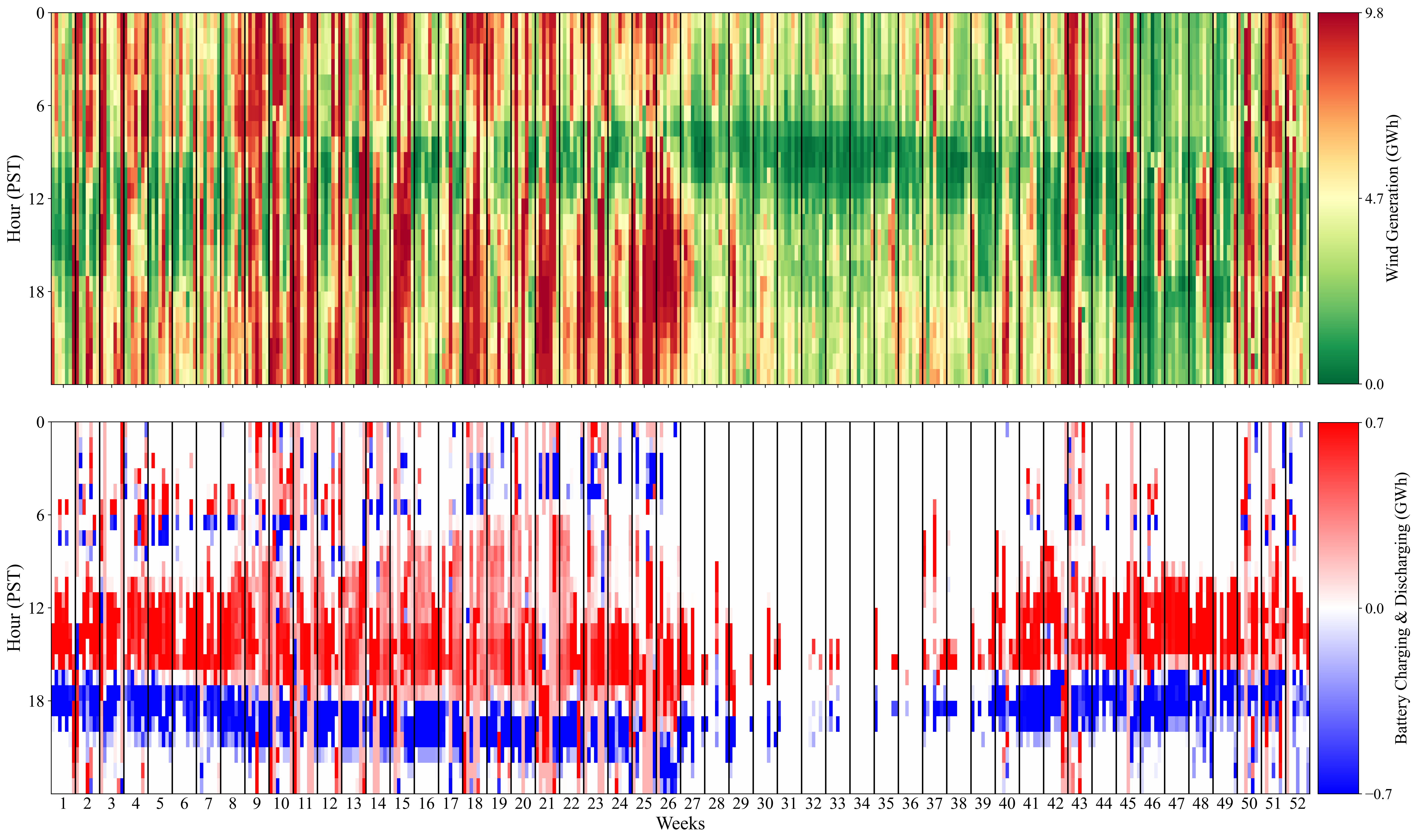}}
            \caption{\textbf{One of the Largest Sites in Western Interconnection (located in Southern Nevada) Wind Generation and Battery Charging/Discharging.}\\
            The two plots showcase the heat maps of the 5 GW of forced system-wide battery capacity, low-cost, co-located storage scenario of (a) Wind generation of one of the largest wind-battery clusters in the Western Interconnection (GWh), and (b) The co-located battery charging (+0.7) and discharging (-0.7) at one of the largest wind-battery clusters (GWh). For Nevada's largest cluster of a co-located wind-battery resource built, 13.1 GW of wind, 0.7 GW of inverter capacity, 9.1 GW of interconnection capacity, and 2.9 GWh of energy storage are built. The bottom plot of battery charging and discharging is normalized between -0.7 to 0.7, where the blue region showcases the discharging of the battery and the red region showcases the charging of the battery.}
            \label{fig:wecc_snv}
        \end{figure}

%%       \citet{<label>} ==> Jones et al. [21]
%%       \citep{<label>} ==> [21]
%%


\newpage
\bibliographystyle{elsarticle-num} 
\bibliography{citations}


%\begin{thebibliography}{00}



%\end{thebibliography}
\end{document}